\documentclass[12pt,twoside,english]{extarticle}
\usepackage{lmodern}
\usepackage{helvet}
\usepackage[T1]{fontenc}
\usepackage[latin9]{inputenc}
\usepackage{geometry}
\usepackage{color}
\usepackage{babel}
\usepackage{float}
\usepackage{mathrsfs}
\usepackage{amsmath}
\usepackage{amsthm}
\usepackage{amssymb}
\usepackage{setspace}
\usepackage[unicode=true,pdfusetitle,
 bookmarks=true,bookmarksnumbered=false,bookmarksopen=false,
 breaklinks=true,pdfborder={0 0 0},pdfborderstyle={},backref=false,colorlinks=true]
 {hyperref}

\makeatletter

\providecommand{\tabularnewline}{\\}

\numberwithin{equation}{section}
\numberwithin{table}{section}
\numberwithin{figure}{section}
\usepackage[natbibapa]{apacite}
\theoremstyle{definition}
\newtheorem{defn}{\protect\definitionname}[section]
\theoremstyle{plain}
\newtheorem{assumption}{\protect\assumptionname}
\theoremstyle{plain}
\newtheorem{thm}{\protect\theoremname}[section]
\theoremstyle{plain}
\newtheorem{lem}{\protect\lemmaname}[section]

\usepackage{graphicx}
\usepackage{amsmath}
\usepackage{dcolumn}
\usepackage{listings}
\usepackage{color}
\usepackage{setspace}
\usepackage[normalem]{ulem}  
\definecolor{hellgelb}{rgb}{1,1,0.8}
\definecolor{colKeys}{rgb}{0,0,1}
\definecolor{colIdentifier}{rgb}{0,0,0}
\definecolor{colComments}{rgb}{1,0,0}
\definecolor{colString}{rgb}{0,0.5,0}

\usepackage{lmodern}
\usepackage[T1]{fontenc}
\usepackage{geometry}
\usepackage{fancyhdr}
\usepackage{amsthm}
\usepackage{thmtools}
\usepackage{amsmath}
\usepackage{amssymb}
\usepackage{esint}
\usepackage{multirow}
\usepackage{mathrsfs} 
\usepackage{textgreek}
\usepackage{amsfonts}
\usepackage{amssymb}
\usepackage{mathrsfs} 

\usepackage[USenglish]{isodate}
\usepackage{chngcntr}

\numberwithin{equation}{section}
\numberwithin{table}{section}
\numberwithin{assumption}{section}
\counterwithout{figure}{section}
\counterwithout{table}{section}

\makeatother

  \providecommand{\assumptionname}{Assumption}

  \providecommand{\definitionname}{Definition}

  \providecommand{\lemmaname}{Lemma}

  \providecommand{\theoremname}{Theorem}
 
 \providecommand{\theoremname}{Theorem}
 
\usepackage{thmtools}

\newtheoremstyle{MyTheoremstyle}
  {\topsep} 
  {\topsep} 
  {} 
  {} 
  {\bfseries} 
  {.} 
  {.90em} 
  {} 
\theoremstyle{MyTheoremstyle} 
\theoremstyle{MyTheoremstyle} 
\theoremstyle{MyTheoremstyle} 
\theoremstyle{MyTheoremstyle} 
\theoremstyle{MyTheoremstyle}

\declaretheoremstyle[
    headfont=\bfseries,
    notefont=\normalfont,
    bodyfont=\itshape,
    headpunct=\newline,
    headformat={%
        \makebox{\NAME\ \NUMBER\ }{\NOTE}%
    },
]{theorem}

\newlength{\spacelength}
\declaretheoremstyle[
    headfont=\bfseries,
    notefont=\normalfont,
    bodyfont=\itshape,
    headpunct=\newline,
    headformat={%
        \makebox[0pt][l]{\NAME\ \NUMBER\ }\hskip-\spacelength{\NOTE}%
    },
]{theore}

 \usepackage{fancyhdr}
\usepackage{booktabs}
\usepackage{float}
\usepackage{graphicx}
\usepackage{epstopdf}
\usepackage{morefloats}
\usepackage[referable]{threeparttablex}
\usepackage{footnote}

\usepackage{setspace} 
\usepackage{graphicx,color}

\usepackage{listings}
\RequirePackage[T1]{fontenc}
\RequirePackage{ae,fancyvrb}
\DefineVerbatimEnvironment{Sinput}{Verbatim}{fontshape=sl}
\DefineVerbatimEnvironment{Soutput}{Verbatim}{}
\DefineVerbatimEnvironment{Scode}{Verbatim}{fontshape=sl}

\title{\bf Simultaneous Bandwidths Determination for DK-HAC Estimators and Long-Run Variance Estimation in Nonparametric Settings}
\author{
\textsc{\textcolor{MyBlue}{Federico Belotti}}\thanks{Dep. of Economics and Finance, University of Rome Tor Vergata, Via Columbia 2, Rome 00133, IT. 
Email: 
\texttt{\textcolor{MyBlue}{{federico.belotti@uniroma2.it}}}.} 
\\
\small{University of Rome Tor Vergata}
\and
\textsc{\textcolor{MyBlue}{Alessandro Casini}}\thanks{Corresponding author at: Dep. of Economics and Finance, University of Rome Tor Vergata, Via Columbia 2, Rome 00133, IT. 
Email: 
\texttt{\textcolor{MyBlue}{{alessandro.casini@uniroma2.it}}}.} 
\\
\small{University of Rome Tor Vergata}
\and
\textsc{\textcolor{MyBlue}{Leopoldo Catania}}\thanks{Dep. of Economics and Business Economics, Aarhus University, Fuglesangs Allé 4, 8210 Aarhus V, DE. 
Email: 
\texttt{\textcolor{MyBlue}{{leopoldo.catania@econ.au.dk}}}.} 
\\
\small{Aarhus University}
\and
\textsc{\textcolor{MyBlue}{Stefano Grassi}}\thanks{Dep. of Economics and Finance, University of Rome Tor Vergata, Via Columbia 2, Rome 00133, IT. 
Email: 
\texttt{\textcolor{MyBlue}{{stefano.grassi@uniroma2.it}}}.} 
\\
\small{University of Rome Tor Vergata}
\and
\textsc{\textcolor{MyBlue}{Pierre Perron}}\thanks{Dep. of Economics, Boston University, 270 Bay State Road, Boston, MA 02215, US. 
Email: 
\texttt{\textcolor{MyBlue}{\mbox{perron@bu.edu}}}.} 
\\
\small{Boston University}
}

 \makeatletter

\numberwithin{equation}{section}
\makeatother

\usepackage{babel}
\addto\captionsenglish{%
}

\usepackage{color}
\usepackage{xcolor}

\renewcommand*{\thesection}{\arabic{section}}

\definecolor{MyRed}{rgb}{0.8,0,0}
\definecolor{MyBlue}{rgb}{0,0,0.7}
\definecolor{Green}{rgb}{0,0.5,0}
\definecolor{hellgelb}{rgb}{1,1,0.8}
\definecolor{colKeys}{rgb}{0,0,1}
\definecolor{colIdentifier}{rgb}{0,0,0}
\definecolor{colComments}{rgb}{1,0,0}
\definecolor{colString}{rgb}{0,0.5,0}

\definecolor{MyLightRed}{rgb}{2.2,0.2,0.4} 
\definecolor{MyLightRed2}{rgb}{0.6,0.2,0.3} 
\definecolor{MyLightRed2temp}{rgb}{0.6,0.2,0.3}
\definecolor{MyLightRed3}{rgb}{0.8,0.1,0.1} 
\definecolor{MyRed}{rgb}{0.7,0.0,0}

\definecolor{MyLigthBlue13}{rgb}{0,0.2,0.7}
 \definecolor{MyLigthBlack}{rgb}{0.2,0.25,0.3} 

\hypersetup{%
  bookmarks=true
  pdftitle = {Title Here},
  pdfsubject = {},
  pdfkeywords = {Keywords},
  pdfauthor = {Alessandro Casini},}
\hypersetup{ 
  colorlinks = {true}, 
  citecolor = {MyBlue},
  linkcolor = {MyRed},
  filecolor= {Green},	
  urlcolor = {MyBlue},
  hyperindex = {true},
  linktocpage = {true},
  linkbordercolor ={1 0 0},
 citebordercolor ={0 1 1},
 urlbordercolor ={0 1 1},
hypertexnames=false
}

\usepackage{bookmark}

\usepackage[bottom]{footmisc}
\makeatother

\providecommand{\assumptionname}{Assumption}
\providecommand{\definitionname}{Definition}
\providecommand{\lemmaname}{Lemma}
\providecommand{\theoremname}{Theorem}

\begin{document}
\pagebreak{}

\thispagestyle{empty}
\setcounter{page}{0}
\raggedbottom
\title{\textbf{\Large{}Simultaneous Bandwidths Determination for DK-HAC Estimators
and Long-Run Variance Estimation in Nonparametric Settings}\textbf{}\thanks{We thank Zhongjun Qu for useful comments.}}
\maketitle
\begin{abstract}
{\footnotesize{}We consider the derivation of data-dependent simultaneous
bandwidths for double kernel heteroskedasticity and autocorrelation
consistent (DK-HAC) estimators. In addition to the usual smoothing
over lagged autocovariances for classical HAC estimators, the DK-HAC
estimator also applies smoothing over the time direction. We obtain
the optimal bandwidths that jointly minimize the global asymptotic
MSE criterion and discuss the trade-off between bias and variance
with respect to smoothing over lagged autocovariances and over time.
Unlike the MSE results of \citet{andrews:91}, we establish how nonstationarity
affects the bias-variance trade-off. We use the plug-in approach to
construct data-dependent bandwidths for the DK-HAC estimators and
compare them with the DK-HAC estimators from \citet{casini_hac} that
use data-dependent bandwidths obtained from a sequential MSE criterion.
The former performs better in terms of size control, especially with
stationary and close to stationary data. Finally, we consider long-run
variance estimation under the assumption that the series is a function
of a nonparametric estimator rather than of a semiparametric estimator
that enjoys the usual $\sqrt{T}$ rate of convergence. Thus, we also
establish the validity of consistent long-run variance estimation
in nonparametric parameter estimation settings.}{\footnotesize\par}
\end{abstract}
{\footnotesize{\indent {\bf{JEL Classification}}: C12, C13, C18, C22, C32, C51}}  

\onehalfspacing
\thispagestyle{empty}

\pagebreak{}

\section{Introduction }

Long-run variance (LRV) estimation has a long history in econometrics
and statistics since it plays a key role for heteroskedasticity and
autocorrelation robust (HAR) inference. The classical approach in
HAR inference relies on consistent estimation of the LRV. \citet{newey/west:87}
and \citet{andrews:91} proposed kernel heteroskedasticity and autocorrelation
consistent (HAC) estimators and showed their consistency. However,
recent work by \citet{casini_hac} showed that, both in the linear
regression model and other contexts, their results do not provide
accurate approximations in that test statistics normalized by classical
HAC estimators may exhibit  size distortions and substantial power
losses. Issues with the power have been shown for a variety of HAR
testing problems outside the regression model {[}e.g., \citet{altissimo/corradi:2003},
Casini (\citeyear{casini_CR_Test_Inst_Forecast}), Casini and Perron
(\citeyear{casini/perron_Oxford_Survey}, \citeyear{casini/perron_Lap_CR_Single_Inf},
\citeyear{casini/perron_SC_BP_Lap}), \citet{chan:2020}, \citet{chang/perron:18},
\citet{crainiceanu/vogelsang:07}, \citet{deng/perron:06}, \citet{juhl/xiao:09},
\citet{kim/perron:09}, \citet{martins/perron:16}, \citet{perron/yamamoto:18}
and \citet{vogeslang:99}{]}. \citet{casini/perron_Low_Frequency_Contam_Nonstat:2020}
showed theoretically that such power issues are generated by low frequency
contamination induced by nonstationarity. More specifically, nonstationarity
biases upward each sample autocovariance. Thus, LRV estimators are
inflated and HAR test statistics lose power. These issues can also
be provoked by misspecification, nonstationary alternative hypotheses
and outliers. They also showed that LRV estimators that rely on fixed-$b$
or versions thereof suffer more from these problems than classical
HAC estimators since the former use a larger number of sample autocovariances.\footnote{The fixed-$b$ literature is extensive. Pioneering contribution of
\citet{Kiefer/vogelsang/bunzel:00} and Kiefer and Vogelsang (\citeyear{Kiefer/vogelsang:02};
\citeyear{kiefer/vogelsang:05}) introduced the fixed-$b$ LRV estimators.
Additional contributions can be found in \citet{dou:18}, \citet{lazarus/lewis/stock:17},
\citet{lazarus/lewis/stock/watson:18}, \citet{Goncalves/vogelsang:11},
\citet{dejong/davidson:00}, \citet{ibragimov/muller:10}, \citet{jansson:04},
\citeauthor{muller:07} (2007, 2014)\nocite{mueller:14}, \citet{phillips:05},
\citet{politis:11}, \citet{preinerstorfer/potscher:16}, \citeauthor{potscher/preinerstorfer:18}
(2018; 2019)\nocite{potscher/preinerstorfer:19}, \citet{robinson:98},
Sun \citeyearpar{sun:13,sun:14,sun:14a}, \citet{velasco/robinson:01}
and \citet{zhang/shao:13}. }

In order to flexibly account for nonstationarity, \citet{casini_hac}
introduced a double kernel HAC (DK-HAC) estimator that applies kernel
smoothing over two directions. In addition to the usual smoothing
over lagged autocovariances used in classical HAC estimators, the
DK-HAC estimator uses a second kernel that applies smoothing over
time. The latter accounts for time variation in the covariance structure
of time series which is a relevant feature in economics and finance.
Since the DK-HAC uses two kernels and bandwidths, one cannot rely
on the theory of \citet{andrews:91} or \citet{newey/west:94} for
selecting the bandwidths. \citet{casini_hac} considered a sequential
MSE criterion that determines the optimal bandwidth controlling the
number of lags as a function of the optimal bandwidth controlling
the smoothing over time. Thus, the latter influences the former but
not viceversa. However, each smoothing affects the bias-variance trade-off
so that the two bandwidths should affect each others optimal value.
Consequently, it is useful to consider an alternative criterion to
select the bandwidths. In this paper, we consider simultaneous bandwidths
determination obtained by jointly minimizing the asymptotic MSE of
the DK-HAC estimator. We obtain the asymptotic optimal formula for
the two bandwidths and use the plug-in approach to replace unknown
quantities by consistent estimates. Our results are established under
the nonstationary framework characterized by segmented locally stationary
processes {[}cf. \citet{casini_hac}{]}. The latter extends the locally
stationary framework of \citet{dahlhaus:96} to allow for discontinuities
in the spectrum. Thus, the class of segmented locally stationary processes
includes structural break models {[}see e.g., \citet{bai/perron:98}
and \citet{casini/perron_CR_Single_Break}{]}, time-varying parameter
models {[}see e.g., \citet{cai:07}{]} and regime switching {[}cf.
\citet{hamilton:89}{]}. 

We establish the consistency, rate of convergence and asymptotic MSE
results for the DK-HAC estimators with data-dependent simultaneous
bandwidths. The optimal bandwidths have the same order $O(T^{-1/6})$
whereas under the sequential criterion the optimal bandwidths smoothing
over time has an order $O(T^{-1/5})$ and the optimal bandwidth smoothing
the lagged autocovariances has an order $O(T^{-4/25})$. Thus, asymptotically,
the joint MSE criterion implies the use of (marginally) more lagged
autocovariances and a longer segment length for the smoothing over
time relative to the sequential criterion. Hence, the former should
control more accurately the variance due to nonstationarity while
the latter should control better the bias. If the degree of nonstationarity
is high then the theory suggests that one should expect the sequential
criterion to perform marginally better. The difference in the smoothing
over lags is very minor between the order of the corresponding bandwidths
implied by the two criteria. Our simulation analysis supports this
view as we show that the joint MSE criterion performs better especially
when the degree of nonstationarity is not too high. 

Overall, we find that HAR tests normalized by DK-HAC estimators strike
the best balance between size and power among the existing LRV estimators
and we also find that using the bandwidths selected from the joint
MSE criterion yields tests that perform better than the sequential
criterion in terms of size control. The optimal rate $O(T^{-1/6})$
is also found by \citet{neumann/von_sachs:1997} and \citet{dahlhaus:12}
in the context of local spectral density estimates under local stationarity.
Under both sequential and joint MSE criterion the optimal kernels
are found to be the same, i.e., the quadratic spectral kernel for
smoothing over autocovariance lags {[}similar to \citet{andrews:91}{]}
and a parabolic kernel {[}cf. \citet{epanechnikov:69}{]} for smoothing
over time. 

Another contribution of the paper is develop asymptotic results for
consistent LRV estimation in nonparametric parameter estimation settings.
\citet{newey/west:87} and \citet{andrews:91} established the consistency
of HAC estimators for the long-run variance of some series $\{V_{t}(\widehat{\beta})\}$
where $\widehat{\beta}$ is a semiparametric estimator of $\beta_{0}$
having the usual parametric rate of convergence $\sqrt{T}$ {[}i.e.,
they assumed that $\sqrt{T}(\widehat{\beta}-\beta_{0})=O_{\mathbb{P}}\left(1\right)${]}.
For example, in the linear regression model estimated by least-squares,
$V_{t}(\widehat{\beta})=\widehat{e}_{t}x_{t}$ where $\{\widehat{e}_{t}\}$
are the least-squares residuals and $\{x_{t}\}$ is a vector of regressors.
Unfortunately, the condition $\sqrt{T}(\widehat{\beta}-\beta_{0})=O_{\mathbb{P}}\left(1\right)$
does not hold for nonparametric estimators $\widehat{\beta}_{\mathrm{np}}$
since they satisfy $T^{\vartheta}(\widehat{\beta}_{\mathrm{np}}-\beta_{0})=O_{\mathbb{P}}\left(1\right)$
for some $\vartheta\in\left(0,\,1/2\right)$. For example, for tests
for forecast evaluation often forecasters use nonparametric kernel
methods to obtain the forecasts {[}i.e., $\{V_{t}(\widehat{\beta})\}=L(e_{t}(\widehat{\beta}_{\mathrm{np}}))$
where $L\left(\cdot\right)$ is a forecast loss, $e_{t}(\cdot)$ is
a forecast error and $\widehat{\beta}_{\mathrm{np}}$ is, e.g., a
rolling window estimate of a parameter that is used to construct the
forecasts{]}. Given the widespread use of nonparametric methods in
applied work, it is useful to extend the theoretical results of HAC
and DK-HAC estimators for these settings. We establish the validity
of HAC and DK-HAC estimators including the validity of the corresponding
estimators based on data-dependent bandwidths. 

The remainder of the paper is organized as follows. Section \ref{Section: Statistical Enviromnent}
introduces the statistical setting and the joint MSE criterion. Section
\ref{Section Simultaneous Bandwiths Determination} presents consistency,
rates of convergence, asymptotic MSE results, and optimal kernels
and bandwidths for the DK-HAC estimators using the joint MSE criterion.
Section \ref{Section Data-Dependent-Bandwidths} develops a data-dependent
method for simultaneous bandwidth parameters selection and its asymptotic
properties are then discussed. Section \ref{Section LRV with nonparametric rate of convergence}
presents theoretical results for LRV estimation in nonparametric parameter
estimation. Section \ref{Section Monte Carlo} presents Monte Carlo
results about the small-sample size and power of HAR tests based on
the DK-HAC estimators using the proposed automatic simultaneous bandwidths.
We also provide comparisons with a variety of other approaches. Section
\ref{Section Conclusions} concludes the paper. The supplemental material
{[}\citet{belotti/casini/catania/grassi/perron_HAC_Sim_Bandws_Supp}{]}
contains the mathematical proofs. The code to implement the proposed
methods is available online in $\mathsf{\mathrm{\mathtt{Matlab}}}$,
$\mathtt{R}$ and $\mathtt{Stata}$ languages. 

\section{The Statistical Environment\label{Section: Statistical Enviromnent}}

We consider the  estimation of the LRV $J\triangleq\mathrm{lim}_{T\rightarrow\infty}J_{T}$
where $J_{T}=T^{-1}\sum_{s=1}^{T}\sum_{t=1}^{T}\mathbb{E}(V_{s}\left(\beta_{0}\right)V_{t}\left(\beta_{0}\right)')$
with $V_{t}\left(\beta\right)$ being a random $p$-vector for each
$\beta\in\Theta$. For example, for the linear model $V_{t}\left(\beta\right)=\left(y_{t}-x'_{t}\beta\right)x_{t}$.
The classical approach for inference in the context of serially correlated
data is based on consistent estimation of $J$. \citet{newey/west:87}
and \citet{andrews:91} considered the class of kernel HAC estimators,
where the subscript Cla stands for classical,
\begin{align*}
\widehat{J}_{\mathrm{Cla},T}=\widehat{J}_{\mathrm{Cla},T}\left(b_{1,T}\right) & \triangleq\frac{T}{T-p}\sum_{k=-T+1}^{T-1}K_{1}\left(b_{1,T}k\right)\widehat{\Gamma}_{\mathrm{Cla}}\left(k\right),\\
\mathrm{with}\qquad\widehat{\Gamma}_{\mathrm{Cla}}\left(k\right) & \triangleq\begin{cases}
T^{-1}\sum_{t=k+1}^{T}\widehat{V}_{t}\widehat{V}'_{t-k}, & k\geq0\\
T^{-1}\sum_{t=-k+1}^{T}\widehat{V}_{t+k}\widehat{V}'_{t}, & k<0,
\end{cases}
\end{align*}
 $\widehat{V}_{t}=V_{t}(\widehat{\beta})$, $K_{1}\left(\cdot\right)$
is a real-valued kernel in the class $\boldsymbol{K}_{1}$ defined
below and $b_{1,T}$ is a bandwidth sequence. The factor $T/\left(T-p\right)$
is an optional small-sample degrees of freedom adjustment. For the
Newey-West estimator $K_{1}$ corresponds to the Bartlett kernel while
for Andrews' \citeyearpar{andrews:91} $K_{1}$ corresponds to the
quadratic spectral (QS) kernel. Data-dependent methods for the selection
of $b_{1,T}$ were proposed by \citet{newey/west:94} and \citet{andrews:91},
respectively. Under appropriate conditions on $b_{1,T}\rightarrow0$
they showed that $\widehat{J}_{\mathrm{Cla},T}\overset{\mathbb{P}}{\rightarrow}J$.
When $\left\{ V_{t}\right\} $ is second-order stationary, $J=2\pi f\left(0\right)$
where $f\left(0\right)$ is the spectral density of $\left\{ V_{t}\right\} $
at frequency zero. Most of the LRV estimation literature has focused
on the stationarity assumption for $\left\{ V_{t}\right\} $ {[}e.g.,
\citet{Kiefer/vogelsang/bunzel:00}, \citet{muller:07} and \citet{lazarus/lewis/stock:17}{]}.
Unlike the HAC estimators, fixed-$b$ (and versions thereof) LRV estimators
require stationarity of $\left\{ V_{t}\right\} $. The latter assumption
is restrictive for economic and financial time series. The properties
of $J$ under nonstationarity were studied recently by \citet{casini_hac}
who showed that if $\left\{ V_{t}\right\} $ is either locally stationary
or segmented locally stationary (SLS), then $J=2\pi\int_{0}^{1}f\left(u,\,0\right)du$
where $f\left(u,\,0\right)$ is the time-varying spectral density
at rescaled time $u=t/T$ and frequency zero. For locally stationary
processes, $f\left(u,\,0\right)$ is smooth in $u$ while for SLS
processes $f\left(u,\,0\right)$ can in addition contain a finite-number
of discontinuities. The number of discontinuities can actually grow
to infinity with unchanged results though at the expense of slightly
more complex derivations. Since the assumption of a finite number
of discontinuities capture well the idea that a finite number of regimes
or structural breaks is enough to account for structural changes (or
big events) in economic time series we maintain this assumption here.
The latter is relaxed by \citet{casini/perron_PrewhitedHAC}. 

Under nonstationarity \citet{casini_hac} argued that an extension
of the classical HAC estimators can actually account flexibly for
the time-varying properties of the data. He proposed the class of
double kernel HAC (DK-HAC) estimators,
\begin{align*}
\widehat{J}_{T}=\widehat{J}_{T}\left(b_{1,T},\,b_{2,T}\right) & \triangleq\frac{T}{T-p}\sum_{k=-T+1}^{T-1}K_{1}\left(b_{1,T}k\right)\widehat{\Gamma}\left(k\right),\quad\mathrm{with}\,\,\\
\widehat{\Gamma}\left(k\right) & \triangleq\frac{n_{T}}{T-n_{T}}\sum_{r=0}^{\left\lfloor \left(T-n_{T}\right)/n_{T}\right\rfloor }\widehat{c}_{T}\left(rn_{T}/T,\,k\right),
\end{align*}
 where $n_{T}\rightarrow\infty$ satisfies the conditions given below,
and 
\begin{align}
\widehat{c}_{T}\left(rn_{T}/T,\,k\right) & \triangleq\begin{cases}
\left(Tb_{2,T}\right)^{-1}\sum_{s=k+1}^{T}K_{2}^{*}\left(\frac{\left(\left(r+1\right)n_{T}-\left(s-k/2\right)\right)/T}{b_{2,T}}\right)\widehat{V}_{s}\widehat{V}'_{s-k}, & k\geq0\\
\left(Tb_{2,T}\right)^{-1}\sum_{s=-k+1}^{T}K_{2}^{*}\left(\frac{\left(\left(r+1\right)n_{T}-\left(s+k/2\right)\right)/T}{b_{2,T}}\right)\widehat{V}_{s+k}\widehat{V}'_{s}, & k<0
\end{cases},\label{eq: Def. chat}
\end{align}
with $K_{2}^{*}$ being a real-valued kernel and $b_{2,T}$ is a
bandwidth sequence. $\widehat{c}_{T}\left(u,\,k\right)$ is an estimate
of the local autocovariance $c\left(u,\,k\right)=\mathbb{E}(V_{\left\lfloor Tu\right\rfloor },\,V'_{\left\lfloor Tu\right\rfloor -k})+O\left(T^{-1}\right)$
{[}under regularity conditions; see \citet{casini_hac}{]} at lag
$k$ and time $u=rn_{T}/T$.  $\widehat{\Gamma}\left(k\right)$
estimates the local autocovariance across blocks of length $n_{T}$
and then takes an average over the blocks. The estimator $\widehat{J}_{T}$
involves two kernels: $K_{1}$ smooths the lagged autocovariances\textemdash akin
to the classical HAC estimators\textemdash while $K_{2}$ applies
smoothing over time. The smoothing over time better account for
nonstationarity and makes $\widehat{J}_{\mathrm{DK,}T}$ robust to
low frequency contamination. See \citet{casini/perron_Low_Frequency_Contam_Nonstat:2020}
who showed theoretically that existing LRV estimators are contaminated
by nonstationarity so that they become inflated with consequent large
power losses when the estimators are used to normalize HAR test statistics. 

\citet{casini_hac} considered adaptive estimators $\widehat{J}_{\mathrm{DK,}T}$
for which $b_{1,T}$ and $b_{2,T}$ are data-dependent. Observe that
the optimal $b_{2,T}$ actually depends on the properties of $\left\{ V_{t,T}\right\} $
in any given block {[}i.e., $b_{2,T}=b_{2,T}\left(t/T\right)${]}.
Let 
\begin{align*}
\mathrm{MSE} & \left(b_{2,T}^{-4},\,\widehat{c}_{T}\left(u_{0},\,k,\,\right),\,\widetilde{W}_{T}\right)\\
 & \triangleq b_{2,T}^{-4}\mathbb{E}\left[\mathrm{vec}\left(\widehat{c}_{T}\left(u_{0},\,k\right)-c\left(u_{0},\,k\right)\right)\right]'\widetilde{W}_{T}\left[\mathrm{vec}\left(\widehat{c}_{T}\left(u_{0},\,k\right)-c\left(u_{0},\,k\right)\right)\right],
\end{align*}
 where $\widetilde{W}_{T}$ is some $p\times p$ positive semidefinite
matrix. He considered a sequential MSE criterion to determine the
optimal kernels and bandwidths. For $K_{1},$ the result states that
the QS kernel minimizes the asymptotic MSE for any $K_{2}\left(\cdot\right)$.
The optimal $b_{1,T}^{\mathrm{opt}}$ and $b_{2,T}^{\mathrm{opt}}$
satisfy the following, 
\begin{align}
\mathrm{MSE} & \left(Tb_{1,T}^{\mathrm{opt}}\overline{b}_{2,T}^{\mathrm{opt}},\,\widehat{J}_{T}\left(b_{1,T}^{\mathrm{opt}},\,\overline{b}_{2,T}^{\mathrm{opt}}\right),\,W_{T}\right)\leq\mathrm{MSE}\left(Tb_{1,T}^{\mathrm{opt}}\overline{b}_{2,T}^{\mathrm{opt}},\,\widehat{J}_{T}\left(b_{1,T},\,\overline{b}_{2,T}^{\mathrm{opt}}\right),\,W_{T}\right)\label{eq (MSE criterio)}\\
\mathrm{where} & \,\,\overline{b}_{2,T}^{\mathrm{opt}}=\int_{0}^{1}b_{2,T}^{\mathrm{opt}}\left(u\right)du\nonumber \\
\mathrm{and} & \,\,b_{2,T}^{\mathrm{opt}}\left(u\right)=\underset{b_{2,T}}{\mathrm{argmin}}\mathrm{MSE}\left(b_{2,T}^{-4},\,\widehat{c}_{T}\left(u_{0},\,k\right)-c\left(u_{0},\,k\right),\,\widetilde{W}_{T}\right).\nonumber 
\end{align}
 $\widehat{J}_{T}(b_{1,T},\,\overline{b}_{2,T}^{\mathrm{opt}})$ indicates
the estimator $\widehat{J}_{T}$ that uses $b_{1,T}^{\mathrm{}}$
and $\overline{b}_{2,T}^{\mathrm{opt}}$. Eq. \eqref{eq (MSE criterio)}
holds as $T\rightarrow\infty$. The above criterion determines the
globally optimal $b_{1,T}^{\mathrm{opt}}$ given the integrated locally
optimal $b_{2,T}^{\mathrm{opt}}\left(u\right)$. Under \eqref{eq (MSE criterio)},
only $b_{2,T}$ affects $b_{1,T}$ but not vice-versa. Intuitively,
this is a limitation because it is likely that in order to minimize
the global MSE the bandwidths $b_{1,T}$ and $b_{2,T}$ affect each
other. 

In this paper, we consider a more theoretically appealing criterion
to determine the optimal bandwidths. That is, we consider bandwidths
$(\widetilde{b}_{1,T}^{\mathrm{opt}},\,\widetilde{b}_{2,T}^{\mathrm{opt}})$
that jointly minimize the global asymptotic relative MSE, denoted
by ReMSE, 
\begin{align}
\lim_{T\rightarrow\infty}\mathrm{ReMSE} & \left(Tb_{1,T}b_{2,T},\,\widehat{J}_{T}\left(b_{1,T},\,b_{2,T}\right)J^{-1},\,W_{T}\right),\label{Eq. (Joint MSE Criterion)}\\
 & =\lim_{T\rightarrow\infty}Tb_{1,T}b_{2,T}\mathbb{E}\left(\mathrm{vec}\left(\widehat{J}_{T}J^{-1}-I_{p}\right)'W_{T}\mathrm{vec}\left(\widehat{J}_{T}J^{-1}-I_{p}\right)\right),\nonumber 
\end{align}
 where $W_{T}$ is $p^{2}\times p^{2}$ weight matrix. Under \eqref{Eq. (Joint MSE Criterion)},
$\widetilde{b}_{1,T}^{\mathrm{opt}}$ and $\widetilde{b}_{2,T}^{\mathrm{opt}}$
affect each other simultaneously. This is a more reasonable property.
In Section \ref{Section Simultaneous Bandwiths Determination} we
solve for the sequences $(\widetilde{b}_{1,T}^{\mathrm{opt}},\,\widetilde{b}_{2,T}^{\mathrm{opt}})$
that minimize \eqref{Eq. (Joint MSE Criterion)}. We propose a data-dependent
method for $(\widetilde{b}_{1,T}^{\mathrm{opt}},\,\widetilde{b}_{2,T}^{\mathrm{opt}})$
in Section \ref{Section Data-Dependent-Bandwidths}. 

The literature on LRV estimation has routinely focused on the case
where $\widehat{V}_{t}$ is a function of a parameter estimate $\widehat{\beta}$
that enjoys a standard $\sqrt{T}$ parametric rate of convergence.
While this is an important case, the recent increasing use of nonparametric
methods suggests that the case where $\widehat{\beta}$ enjoys a
nonparametric rate of convergence slower than $\sqrt{T}$ is of potential
interest. Hence, in Section \ref{Section LRV with nonparametric rate of convergence}
we consider consistent LRV estimation under the latter framework and
develop corresponding results for the classical HAC as well as the
DK-HAC estimators. 

We consider the following standard classes of kernels {[}cf. \citet{andrews:91}{]},
\begin{align}
\boldsymbol{K}_{1} & =\left\{ K_{1}\left(\cdot\right):\,\mathbb{R}\rightarrow\left[-1,\,1\right]:\,K_{1}\left(0\right)=1,\,K_{1}\left(x\right)=K_{1}\left(-x\right),\,\forall x\in\mathbb{R}\right.\label{Eq. (2.6) K1 Kernel class}\\
 & \quad\left.\int_{-\infty}^{\infty}K_{1}^{2}\left(x\right)dx<\infty,\,K_{1}\left(\cdot\right)\,\mathrm{is\,continuous\,at\,0\,and\,at\,all\,but\,finite\,numbers\,of\,points}\right\} .\nonumber \\
\boldsymbol{K}_{2} & =\biggl\{ K_{2}\left(\cdot\right):\,\mathbb{R}\rightarrow\left[0,\,\infty\right]:\,K_{2}\left(x\right)=K_{2}\left(1-x\right),\,\int K_{2}\left(x\right)dx=1,\label{Eq. K2 Kernel class}\\
 & \qquad\qquad K_{2}\left(x\right)=0,\,\mathrm{for\,}\,x\notin\left[0,\,1\right],\,K_{2}\left(\cdot\right)\,\mathrm{is\,continuous}\biggr\}.\nonumber 
\end{align}
The class $\boldsymbol{K}_{1}$ was also considered by \citet{andrews:91}.
Examples of kernels in $\boldsymbol{K}_{1}$ include the Truncated,
Bartlett, Parzen, Quadratic Spectral (QS) and Tukey-Hanning kernels.
The QS kernel was shown to be optimal for $\widehat{J}_{\mathrm{Cla},T}$
under the MSE criterion by \citet{andrews:91} and for $\widehat{J}_{T}$
under a sequential MSE criterion by \citet{casini_hac},
\begin{align*}
K_{1}^{\mathrm{QS}}\left(x\right) & =\frac{25}{12\pi^{2}x^{2}}\left(\frac{\sin\left(6\pi x/5\right)}{6\pi x/5}-\cos\left(6\pi x/5\right)\right).
\end{align*}
 The class $\boldsymbol{K}_{2}$ was also considered by, for example,
\citet{Dahlhaus/Giraitis:98}.

Throughout we adopt the following notational conventions.  The $j$th
element of a vector $x$ is indicated by $x^{\left(j\right)}$ while
the $\left(j,\,l\right)$th element of a matrix $X$ is indicated
as $X^{\left(j,\,l\right)}$. $\mathrm{tr}\left(\cdot\right)$ denotes
the trace function and $\otimes$ denotes the tensor (or Kronecker)
product operator. The $p^{2}\times p^{2}$ matrix $C_{pp}$ is a commutation
matrix that transforms $\mathrm{vec}\left(A\right)$ into $\mathrm{vec}\left(A'\right)$,
i.e., $C_{pp}=\sum_{j=1}^{p}\sum_{l=1}^{p}\iota_{j}\iota_{l}'\otimes\iota_{l}\iota_{j}'$,
where $\iota_{j}$ is the $j$th elementary $p$-vector. $\lambda_{\max}\left(A\right)$
denotes the largest eigenvalue of the matrix $A$. $W$ and $\widetilde{W}$
are used for $p^{2}\times p^{2}$ weight matrices. $\mathbb{C}$
is used for the set of complex numbers and $\overline{A}$ for the
complex conjugate of $A\in\mathbb{C}$. Let $0=\lambda_{0}<\lambda_{1}<\ldots<\lambda_{m}<\lambda_{m+1}=1$.
A function $G\left(\cdot,\,\cdot\right):\,\left[0,\,1\right]\times\mathbb{R}\rightarrow\mathbb{C}$
is said to be piecewise (Lipschitz) continuous with $m+1$ segments
if it is (Lipschitz) continuous within each segment. For example,
it is piecewise Lipschitz continuous if for each segment $j=1,\ldots,\,m+1$
it satisfies $\sup_{u\neq v}\left|G\left(u,\,\omega\right)-G\left(v,\,\omega\right)\right|\leq K\left|u-v\right|$
for any $\omega\in\mathbb{R}$ with $\lambda_{j-1}<u,\,v\leq\lambda_{j}$
for some $K<\infty.$ We define $G_{j}\left(u,\,\omega\right)=G\left(u,\,\omega\right)$
for $\lambda_{j-1}<u\leq\lambda_{j}$, so $G_{j}\left(u,\,\omega\right)$
is Lipschitz continuous for each $j.$ If we say piecewise Lipschitz
continuous with index $\vartheta>0$, then the above inequality is
replaced by $\sup_{u\neq v}\left|G\left(u,\,\omega\right)-G\left(v,\,\omega\right)\right|\leq K\left|u-v\right|^{\vartheta}$.
A function $G\left(\cdot,\,\cdot\right):\,\left[0,\,1\right]\times\mathbb{R}\rightarrow\mathbb{C}$
is said to be left-differentiable at $u_{0}$ if $\partial G\left(u_{0},\omega\right)/\partial_{-}u\triangleq\lim_{u\rightarrow u_{0}^{-}}\left(G\left(u_{0},\,\omega\right)-G\left(u,\,\omega\right)\right)/\left(u_{0}-u\right)$
exists for any $\omega\in\mathbb{R}$. We use $\left\lfloor \cdot\right\rfloor $
to denote the largest smaller integer function. The symbol ``$\triangleq$''
is for definitional equivalence. 

\section{\label{Section Simultaneous Bandwiths Determination}Simultaneous
Bandwidths Determination for DK-HAC Estimators}

In Section \ref{Subsection: Asymptotic-MSE-Properties} we present
the consistency, rate of convergence and asymptotic MSE properties
of predetermined bandwidths for the DK-HAC estimators.  We use the
MSE results to determine the optimal bandwidths and kernels in Section
\ref{Subsection: Optimal-Bandwidths-and}. We use the framework for
nonstationarity introduced in \citet{casini_hac}. That is, we assume
that $\left\{ V_{t,T}\right\} $ is segmented locally stationary (SLS).
Suppose $\left\{ V_{t}\right\} _{t=1}^{T}$ is defined on an abstract
probability space $\left(\Omega,\,\mathscr{F},\,\mathbb{P}\right)$,
where $\Omega$ is the sample space, $\mathscr{F}$ is the $\sigma$-algebra
and $\mathbb{P}$ is a probability measure. We use an infill asymptotic
setting and rescale the original discrete time horizon $\left[1,\,T\right]$
by dividing each $t$ by $T.$ Letting $u=t/T$ and $T\rightarrow\infty,$
this defines a new time scale $u\in\left[0,\,1\right]$. Let $i\triangleq\sqrt{-1}$.
\begin{defn}
\label{Definition Segmented-Locally-Stationary}A sequence of stochastic
processes $\{V_{t,T}\}_{t=1}^{T}$ is called Segmented Locally Stationarity
(SLS) with $m_{0}+1$ regimes, transfer function $A^{0}$  and trend
$\mu_{\cdot}$ if there exists a representation 
\begin{align}
V_{t,T} & =\mu_{j}\left(t/T\right)+\int_{-\pi}^{\pi}\exp\left(i\omega t\right)A_{j,t,T}^{0}\left(\omega\right)d\xi\left(\omega\right),\qquad\qquad\left(t=T_{j-1}^{0}+1,\ldots,\,T_{j}^{0}\right),\label{Eq. Spectral Rep of SLS}
\end{align}
for $j=1,\ldots,\,m_{0}+1$, where by convention $T_{0}^{0}=0$ and
$T_{m_{0}+1}^{0}=T$ and the following holds: 

(i) $\xi\left(\omega\right)$ is a stochastic process on $\left[-\pi,\,\pi\right]$
with $\overline{\xi\left(\omega\right)}=\xi\left(-\omega\right)$
and 
\begin{align*}
\mathrm{cum}\left\{ d\xi\left(\omega_{1}\right),\ldots,\,d\xi\left(\omega_{r}\right)\right\}  & =\varphi\left(\sum_{j=1}^{r}\omega_{j}\right)g_{r}\left(\omega_{1},\ldots,\,\omega_{r-1}\right)d\omega_{1}\ldots d\omega_{r},
\end{align*}
 where $\mathrm{cum}\left\{ \cdot\right\} $ is the cumulant  of
$r$th order, $g_{1}=0,\,g_{2}\left(\omega\right)=1$, $\left|g_{r}\left(\omega_{1},\ldots,\,\omega_{r-1}\right)\right|\leq M_{r}<\infty$
  and $\varphi\left(\omega\right)=\sum_{j=-\infty}^{\infty}\delta\left(\omega+2\pi j\right)$
is the period $2\pi$ extension of the Dirac delta function $\delta\left(\cdot\right)$.

(ii) There exists a constant $K>0$  and a piecewise continuous function
$A:\,\left[0,\,1\right]\times\mathbb{R}\rightarrow\mathbb{C}$ such
that, for each $j=1,\ldots,\,m_{0}+1$, there exists a $2\pi$-periodic
function $A_{j}:\,(\lambda_{j-1},\,\lambda_{j}]\times\mathbb{R}\rightarrow\mathbb{C}$
with $A_{j}\left(u,\,-\omega\right)=\overline{A_{j}\left(u,\,\omega\right)}$,
$\lambda_{j}^{0}\triangleq T_{j}^{0}/T$ and for all $T,$
\begin{align}
A\left(u,\,\omega\right) & =A_{j}\left(u,\,\omega\right)\,\mathrm{\,for\,}\,\lambda_{j-1}^{0}<u\leq\lambda_{j}^{0},\label{Eq A(u) =00003D Ai}\\
\sup_{1\leq j\leq m_{0}+1} & \sup_{T_{j-1}^{0}<t\leq T_{j}^{0},\,\omega}\left|A_{j,t,T}^{0}\left(\omega\right)-A_{j}\left(t/T,\,\omega\right)\right|\leq KT^{-1}.\label{Eq. 2.4 Smothenss Assumption on A}
\end{align}
(iii) $\mu_{j}\left(t/T\right)$ is piecewise continuous. 
\end{defn}
Observe that this representation is similar to the spectral representation
of stationary processes {[}see \citet{anderson:71}, \citet{brillinger:75},
\citet{hannan:70} and \citet{priestley:85} for  introductory concepts{]}.
The main difference is that $A\left(t/T,\,\omega\right)$ and $\mu\left(t/T\right)$
are not constant in $t$. \citet{dahlhaus:96} used the time-varying
spectral representation to define the so-called locally stationary
processes which are characterized, broadly speaking, by smoothness
conditions on $\mu\left(\cdot\right)$ and $A\left(\cdot,\,\cdot\right)$.
Locally stationary processes are often referred to as time-varying
parameter processes {[}see e.g., \citet{cai:07} and \citet{chen/hong:12}{]}.
However, the smoothness restrictions exclude many prominent models
that account for time variation in the parameters. For example,
structural change and regime switching-type models do not belong to
this class because parameter changes occur suddenly at a particular
time. Thus, the class of SLS processes is more general and likely
to be more useful. Stationarity and local stationarity are recovered
as special cases of the SLS definition.

Let $\mathcal{T}\triangleq\{T_{1}^{0},\,\ldots,\,T_{m_{0}}^{0}\}$.
The spectrum of $V_{t,T}$ is defined (for fixed $T$) as 
\begin{align*}
f_{j,T} & \left(u,\,\omega\right)\\
 & \triangleq\begin{cases}
\left(2\pi\right)^{-1}\sum_{s=-\infty}^{\infty}\mathrm{Cov}\left(V_{\left\lfloor uT-3\left|s\right|/2\right\rfloor ,T},\,V_{\left\lfloor uT-\left|s\right|/2\right\rfloor ,T}\right)\exp\left(-i\omega s\right), & Tu\in\mathcal{T},\,u=T_{j}^{0}/T\\
\left(2\pi\right)^{-1}\sum_{s=-\infty}^{\infty}\mathrm{Cov}\left(V_{\left\lfloor uT-s/2\right\rfloor ,T},\,V_{\left\lfloor uT+s/2\right\rfloor ,T}\right)\exp\left(-i\omega s\right), & Tu\notin\mathcal{T},\,T_{j-1}^{0}/T<u<T_{j}^{0}/T
\end{cases},
\end{align*}
with $A_{1,t,T}^{0}\left(\omega\right)=A_{1}\left(0,\,\omega\right)$
for $t<1$ and $A_{m+1,t,T}^{0}\left(\omega\right)=A_{m+1}\left(1,\,\omega\right)$
for $t>T$. \citet{casini_hac} showed that $f_{j,T}\left(u,\,\omega\right)$
tends in mean-squared to $f_{j}\left(u,\,\omega\right)\triangleq\left|A_{j}\left(u,\,\omega\right)\right|^{2}$
for $T_{j-1}^{0}/T<u=t/T\leq T_{j}^{0}/T$, which is the spectrum
that corresponds to the spectral representation. Therefore, we call
$f_{j}\left(u,\,\omega\right)$ the time-varying spectral density
matrix of the process.  Given $f\left(u,\,\omega\right),$ we can
define the local covariance of $V_{t,T}$ at rescaled time $u$ with
$Tu\notin\mathcal{T}$ and lag $k\in\mathbb{Z}$ as $c\left(u,\,k\right)\triangleq\int_{-\pi}^{\pi}e^{i\omega k}f\left(u,\,\omega\right)d\omega$.
The same definition is also used when $Tu\in\mathcal{T}$ and $k\geq0$.
For $Tu\in\mathcal{T}$ and $k<0$ it is defined as $c\left(u,\,k\right)\triangleq\int_{-\pi}^{\pi}e^{i\omega k}A\left(u,\,\omega\right)A\left(u-k/T,\,-\omega\right)d\omega$.

\subsection{\label{Subsection: Asymptotic-MSE-Properties}Asymptotic MSE Properties
of DK-HAC estimators}

Let $\widetilde{J}_{T}$ denote the pseudo-estimator identical to
$\widehat{J}_{T}$ but based on  $\{V_{t,T}\}=\{V_{t,T}\left(\beta_{0}\right)\}$
rather than on $\{\widehat{V}_{t,T}\}=\{V_{t,T}(\widehat{\beta})\}$. 
\begin{assumption}
\label{Assumption Smothness of A (for HAC)}(i) $\{V_{t,T}\}$ is
a mean-zero SLS process with $m_{0}+1$ regimes; (ii) $A\left(u,\,\omega\right)$
is \textcolor{red}{ }twice continuously differentiable in $u$ at
all $u\neq\lambda_{j}^{0}$ $(j=1,\ldots,\,m_{0}+1)$ with uniformly
bounded derivatives $\left(\partial/\partial u\right)A\left(u,\,\cdot\right)$
and $\left(\partial^{2}/\partial u^{2}\right)A\left(u,\,\cdot\right)$,
and Lipschitz continuous in the second component with index $\vartheta=1$;
(iii) $\left(\partial^{2}/\partial u^{2}\right)A\left(u,\,\cdot\right)$
is Lipschitz continuous at all $u\neq\lambda_{j}^{0}$ $(j=1,\ldots,\,m_{0}+1)$;
(iv) $A\left(u,\,\omega\right)$ is twice left-differentiable in $u$
at $u=\lambda_{j}^{0}$ $(j=1,\ldots,\,m_{0}+1)$ with uniformly bounded
derivatives $\left(\partial/\partial_{-}u\right)A\left(u,\,\cdot\right)$
and $\left(\partial^{2}/\partial_{-}u^{2}\right)A\left(u,\,\cdot\right)$
and has piecewise Lipschitz continuous derivative $\left(\partial^{2}/\partial_{-}u^{2}\right)A\left(u,\,\cdot\right)$.
\end{assumption}
We also need to impose conditions on the temporal dependence of $V_{t}=V_{t,T}$.
Let 
\begin{align*}
\kappa_{V,t}^{\left(a,b,c,d\right)}\left(u,\,v,\,w\right) & \triangleq\kappa^{\left(a,b,c,d\right)}\left(t,\,t+u,\,t+v,\,t+w\right)-\kappa_{\mathscr{N}}^{\left(a,b,c,d\right)}\left(t,\,t+u,\,t+v,\,t+w\right)\\
 & \triangleq\mathbb{E}\left(V_{t}^{\left(a\right)}V_{t+u}^{\left(b\right)}V_{t+v}^{\left(c\right)}V_{t+w}^{\left(d\right)}\right)-\mathbb{E}\left(V_{\mathscr{N},t}^{\left(a\right)}V_{\mathscr{N},t+u}^{\left(b\right)}V_{\mathscr{N},t+v}^{\left(c\right)}V_{\mathscr{N},t+w}^{\left(d\right)}\right),
\end{align*}
where $\left\{ V_{\mathscr{N},t}\right\} $ is a Gaussian sequence
with the same mean and covariance structure as $\left\{ V_{t}\right\} $.
$\kappa_{V,t}^{\left(a,b,c,d\right)}\left(u,\,v,\,w\right)$ is the
time-$t$ fourth-order cumulant of $(V_{t}^{\left(a\right)},\,V_{t+u}^{\left(b\right)},\,V_{t+v}^{\left(c\right)},$
$\,V_{t+w}^{\left(d\right)})$ while $\kappa_{\mathscr{N}}^{\left(a,b,c,d\right)}(t,\,t+u,$
$\,t+v,\,t+w)$ is the time-$t$ centered fourth moment of $V_{t}$
if $V_{t}$ were Gaussian.
\begin{assumption}
\label{Assumption A - Dependence}(i) $\sum_{k=-\infty}^{\infty}\sup_{u\in\left[0,\,1\right]}$
$\left\Vert c\left(u,\,k\right)\right\Vert <\infty$, $\sum_{k=-\infty}^{\infty}\sup_{u\in\left[0,\,1\right]}\left\Vert \left(\partial^{2}/\partial u^{2}\right)c\left(u,\,k\right)\right\Vert <\infty$
and $\sum_{k=-\infty}^{\infty}\sum_{j=-\infty}^{\infty}\sum_{l=-\infty}^{\infty}\sup_{u\in\left[0,\,1\right]}|\kappa_{V,\left\lfloor Tu\right\rfloor }^{\left(a,b,c,d\right)}$
$\left(k,\,j,\,l\right)|<\infty$ for all $a,\,b,\,c,\,d\leq p$.
(ii) For all $a,\,b,\,c,\,d\leq p$ there exists a function $\widetilde{\kappa}_{a,b,c,d}:\,\left[0,\,1\right]\times\mathbb{Z}\times\mathbb{Z}\times\mathbb{Z}\rightarrow\mathbb{R}$
such that $\sup_{u\in\left(0,\,1\right)}|\kappa_{V,\left\lfloor Tu\right\rfloor }^{\left(a,b,c,d\right)}\left(k,\,s,\,l\right)$
$-\widetilde{\kappa}_{a,b,c,d}\left(u,\,k,\,s,\,l\right)|\leq KT^{-1}$
for some constant $K$; the function $\widetilde{\kappa}_{a,b,c,d}\left(u,\,k,\,s,\,l\right)$
is twice differentiable in $u$ at all $u\neq\lambda_{j}^{0}$, $(j=1,\ldots,\,m_{0}+1)$,
with uniformly bounded derivatives $\left(\partial/\partial u\right)\widetilde{\kappa}_{a,b,c,d}\left(u,\cdot,\cdot,\cdot\right)$
and $\left(\partial^{2}/\partial u^{2}\right)\widetilde{\kappa}_{a,b,c,d}\left(u,\cdot,\cdot,\cdot\right)$,
and twice left-differentiable in $u$ at $u=\lambda_{j}^{0}$ $(j=1,\ldots,\,m_{0}+1)$
with uniformly bounded derivatives $\left(\partial/\partial_{-}u\right)\widetilde{\kappa}_{a,b,c,d}\left(u,\cdot,\cdot,\cdot\right)$
and $\left(\partial^{2}/\partial_{-}u^{2}\right)\widetilde{\kappa}_{a,b,c,d}$
$\left(u,\cdot,\cdot,\cdot\right)$ and piecewise Lipschitz continuous
derivative $\left(\partial^{2}/\partial_{-}u^{2}\right)\widetilde{\kappa}_{a,b,c,d}\left(u,\cdot,\cdot,\cdot\right)$.
\end{assumption}
We do not require fourth-order stationarity but only that the time-$t=Tu$
fourth order cumulant is locally constant in a neighborhood of $u$.

Following \citet{parzen:57}, we define $K_{1,q}\triangleq\lim_{x\downarrow0}\left(1-K_{1}\left(x\right)\right)/\left|x\right|^{q}$
for $q\in[0,\,\infty);$ $q$ increases with the smoothness of $K_{1}\left(\cdot\right)$
with the largest value being such that $K_{1,q}<\infty$. When $q$
is an even integer, $K_{1,q}=-\left(d^{q}K_{1}\left(x\right)/dx^{q}\right)|_{x=0}/q!$
and $K_{1,q}<\infty$ if and only if $K_{1}\left(x\right)$ is $q$
times differentiable at zero.  We define the index of smoothness
of $f\left(u,\,\omega\right)$ at $\omega=0$ by $f^{\left(q\right)}\left(u,\,0\right)\triangleq\left(2\pi\right)^{-1}\sum_{k=-\infty}^{\infty}\left|k\right|^{q}c\left(u,\,k\right)$,
for $q\in[0,\,\infty)$. If $q$ is even, then $f^{\left(q\right)}\left(u,\,0\right)=\left(-1\right)^{q/2}\left(d^{q}f\left(u,\,\omega\right)/d\omega^{q}\right)|_{\omega=0}$.
Further, $||f^{\left(q\right)}\left(u,\,0\right)||<\infty$ if and
only if $f\left(u,\,\omega\right)$ is $q$ times differentiable at
$\omega=0$. We define 
\begin{align}
\mathrm{MSE}\left(Tb_{1,T}b_{2,T},\,\widetilde{J}_{T},\,W\right) & =Tb_{1,T}b_{2,T}\mathbb{E}\left[\mathrm{vec}\left(\widetilde{J}_{T}-J_{T}\right)'W\mathrm{vec}\left(\widetilde{J}_{T}-J_{T}\right)\right].\label{Eq: 3.5 MSE}
\end{align}
\begin{thm}
\label{Theorem MSE J}Suppose $K_{1}\left(\cdot\right)\in\boldsymbol{K}_{1}$,
$K_{2}\left(\cdot\right)\in\boldsymbol{K}_{2}$, Assumption \ref{Assumption Smothness of A (for HAC)}-\ref{Assumption A - Dependence}
hold, $b_{1,T},\,b_{2,T}\rightarrow0$, $n_{T}\rightarrow\infty,\,n_{T}/T\rightarrow0$
and $1/Tb_{1,T}b_{2,T}\rightarrow0$. We have: (i)~
\begin{align*}
\lim_{T\rightarrow\infty} & Tb_{1,T}b_{2,T}\mathrm{Var}\left[\mathrm{vec}\left(\widetilde{J}_{T}\right)\right]\\
 & =4\pi^{2}\int K_{1}^{2}\left(y\right)dy\int_{0}^{1}K_{2}^{2}\left(x\right)dx\left(I+C_{pp}\right)\left(\int_{0}^{1}f\left(u,\,0\right)du\right)\otimes\left(\int_{0}^{1}f\left(v,\,0\right)dv\right).
\end{align*}

(ii) If $1/Tb_{1,T}^{q}b_{2,T}\rightarrow0$, $n_{T}/Tb_{1,T}^{q}\rightarrow0$
and $b_{2,T}^{2}/b_{1,T}^{q}\rightarrow\nu\in\left(0,\,\infty\right)$
for some $q\in[0,\,\infty)$ for which $K_{1,q},$ $||\int_{0}^{1}f^{\left(q\right)}\left(u,\,0\right)du||\in[0,\,\infty)$
then $\lim_{T\rightarrow\infty}b_{1,T}^{-q}\mathbb{E}(\widetilde{J}_{T}-J_{T})=\mathsf{B}_{1}+\mathsf{B}_{2}$
where $\mathsf{B}_{1}=-2\pi K_{1,q}\int_{0}^{1}f^{\left(q\right)}\left(u,\,0\right)du$
and $\mathrm{\mathsf{B}}_{2}=2^{-1}\nu\int_{0}^{1}x^{2}K_{2}\left(x\right)\sum_{k=-\infty}^{\infty}\int_{0}^{1}\left(\partial^{2}/\partial u^{2}\right)c\left(u,\,k\right)du.$ 

(iii) If $n_{T}/Tb_{1,T}^{q}\rightarrow0$, $b_{2,T}^{2}/b_{1,T}^{q}\rightarrow\nu$
and $Tb_{1,T}^{2q+1}b_{2,T}\rightarrow\gamma\in\left(0,\,\infty\right)$
for some $q\in[0,\,\infty)$ for which $K_{1,q},\,||\int_{0}^{1}f^{\left(q\right)}\left(u,\,0\right)du||\in[0,\,\infty)$
, then 
\begin{align*}
\lim_{T\rightarrow\infty} & \mathrm{MSE}\left(Tb_{1,T}b_{2,T},\,\widetilde{J}_{T},\,W\right)=4\pi^{2}\left[\gamma\left(4\pi^{2}\right)^{-1}\mathrm{vec}\left(\mathsf{B}_{1}+\mathsf{B}_{2}\right)'W\mathrm{vec}\left(\mathsf{B}_{1}+\mathsf{B}_{2}\right)\right.\\
 & \quad\left.+\int K_{1}^{2}\left(y\right)dy\int K_{2}^{2}\left(x\right)dx\,\mathrm{tr}W\left(I_{p^{2}}+C_{pp}\right)\left(\int_{0}^{1}f\left(u,\,0\right)du\right)\otimes\left(\int_{0}^{1}f\left(v,\,0\right)dv\right)\right].
\end{align*}
  
\end{thm}
The bias expression in part (ii) of Theorem \ref{Theorem MSE J} is
different from the corresponding one in \citet{casini_hac} because
$b_{2,T}^{2}/b_{1,T}^{q}\rightarrow\nu\in\left(0,\,\infty\right)$
replaces $b_{2,T}^{2}/b_{1,T}^{q}\rightarrow0$ there. The extra term
is $\mathsf{B}_{2}$. This means that both $b_{1,T}$ and $b_{2,T}$
affect the bias as well as the variance. It is therefore possible
to consider a joint minimization of the asymptotic MSE with respect
to $b_{1,T}$ and $b_{2,T}$. Note that $\mathsf{B}_{2}=0$ when $\int_{0}^{1}\left(\partial^{2}/\partial^{2}u\right)c\left(u,\,k\right)du=0$.
The latter occurs when the process is stationary.  We now move
to the results concerning $\widehat{J}_{T}$.
\begin{assumption}
\label{Assumption B}(i) $\sqrt{T}(\widehat{\beta}-\beta_{0})=O_{\mathbb{P}}\left(1\right)$;
(ii) $\sup_{u\in\left[0,\,1\right]}\mathbb{E}||V_{\left\lfloor Tu\right\rfloor }||^{2}<\infty$;
(iii) $\sup_{u\in\left[0,\,1\right]}\mathbb{E}\sup_{\beta\in\Theta}$
$||\left(\partial/\partial\beta'\right)V_{\left\lfloor Tu\right\rfloor }\left(\beta\right)||^{2}<\infty$;
(iv) $\int_{-\infty}^{\infty}\left|K_{1}\left(y\right)\right|dy,$
$\int_{0}^{1}\left|K_{2}\left(x\right)\right|dx<\infty.$
\end{assumption}
Assumption \ref{Assumption B}(i)-(iii) is the same as Assumption
B in \citet{andrews:91}.  Part (i) is satisfied by standard (semi)parametric
estimators. In Section \ref{Section LRV with nonparametric rate of convergence}
we relax this assumption and consider nonparametric estimators that
satisfy $T^{\vartheta}(\widehat{\beta}-\beta_{0})=O_{\mathbb{P}}\left(1\right)$
where $\vartheta\in\left(0,\,1/2\right)$.  In order to obtain 
rate of convergence results we replace Assumption \ref{Assumption A - Dependence}
with the following assumptions. 
\begin{assumption}
\label{Assumption C Andrews 91}(i) Assumption \ref{Assumption A - Dependence}
holds with $V_{t,T}$ replaced by 
\begin{align*}
\left(V'_{\left\lfloor Tu\right\rfloor },\,\mathrm{vec}\left(\left(\frac{\partial}{\partial\beta'}V_{\left\lfloor Tu\right\rfloor }\left(\beta_{0}\right)\right)-\mathbb{E}\left(\frac{\partial}{\partial\beta'}V_{\left\lfloor Tu\right\rfloor }\left(\beta_{0}\right)\right)\right)'\right)' & .
\end{align*}
(ii) $\sup_{u\in\left[0,\,1\right]}\mathbb{E}(\sup_{\beta\in\Theta}||\left(\partial^{2}/\partial\beta\partial\beta'\right)V_{\left\lfloor Tu\right\rfloor }^{\left(a\right)}\left(\beta\right)||^{2})<\infty$
for all $a=1,\ldots,\,p$.
\end{assumption}
\begin{assumption}
\label{Assumption W_T and unbounded kernel and Cumulant 8}Let $W_{T}$
denote a $p^{2}\times p^{2}$ weight matrix such that $W_{T}\overset{\mathbb{P}}{\rightarrow}W$.
\end{assumption}
\begin{thm}
\label{Theorem 1 -Consistency and Rate}Suppose $K_{1}\left(\cdot\right)\in\boldsymbol{K}_{1}$,
$K_{2}\left(\cdot\right)\in\boldsymbol{K}_{2}$, $b_{1,T},\,b_{2,T}\rightarrow0$,\textbf{
}$n_{T}\rightarrow\infty,\,n_{T}/Tb_{1,T}\rightarrow0,$ and $1/Tb_{1,T}b_{2,T}\rightarrow0$.
We have:

(i) If Assumption \ref{Assumption Smothness of A (for HAC)}-\ref{Assumption B}
hold, $\sqrt{T}b_{1,T}\rightarrow\infty$, $b_{2,T}/b_{1,T}\rightarrow\nu\in[0,\,\infty)$
then $\widehat{J}_{T}-J_{T}\overset{\mathbb{P}}{\rightarrow}0$ and
$\widehat{J}_{T}-\widetilde{J}_{T}\overset{\mathbb{P}}{\rightarrow}0$. 

(ii) If Assumption \ref{Assumption Smothness of A (for HAC)}, \ref{Assumption B}-\ref{Assumption C Andrews 91}
hold,  $n_{T}/Tb_{1,T}^{q}\rightarrow0$, $1/Tb_{1,T}^{q}b_{2,T}\rightarrow0$,
$b_{2,T}^{2}/b_{1,T}^{q}\rightarrow\nu\in[0,\,\infty)$ and $Tb_{1,T}^{2q+1}b_{2,T}\rightarrow\gamma\in\left(0,\,\infty\right)$
for some $q\in[0,\,\infty)$ for which $K_{1,q},\,||\int_{0}^{1}f^{\left(q\right)}\left(u,\,0\right)du||\in[0,\,\infty)$,
then $\sqrt{Tb_{1,T}b_{2,T}}(\widehat{J}_{T}-J_{T})=O_{\mathbb{P}}\left(1\right)$
and $\sqrt{Tb_{1,T}}(\widehat{J}_{T}-\widetilde{J}_{T})=o_{\mathbb{P}}\left(1\right).$ 

(iii) Under the conditions of part (ii) with $\nu\in\left(0,\,\infty\right)$
and Assumption \ref{Assumption W_T and unbounded kernel and Cumulant 8},
\begin{align*}
\lim_{T\rightarrow\infty}\mathrm{MSE}\left(Tb_{1,T}b_{2,T},\,\widehat{J}_{T},\,W_{T}\right)=\lim_{T\rightarrow\infty}\mathrm{MSE}\left(Tb_{1,T}b_{2,T},\,\widetilde{J}_{T},\,W\right) & .
\end{align*}
\end{thm}
Part (ii) yields the consistency of $\widehat{J}_{T}$ with $b_{1,T}$
only required to be $o\left(Tb_{2,T}\right)$. This rate is slower
than the corresponding rate $o\left(T\right)$ of the classical kernel
HAC estimators as shown by \citet{andrews:91} in his Theorem 1-(b).
However, this property is of little practical import because optimal
growth rates typically are less than $T^{1/2}$\textemdash for the
QS kernel the optimal growth rate is $T^{1/5}$ while it is $T^{1/3}$
for the Barteltt. Part (ii) of the theorem presents the rate of convergence
of $\widehat{J}_{T}$ which is $\sqrt{Tb_{2,T}b_{1,T}}$, the same
rate shown by \citet{casini_hac} when $b_{2,T}^{2}/b_{1,T}^{q}\rightarrow0$.
Thus, the presence of the bias term $\mathsf{B}_{2}$ does not alter
the rate of convergence. In Section \ref{Subsection: Optimal-Bandwidths-and},
we compare the rate of convergence of $\widehat{J}_{T}$ with optimal
bandwidths $(\widetilde{b}_{1,T}^{\mathrm{opt}},\,\widetilde{b}_{2,T}^{\mathrm{opt}})$
from the joint MSE criterion \eqref{Eq. (Joint MSE Criterion)} with
that using $(b_{1,T}^{\mathrm{opt}},\,b_{2,T}^{\mathrm{opt}})$ from
the sequential MSE criterion \eqref{eq (MSE criterio)}, and with
that of the classical HAC estimators when the corresponding optimal
bandwidths are used. 

\subsection{\label{Subsection: Optimal-Bandwidths-and}Optimal Bandwidths and
Kernels}

We consider the optimal bandwidths $\left(\widetilde{b}_{1,T}^{\mathrm{opt}},\,\widetilde{b}_{2,T}^{\mathrm{opt}}\right)$
and kernels $\widetilde{K}_{1}^{\mathrm{opt}}$ and $\widetilde{K}_{2}^{\mathrm{opt}}$
that  minimize the global asymptotic relative MSE \eqref{Eq. (Joint MSE Criterion)}
given by 
\begin{align*}
\lim_{T\rightarrow\infty}\mathrm{ReMSE} & \left(\widehat{J}_{T}\left(b_{1,T},\,b_{2,T}\right)J^{-1},\,W_{T}\right),\\
 & =\mathbb{E}\left(\mathrm{vec}\left(\widehat{J}_{T}J^{-1}-I_{p}\right)'W_{T}\mathrm{vec}\left(\widehat{J}_{T}J^{-1}-I_{p}\right)\right).
\end{align*}

 Let  $\Xi_{1,1}=-K_{1,q}$, $\Xi_{1,2}=\left(4\pi\right)^{-1}\int_{0}^{1}x^{2}K_{2}\left(x\right)dx,$
$\Xi_{2}=\int K_{1}^{2}\left(y\right)dy\int_{0}^{1}K_{2}^{2}\left(x\right)dx,$
 
\begin{align*}
\Delta_{1,1,0} & \triangleq\int_{0}^{1}f^{\left(q\right)}\left(u,\,0\right)du\left(\int_{0}^{1}f\left(u,\,0\right)du\right)^{-1}\quad\quad\mathrm{and}\\
\Delta_{1,2} & \triangleq\sum_{k=-\infty}^{\infty}\int_{0}^{1}\left(\partial^{2}/\partial u^{2}\right)c\left(u,\,k\right)du(\int_{0}^{1}f\left(u,\,0\right)du)^{-1}.
\end{align*}

\begin{thm}
\label{Theorem Optimal Kernels}Suppose Assumption \ref{Assumption Smothness of A (for HAC)},
\ref{Assumption B}-\ref{Assumption W_T and unbounded kernel and Cumulant 8}
hold, $\int_{0}^{1}||f^{\left(2\right)}\left(u,\,0\right)||du<\infty$,
$\mathrm{vec}\left(\Delta_{1,1,0}\right)'W$ $\mathrm{vec}\left(\Delta_{1,1,0}\right)>0$,
$\mathrm{vec}\left(\Delta_{1,2}\right)'W\mathrm{vec}\left(\Delta_{1,2}\right)>0$
and $W$ is positive definite. Then, $\lim_{T\rightarrow\infty}\mathrm{ReMSE}$
$(\widehat{J}_{T}\left(b_{1,T},\,b_{2,T}\right)J^{-1},\,W_{T})$ is
jointly minimized by 
\begin{align*}
\widetilde{b}_{1,T}^{\mathrm{opt}} & =0.46\left(\frac{\mathrm{vec}\left(\Delta_{1,2}\right)'W\mathrm{vec}\left(\Delta_{1,2}\right)}{\left(\mathrm{vec}\left(\Delta_{1,1,0}\right)'W\mathrm{vec}\left(\Delta_{1,1,0}\right)\right)^{5}}\right)^{1/24}T^{-1/6},\\
\widetilde{b}_{2,T}^{\mathrm{opt}} & =3.56\left(\frac{\mathrm{vec}\left(\Delta_{1,1,0}\right)'W\mathrm{vec}\left(\Delta_{1,1,0}\right)}{\left(\mathrm{vec}\left(\Delta_{1,2}\right)'W\mathrm{vec}\left(\Delta_{1,2}\right)\right)^{5}}\right)^{1/24}T^{-1/6}.
\end{align*}
Furthermore, the optimal kernels are given by $K_{1}^{\mathrm{opt}}=K_{1}^{\mathrm{QS}}$
and $K_{2}^{\mathrm{opt}}\left(x\right)=6x\left(1-x\right)$ for $x\in\left[0,\,1\right]$. 
\end{thm}
The requirement $\int_{0}^{1}||f^{\left(2\right)}\left(u,\,0\right)||du<\infty$
is not stringent and reduces to the one used by \citet{andrews:91}
when $\left\{ V_{t,T}\right\} $ is stationary. Note that $\Delta_{1,1,0}$
accounts for the relative variation of $\int_{0}^{1}f\left(u,\,\omega\right)$
around $\omega=0$ whereas $\Delta_{1,2}$ accounts for the relative
time variation (i.e., nonstationarity). The theorem states that as
$\Delta_{1,1,0}$ increases $\widetilde{b}_{1,T}^{\mathrm{opt}}$
becomes smaller while $\widetilde{b}_{2,T}^{\mathrm{opt}}$ becomes
larger. This is intuitive. With more variation around the zero frequency,
more smoothing is required over the frequency direction and less over
the time direction. Conversely, the more nonstationary is the data
the more smoothing is required over the time direction (i.e., $\widetilde{b}_{2,T}^{\mathrm{opt}}$
is smaller and the optimal block length $T\widetilde{b}_{2,T}^{\mathrm{opt}}$
smaller) relative to the frequency direction. Both optimal bandwidths
$(\widetilde{b}_{1,T}^{\mathrm{opt}},\,\widetilde{b}_{2,T}^{\mathrm{opt}})$
have the same order  $O(T^{-1/6}).$ We can compare it with $b_{1,T}^{\mathrm{opt}}=O(T^{-4/25})$
and $\overline{b}_{2,T}^{\mathrm{opt}}=O(T^{-1/5})$ resulting from
the sequential MSE criterion in \citet{casini_hac}. The latter leads
to a slightly smaller block length relative to the global criterion
\eqref{Eq. (Joint MSE Criterion)} {[}i.e., $O(T\overline{b}_{2,T}^{\mathrm{opt}})<O(T\widetilde{b}_{2,T}^{\mathrm{opt}})${]}.
Since $K_{2}$ applies overlapping smoothing, a smaller block length
is beneficial if there is substantial nonstationarity. On the same
note, a smaller block length is less exposed to low frequency contamination
since it allows to better account for nonstationarity. The rate of
convergence when the optimal bandwidths are used is $O(T^{1/3})$
which is sightly faster than the corresponding rate of convergence
when $(b_{1,T}^{\mathrm{opt}},\,b_{2,T}^{\mathrm{opt}})$. The latter
is $O(T^{0.32})$, so the difference is small. 

\section{Data-Dependent Bandwidths\label{Section Data-Dependent-Bandwidths}}

In this section we consider estimators $\widehat{J}_{T}$ that use
bandwidths $b_{1,T}$ and $b_{2,T}$ whose values are determined via
data-dependent methods.  We use the ``plug-in'' method which
is characterized by plugging-in estimates of unknown quantities into
a formula for an optimal bandwidth parameter (i.e., the expressions
for $\widetilde{b}_{1,T}^{\mathrm{opt}}$  and $\widetilde{b}_{2,T}^{\mathrm{opt}}$).
 Section \ref{subsec:Implementation} discusses the implementation
of the automatic bandwidths, while Section \ref{subsec:Theoretical-Results}
presents the corresponding theoretical results. 

\subsection{\label{subsec:Implementation}Implementation}

 The first step for the construction of data-dependent bandwidth
parameters is to specify $p$ univariate  parametric models for 
$V_{t}=(V_{t}^{\left(1\right)},\ldots,\,V_{t}^{\left(p\right)})$.
The second step involves the estimation of the parameters of the
parametric models.  Here standard estimation methods are local least-squares
(LS) (i.e., LS method applied to rolling windows) and nonparametric
kernel methods. Let 
\begin{align*}
\phi_{1}\triangleq\frac{\mathrm{vec}\left(\Delta_{1,2}\right)'W\mathrm{vec}\left(\Delta_{1,2}\right)}{\left(\mathrm{vec}\left(\Delta_{1,1,0}\right)'W\mathrm{vec}\left(\Delta_{1,1,0}\right)\right)^{5}}, & \qquad\phi_{2}\triangleq\frac{\mathrm{vec}\left(\Delta_{1,1,0}\right)'W\mathrm{vec}\left(\Delta_{1,1,0}\right)}{\left(\mathrm{vec}\left(\Delta_{1,2}\right)'W\mathrm{vec}\left(\Delta_{1,2}\right)\right)^{5}}.
\end{align*}
In a third step, we replace the unknown parameters in $\phi_{1}$
and $\phi_{2}$ with corresponding estimates. Such estimates $\widehat{\phi}_{1}$
and $\widehat{\phi}_{2}$ are then substituted into the expression
for $\widetilde{b}_{1,T}^{\mathrm{opt}}$ and $\widetilde{b}_{2,T}^{\mathrm{opt}}$
to yield  
\begin{align}
\widehat{b}_{1,T}=0.46\phi_{1}^{1/24}T^{-1/6}, & \qquad\widehat{b}_{2,T}=3.56\phi_{2}^{1/24}T^{-1/6}.\label{Eq. (6.1) Andrews 91}
\end{align}
In practice, a reasonable candidate to be used as an approximating
parametric model is the first order autoregressive {[}AR(l){]} model
for $\{V_{t}^{\left(r\right)}\},\,r=1,\ldots,\,p$ (with different
parameters for each $r$) or a first order vector autoregressive {[}VAR(l){]}
model for $\{V_{t}\}$ {[}see \citet{andrews:91}{]}. However, in
our context it is reasonable to allow the parameters to be time-varying.
For parsimony, we consider time-varying AR(1) models with no break
points in the spectrum (i.e., $V_{t}^{\left(r\right)}=a_{1}\left(t/T\right)V_{t-1}^{\left(r\right)}+u_{t}^{\left(r\right)}$).

The use of $p$ univariate parametric models requires $W$ to be
a diagonal matrix. This leads to $\phi_{1}=\phi_{1,1}/\phi_{1,2}^{5}$
and $\phi_{2}=\phi_{1,2}/\phi_{1,1}^{5}$ where 
\begin{align*}
\phi_{1,1} & =\sum_{r=1}^{p}W^{\left(r,r\right)}\left(\sum_{k=-\infty}^{\infty}\int_{0}^{1}\frac{\partial^{2}}{\partial u^{2}}c^{\left(r,r\right)}\left(u,\,k\right)du\right)^{2}/\left(\int_{0}^{1}f^{\left(r,r\right)}\left(u,\,0\right)du\right)^{2}.\\
\phi_{1,2} & =\sum_{r=1}^{p}W^{\left(r,r\right)}\left(\int_{0}^{1}f^{\left(q\right)\left(r,r\right)}\left(u,\,0\right)du\right)^{2}/\left(\int_{0}^{1}f^{\left(r,r\right)}\left(u,\,0\right)du\right)^{2}.
\end{align*}
The usual choice is $W^{\left(r,r\right)}=1$ for all $r$. 
An estimate of $f^{\left(r,r\right)}\left(u,\,0\right)$ $\left(r=1,\ldots,\,p\right)$
is $\widehat{f}^{\left(r,r\right)}\left(u,\,0\right)=\left(2\pi\right)^{-1}(\widehat{\sigma}^{\left(r\right)}\left(u\right))^{2}(1-\widehat{a}_{1}^{\left(r\right)}\left(u\right))^{-2}$
while $f^{\left(2\right)\left(r,r\right)}\left(u,\,0\right)$ can
be estimated by $\widehat{f}^{\left(2\right)\left(r,r\right)}\left(u,\,0\right)=3\pi^{-1}$
$((\widehat{\sigma}^{\left(r\right)}\left(u\right))^{2}\widehat{a}_{1}^{\left(r\right)}\left(u\right))(1-\widehat{a}_{1}^{\left(r\right)}\left(u\right))^{-4}$
where $\widehat{a}_{1}^{\left(r\right)}\left(u\right)$ and $\widehat{\sigma}^{\left(r\right)}\left(u\right)$
are the LS estimates computed using local data to the left of $u=t/T$:
\begin{align}
\widehat{a}_{1}^{\left(r\right)}\left(u\right) & =\frac{\sum_{j=t-n_{2,T}+1}^{t}\widehat{V}_{j}^{\left(r\right)}\widehat{V}_{j-1}^{\left(r\right)}}{\sum_{j=t-n_{2,T}+1}^{t}\left(\widehat{V}_{j-1}^{\left(r\right)}\right)^{2}},\qquad\widehat{\sigma}^{\left(r\right)}\left(u\right)=\left(\sum_{j=t-n_{2,T}+1}^{t}\left(\widehat{V}_{j}^{\left(r\right)}-\widehat{a}_{1}^{\left(r\right)}\left(u\right)\widehat{V}_{j-1}^{\left(r\right)}\right)^{2}\right)^{1/2},\label{eq: alpha and sigma}
\end{align}
where $n_{2,T}\rightarrow\infty$. More complex is the estimation
of $\overline{\Delta}_{1,2,1}\triangleq\sum_{k=-\infty}^{\infty}\int_{0}^{1}\left(\partial^{2}/\partial u^{2}\right)c\left(u,\,k\right)$
because it involves the second partial derivative of $c\left(u,\,k\right).$
We need a further parametric assumption. We assume  that the parameters
of the approximating time-varying AR(1) models change slowly such
that the smoothness of $f\left(\cdot,\,\omega\right)$ and thus of
$c\left(\cdot,\,\cdot\right)$ is the same to the one that would arise
if $a_{1}\left(u\right)=0.8\left(\cos1.5+\cos4\pi u\right)$ and $\sigma\left(u\right)=\sigma=1$
for all $u\in\left[0,\,1\right]$ {[}cf. \citet{dahlhaus:12}{]}.
Then,  $\Delta_{1,2,1}^{\left(r,r\right)}\left(u,\,k\right)\triangleq\left(\partial^{2}/\partial u^{2}\right)c^{\left(r,r\right)}\left(u,\,k\right)$
can be computed analytically: 
\begin{align*}
\Delta_{1,2,1}^{\left(r,r\right)}\left(u,\,k\right) & =\int_{-\pi}^{\pi}e^{ik\omega}\left[\frac{3}{\pi}\left(1+0.8\left(\cos1.5+\cos4\pi u\right)\exp\left(-i\omega\right)\right)^{-4}\left(0.8\left(-4\pi\sin\left(4\pi u\right)\right)\right)\exp\left(-i\omega\right)\right.\\
 & \quad\left.-\frac{1}{\pi}\left|1+0.8\left(\cos1.5+\cos4\pi u\right)\exp\left(-i\omega\right)\right|^{-3}\left(0.8\left(-16\pi^{2}\cos\left(4\pi u\right)\right)\right)\exp\left(-i\omega\right)\right]d\omega).
\end{align*}
 An estimate of $\Delta_{1,2,1}^{\left(r,r\right)}\left(u,\,k\right)$
is given by
\begin{align*}
\widehat{\Delta}_{1,2,1}^{\left(r,r\right)} & \left(u,\,k\right)\triangleq\\
 & \left[S_{\omega}\right]^{-1}\sum_{s\in S_{\omega}}e^{ik\omega_{s}}\left[\frac{3}{\pi}\left(1+0.8\left(\cos1.5+\cos4\pi u\right)\exp\left(-i\omega_{s}\right)\right)^{-4}\left(0.8\left(-4\pi\sin\left(4\pi u\right)\right)\right)\exp\left(-i\omega_{s}\right)\right.\\
 & \left.-\frac{1}{\pi}\left|1+0.8\left(\cos1.5+\cos4\pi u\right)\exp\left(-i\omega_{s}\right)\right|^{-3}\left(0.8\left(-16\pi^{2}\cos\left(4\pi u\right)\right)\right)\exp\left(-i\omega_{s}\right)\right],
\end{align*}
where $\left[S_{\omega}\right]$ is the cardinality of $S_{\omega}$
and $\omega_{s+1}>\omega_{s}$ with $\omega_{1}=-\pi,\,\omega_{\left[S_{\omega}\right]}=\pi.$
In our simulations, we use $S_{\omega}=\left\{ -\pi,\,-3,\,-2,\,-1,\,0,\,1,\,2,\,3,\,\pi\right\} $.
 We can average over $u$ and sum over $k$ to obtain an estimate
of $\overline{\Delta}_{1,2,1}^{\left(r,r\right)}:$ $\widehat{\overline{\Delta}}_{1,2,1}^{\left(r,r\right)}=\sum_{k=-\left\lfloor T^{1/6}\right\rfloor }^{\left\lfloor T^{1/6}\right\rfloor }\frac{n_{3,T}}{T}\sum_{j=0}^{\left\lfloor T/n_{3,T}\right\rfloor }\widehat{\Delta}_{1,2,1}^{\left(r,r\right)}\left(jn_{T}/T,\,k\right)$
where the number of summands over $k$ grows at the same rate as
 $1/\widetilde{b}_{1,T}^{\mathrm{opt}}$; a different choice is allowed
as long as it grows at a slower rate than $T^{2/5}$ but our sensitivity
analysis does not indicate significant changes. 

  Then, $\widehat{b}_{1,T}=0.46\widehat{\phi}_{1}^{1/24}T^{-1/6}$
and $\widehat{b}_{2,T}=3.56\widehat{\phi}_{2}^{1/24}T^{-1/6}$ where
$\widehat{\phi}_{1}=\widehat{\phi}_{1,1}/\widehat{\phi}_{1,2}^{5}$,
$\widehat{\phi}_{2}=\widehat{\phi}_{1,2}/\widehat{\phi}_{1,1}^{5},$
 
\begin{align*}
\widehat{\phi}_{1,1} & =\left(4\pi\right)^{-2}\sum_{r=1}^{p}W^{\left(r,r\right)}\left(\widehat{\overline{\Delta}}_{1,2,1}^{\left(r,r\right)}\right)^{2}/\left(\frac{n_{3,T}}{T}\sum_{j=0}^{\left\lfloor T/n_{3,T}\right\rfloor -1}(\widehat{\sigma}^{\left(r\right)}\left(jn_{3,T}+1\right))^{2}(1-\widehat{a}_{1}^{\left(r\right)}\left(jn_{3,T}+1\right))^{-2}\right)^{2},\\
\widehat{\phi}_{1,2} & =36\sum_{r=1}^{p}W^{\left(r,r\right)}\left(\frac{\frac{n_{3,T}}{T}\sum_{j=0}^{\left\lfloor T/n_{3,T}\right\rfloor -1}((\widehat{\sigma}^{\left(r\right)}\left(jn_{3,T}+1\right))^{2}\widehat{a}_{1}^{\left(r\right)}\left(jn_{3,T}+1\right))(1-\widehat{a}_{1}^{\left(r\right)}\left(jn_{3,T}+1\right))^{-4}}{\frac{n_{3,T}}{T}\sum_{j=0}^{\left\lfloor T/n_{3,T}\right\rfloor -1}(\widehat{\sigma}^{\left(r\right)}\left(jn_{3,T}+1\right))^{2}(1-\widehat{a}_{1}^{\left(r\right)}\left(jn_{3,T}+1\right))^{-2}}\right)^{2}.
\end{align*}
 For most of the results below we can take $n_{3,T}=n_{2,T}=n_{T}.$ 

\subsection{\label{subsec:Theoretical-Results}Theoretical Results}

We establish  results corresponding to Theorem \ref{Theorem 1 -Consistency and Rate}
for the estimator $\widehat{J}_{T}(\widehat{b}_{1,T},\,\widehat{b}_{2,T})$
that uses  $\widehat{b}_{1,T}$ and $\widehat{b}_{2,T}$. We restrict
the class of admissible kernels to the following, 
\begin{align*}
\boldsymbol{K}_{3} & =\left\{ K_{3}\left(\cdot\right)\in\boldsymbol{K}_{1}:\,\left(i\right)\,\left|K_{1}\left(x\right)\right|\leq C_{1}\left|x\right|^{-b}\,\mathrm{with\,}b>\max\left(1+1/q,\,3\right)\,\mathrm{for}\,\left|x\right|\in\left[\overline{x}_{L},\,D_{T}h_{T}\overline{x}_{U}\right],\right.\\
 & \quad\,b_{1,T}^{2}h_{T}\rightarrow\infty,\,D_{T}>0,\,\overline{x}_{L},\,\overline{x}_{U}\in\mathbb{R},\,1\leq\overline{x}_{L}<\overline{x}_{U},\,\mathrm{and}\,\\
 & \quad\mathrm{with\,}b>1+1/q\,\mathrm{for}\,\left|x\right|\notin\left[\overline{x}_{L},\,D_{T}h_{T}\overline{x}_{U}\right],\,\mathrm{and\,some\,}C_{1}<\infty,\\
 & \quad\mathrm{where}\,q\in\left(0,\,\infty\right)\,\mathrm{is\,such\,that\,}K_{1,q}\in\left(0,\,\infty\right),\,\left(ii\right)\,\left|K_{1}\left(x\right)-K_{1}\left(y\right)\right|\leq C_{2}\left|x-y\right|\,\forall x,\\
 & \quad\left.y\in\mathbb{R}\,\mathrm{for\,some\,costant\,}C_{2}<\infty\right\} .
\end{align*}
Let $\widehat{\theta}$ denote the estimator of the parameter of the
approximate (time-varying) parametric model(s) introduced above {[}i.e.,
$\widehat{\theta}=(\int_{0}^{1}\widehat{a}_{1}\left(u\right)du,\,\int_{0}^{1}\widehat{\sigma}_{1}^{2}\left(u\right)du,\ldots,\,\int_{0}^{1}\widehat{a}_{p}^{2}\left(u\right)du,\,\int_{0}^{1}\widehat{\sigma}_{p}^{2}\left(u\right)du)'${]}.
Let $\theta^{*}$ denote the probability limit of $\widehat{\theta}$.
$\widehat{\phi}_{1}$ and $\widehat{\phi}_{2}$ are the values of
$\phi_{1}$ and $\phi_{2}$, receptively, with $\widehat{\theta}$
instead of $\theta$. The probability limits of $\widehat{\phi}_{1}$
and $\widehat{\phi}_{2}$ are denoted by $\phi_{1,\theta^{*}}$ and
$\phi_{2,\theta^{*}},$ respectively.
\begin{assumption}
\label{Assumption E-F-G}(i) $\widehat{\phi}_{1}=O\mathbb{_{P}}\left(1\right),$
$1/\widehat{\phi}_{1}=O\mathbb{_{P}}\left(1\right),$ $\widehat{\phi}_{2}=O\mathbb{_{P}}\left(1\right),$
and $1/\widehat{\phi}_{2}=O\mathbb{_{P}}\left(1\right)$; (ii) $\inf\bigl\{ T/n_{3,T},\,\sqrt{n_{2,T}}\bigr\}((\widehat{\phi}_{1}-\phi_{1,\theta^{*}}),\,(\widehat{\phi}_{2}-\phi_{2,\theta^{*}}))'=O_{\mathbb{P}}\left(1\right)$
for some $\phi_{1,\theta^{*}},\,\phi_{2,\theta^{*}}\in\left(0,\,\infty\right)$
where $n_{2,T}/T+n_{3,T}/T\rightarrow0,$ $n_{2,T}^{10/6}/T\rightarrow[c_{2},\,\infty),$
$n_{3,T}^{10/6}/T\rightarrow[c_{3},\,\infty)$ with $0<c_{2},\,c_{3}<\infty$;
(iii) $\sup_{u\in\left[0,\,1\right]}\lambda_{\max}(\Gamma_{u}\left(k\right))$
$\leq C_{3}k^{-l}$ for all $k\geq0$ for some $C_{3}<\infty$ and
some $l>3$, where $q$ is as in $\boldsymbol{K}_{3}$; (iv) $|\omega_{s+1}-\omega_{s}|=O\left(T^{-1}\right)$
and $\left[S_{\omega}\right]=O\left(T\right)$; (v) $\boldsymbol{K}_{2}$
includes kernels that satisfy $|K_{2}\left(x\right)-K_{2}\left(y\right)|\leq C_{4}\left|x-y\right|$
for all $x,\,y\in\mathbb{R}$ and some constant $C_{4}<\infty$.
\end{assumption}
Parts (i) and (v) are sufficient for the consistency of $\widehat{J}_{T}(\widehat{b}_{1,T},\,\widehat{b}_{2,T}).$
Parts (ii)-(iii) and (iv)-(v) are required for the rate of convergence
and  MSE results. Note that $\phi_{1,\theta^{*}}$ and $\phi_{2,\theta^{*}}$
coincide with the optimal values $\phi_{1}$ and $\phi_{2}$, respectively,
only when the approximate parametric model indexed by $\theta^{*}$
corresponds to the true data-generating mechanism.  

Let $b_{\theta_{1},T}=0.46\phi_{1,\theta^{*}}^{1/24}T^{-1/6}$ and
$b_{\theta_{2},T}=3.56\phi_{2,\theta^{*}}^{1/24}T^{-1/6}$. The asymptotic
properties of $\widehat{J}_{T}(\widehat{b}_{1,T},\,\widehat{b}_{2,T})$
are shown to be equivalent to those of $\widehat{J}_{T}(b_{\theta_{1},T},\,b_{\theta_{2},T})$
where the theoretical properties of the latter follow from Theorem
\ref{Theorem 1 -Consistency and Rate}.
\begin{thm}
\label{Theorem 3 Andrews 91}Suppose $K_{1}\left(\cdot\right)\in\boldsymbol{K}_{3}$,
$q$ is as in $\boldsymbol{K}_{3}$, $K_{2}\left(\cdot\right)\in\boldsymbol{K}_{2}$,\textbf{
}$n_{T}\rightarrow\infty,\,n_{T}/Tb_{\theta_{1},T}\rightarrow0,$
and $||\int_{0}^{1}f^{\left(q\right)}\left(u,\,0\right)du||<\infty$.
Then, we have:

(i) If Assumption \ref{Assumption Smothness of A (for HAC)}-\ref{Assumption B}
and \ref{Assumption E-F-G}-(i,v) hold and $n_{3,T}=n_{2,T}=n_{T},$
 then $\widehat{J}_{T}(\widehat{b}_{1,T},\,\widehat{b}_{2,T})-J_{T}\overset{\mathbb{P}}{\rightarrow}0$.

(ii) If Assumption \ref{Assumption Smothness of A (for HAC)}, \ref{Assumption B}-\ref{Assumption C Andrews 91}
and \ref{Assumption E-F-G}-(ii,iii,iv,v) hold $q\leq2$ and $n_{T}/Tb_{\theta_{1},T}^{2}\rightarrow0,$
then $\sqrt{Tb_{\theta_{1},T}b_{\theta_{2},T}}$ $(\widehat{J}_{T}(\widehat{b}_{1,T},\,\widehat{b}_{2,T})-J_{T})=O_{\mathbb{P}}\left(1\right)$
and $\sqrt{Tb_{\theta_{1},T}b_{\theta_{2},T}}(\widehat{J}_{T}(\widehat{b}_{1,T},\,\widehat{b}_{2,T})-\widehat{J}_{T}(b_{\theta_{1},T},\,b_{\theta_{2},T}))=o_{\mathbb{P}}\left(1\right)$.

(iii)  If Assumption \ref{Assumption Smothness of A (for HAC)},
\ref{Assumption B}-\ref{Assumption W_T and unbounded kernel and Cumulant 8}
and \ref{Assumption E-F-G}-(ii,iii,iv,v) hold, then
\begin{align*}
\lim_{T\rightarrow\infty} & \mathrm{MSE}\left(T^{2/3},\,\widehat{J}_{T}\left(\widehat{b}_{1,T},\,\widehat{b}_{2,T}\right),\,W_{T}\right)\\
 & =\lim_{T\rightarrow\infty}\mathrm{MSE}\left(Tb_{\theta_{1},T}b_{\theta_{2},T},\,\widehat{J}_{T}\left(b_{\theta_{1},T},\,b_{\theta_{2},T}\right),\,W_{T}\right).
\end{align*}
\end{thm}
When the chosen parametric model indexed by $\theta$ is correct,
it follows that $\phi_{1,\theta^{*}}=\phi_{1}$, $\phi_{2,\theta^{*}}=\phi_{2}$,
$\widehat{\phi}_{1}\overset{\mathbb{P}}{\rightarrow}\phi_{1}$ and
$\widehat{\phi}_{2}\overset{\mathbb{P}}{\rightarrow}\phi_{2}$. The
theorem then implies that $\widehat{J}_{T}(\widehat{b}_{1,T},\,\widehat{b}_{2,T})$
exhibits the same optimality properties presented in Theorem \ref{Theorem Optimal Kernels}.

\section{\label{Section LRV with nonparametric rate of convergence}Consistent
LRV in the Context of Nonparametric Parameter Estimates}

We relax the assumption that $\{V_{t}(\widehat{\beta})\}$ is a function
of a semiparametric estimator $\widehat{\beta}$ satisfying $\sqrt{T}(\widehat{\beta}-\beta_{0})=O_{\mathbb{P}}\left(1\right)$.
This holds, for example, in the linear regression model estimated
by least-squares where $V_{t}(\widehat{\beta})=\widehat{e}_{t}x_{t}$
with $\{\widehat{e}_{t}\}$ being the fitted residuals and $\{x_{t}\}$
being a vector of regressors. However, there are many HAR inference
contexts where one needs an estimate of the LRV based on a sequence
of observations $\{V_{t}(\widehat{\beta}_{\mathrm{np}})\}$ where
$\widehat{\beta}_{\mathrm{np}}$ is a nonparametric estimator that
satisfies $T^{\vartheta}(\widehat{\beta}_{\mathrm{np}}-\beta_{0})=O_{\mathbb{P}}\left(1\right)$
for some $\vartheta\in\left(0,\,1/2\right)$. For example, in forecasting
one needs an estimate of the LRV to obtain a pivotal asymptotic distribution
for forecast evaluation tests while one has access to a sequence $\{V_{t}(\widehat{\beta}_{\mathrm{np}})\}$
obtained from nonparametric estimation using some in-sample. Given
that nonparametric methods have received a great deal of attention
in applied work lately, it is useful to extend the theory of HAC and
DK-HAC estimators to these settings. We consider the HAC estimators
in Section \ref{Subsection Classical-HAC-Estimators} and the DK-HAC
estimators in Section \ref{Subsection: DK-HAC-Estimators}.

\subsection{Classical HAC Estimators\label{Subsection Classical-HAC-Estimators}}

We show that the classical HAC estimators that use the data-dependent
bandwidths suggested in \citet{andrews:91} remain valid when $\sqrt{T}(\widehat{\beta}-\beta_{0})=O_{\mathbb{P}}\left(1\right)$
is replaced by $T^{\vartheta}(\widehat{\beta}_{\mathrm{np}}-\beta_{0})=O_{\mathbb{P}}\left(1\right)$
for some $\vartheta\in\left(0,\,1/2\right)$. We work under the same
assumptions as in \citet{andrews:91}. Under stationarity we have
$\Gamma_{u}\left(k\right)=\Gamma\left(k\right)$ and $\kappa_{V,\left\lfloor Tu\right\rfloor }^{\left(a,b,c,d\right)}\left(k,\,s,\,l\right)=\kappa_{V,0}^{\left(a,b,c,d\right)}\left(k,\,s,\,l\right)$
for any $u\in\left[0,\,1\right]$. 
\begin{assumption}
\label{Assumption A - Dependence Andrews91}$\left\{ V_{t}\right\} $
is a mean-zero, fourth-order stationary sequence with $\sum_{k=-\infty}^{\infty}\left\Vert \Gamma\left(k\right)\right\Vert <\infty$
and $\sum_{k=-\infty}^{\infty}\sum_{s=-\infty}^{\infty}\sum_{l=-\infty}^{\infty}|\kappa_{V,0}^{\left(a,b,c,d\right)}\left(k,\,s,\,l\right)|<\infty$
$\forall a,\,b,\,c,\,d\leq p$. 
\end{assumption}
\begin{assumption}
\label{Assumption B Andrews91}(i) $T^{\vartheta}(\widehat{\beta}_{\mathrm{np}}-\beta_{0})=O_{\mathbb{P}}\left(1\right)$
for some $\vartheta\in\left(0,\,1/2\right)$; (ii) $\sup_{t\geq1}\mathbb{E}\left\Vert V_{t}\right\Vert ^{2}<\infty$;
(iii) $\sup_{t\geq1}\mathbb{E}\sup_{\beta\in\Theta}\left\Vert \left(\partial/\partial\beta\right)V_{t}\left(\beta\right)\right\Vert ^{2}<\infty$;
(iv) $\int\left|K_{1}\left(y\right)\right|dy<\infty$. 
\end{assumption}
\begin{assumption}
\label{Assumption C Andrews 91 - Andrews91}(i) Assumption \ref{Assumption A - Dependence Andrews91}
holds with $V_{t}$ replaced by 
\begin{align*}
\left(V'_{t},\,\mathrm{vec}\left(\left(\frac{\partial}{\partial\beta'}V_{t}\left(\beta_{0}\right)\right)-\mathbb{E}\left(\frac{\partial}{\partial\beta'}V_{t}\left(\beta_{0}\right)\right)\right)'\right)' & .
\end{align*}
(ii) $\sup_{t\geq1}\mathbb{E}(\sup_{\beta\in\Theta}||\left(\partial^{2}/\partial\beta\partial\beta'\right)V_{t}^{\left(a\right)}\left(\beta\right)||^{2})<\infty$
for all $a=1,\ldots,\,p$.
\end{assumption}
Let $\widehat{b}_{\mathrm{Cla,}1,T}=\left(qK_{1,q}^{2}\widehat{\alpha}\left(q\right)T/\int K_{1}^{2}\left(x\right)dx\right)^{-1/\left(2q+1\right)}$.
The form of $\widehat{\alpha}\left(q\right)$ depends on the approximating
parametric model for $\{V_{t}^{\left(r\right)}\}$. \citet{andrews:91}
considered stationarity AR(1) models for $\{V_{t}^{\left(r\right)}\}$,
which result in
\begin{align}
\widehat{\alpha}\left(2\right) & =\sum_{r=1}^{p}W^{\left(r,r\right)}\frac{4\left(\widehat{a}_{1}^{\left(r\right)}\right)^{2}\left(\widehat{\sigma}^{r}\right)^{4}}{\left(1-\widehat{a}_{1}^{\left(r\right)}\right)^{8}}/\sum_{r=1}^{p}W^{\left(r,r\right)}\frac{\left(\widehat{\sigma}^{r}\right)^{4}}{\left(1-\widehat{a}_{1}^{\left(r\right)}\right)^{4}},\quad\mathrm{and}\label{Eq. (6.4) in Andrews91}\\
\widehat{\alpha}\left(1\right) & =\sum_{r=1}^{p}W^{\left(r,r\right)}\frac{4\left(\widehat{a}_{1}^{\left(r\right)}\right)^{2}\left(\widehat{\sigma}^{r}\right)^{4}}{\left(1-\widehat{a}_{1}^{\left(r\right)}\right)^{6}\left(1+\widehat{a}_{1}^{\left(r\right)}\right)^{2}}/\sum_{r=1}^{p}W^{\left(r,r\right)}\frac{\left(\widehat{\sigma}^{r}\right)^{4}}{\left(1-\widehat{a}_{1}^{\left(r\right)}\right)^{4}}.\nonumber 
\end{align}
 Let 
\begin{align}
\boldsymbol{K}_{\mathrm{Cla,}3} & =\left\{ K_{1}\left(\cdot\right)\in\boldsymbol{K}_{1}:\,\left(i\right)\,\left|K_{1}\left(y\right)\right|\leq C_{1}\left|y\right|^{-b}\,\mathrm{for\,some\,\mathit{b}>1+1/}q\,\mathrm{and\,some\,}C_{1}<\infty,\,\right.\label{Eq. (2.6) K3 Kernel class}\\
 & \quad\mathrm{where}\,q\in\left(0,\,\infty\right)\,\mathrm{is\,such\,that\,}K_{1,q}\in\left(0,\,\infty\right),\,\mathrm{and\,}\left(ii\right)\,\left|K_{1}\left(x\right)-K_{1}\left(y\right)\right|\leq C_{2}\left|x-y\right|\,\forall x,\nonumber \\
 & \quad\left.y\in\mathbb{R}\,\mathrm{for\,some\,costant\,}C_{2}<\infty\right\} .\nonumber 
\end{align}
 
\begin{assumption}
\label{Assumption E Andrews91}$\widehat{\alpha}\left(q\right)=O_{\mathbb{P}}\left(1\right)$
and $1/\widehat{\alpha}\left(q\right)=O_{\mathbb{P}}\left(1\right)$. 
\end{assumption}
\begin{thm}
\label{Theorem 3 Andrews91}Suppose $K_{1}\left(\cdot\right)\in\boldsymbol{K}_{3,\mathrm{Cla}}$,
$||f^{\left(q\right)}||<\infty$, $q>\left(1/\vartheta-1\right)/2$,
and Assumption \ref{Assumption A - Dependence Andrews91}-\ref{Assumption E Andrews91}
hold, then $\widehat{J}_{\mathrm{Cla,}T}(\widehat{b}_{\mathrm{Cla,1,}T})-J_{T}\overset{\mathbb{P}}{\rightarrow}0$. 
\end{thm}

\subsection{DK-HAC Estimators\label{Subsection: DK-HAC-Estimators}}

We extend the consistency result in Theorem \ref{Theorem 3 Andrews 91}-(i)
assuming that $\widehat{V}_{t}=V_{t}(\widehat{\beta}_{\mathrm{np}}).$
Thus, we replace Assumption \ref{Assumption B} by the following.
\begin{assumption}
\label{Assumption B' Nonparametric}(i) $T^{\vartheta}(\widehat{\beta}_{\mathrm{np}}-\beta_{0})=O_{\mathbb{P}}\left(1\right)$
for some $\vartheta\in\left(0,\,1/2\right)$; (ii)-(iv) from Assumption
\ref{Assumption B} continue to hold.
\end{assumption}
\begin{thm}
\label{Theorem 3 DK-HAC Nonparametric}Suppose $K_{1}\left(\cdot\right)\in\boldsymbol{K}_{3}$,
$K_{2}\left(\cdot\right)\in\boldsymbol{K}_{2}$, $T^{\vartheta}b_{\theta_{1},T}b_{\theta_{2},T}\rightarrow\infty$,\textbf{
}$n_{T}\rightarrow\infty,\,n_{T}/Tb_{\theta_{1},T}\rightarrow0,$
and $||\int_{0}^{1}f^{\left(q\right)}\left(u,\,0\right)du||<\infty$.
If Assumption \ref{Assumption Smothness of A (for HAC)}-\ref{Assumption A - Dependence},
\ref{Assumption E-F-G}-(i,v), \ref{Assumption B' Nonparametric}
hold and $n_{3,T}=n_{2,T}=n_{T},$  then $\widehat{J}_{T}(\widehat{b}_{1,T},\,\widehat{b}_{2,T})-J_{T}\overset{\mathbb{P}}{\rightarrow}0$. 
\end{thm}
Theorem \ref{Theorem 3 Andrews91}-\ref{Theorem 3 DK-HAC Nonparametric}
require different conditions on the parameter $\vartheta$ that controls
the rate of convergence of the nonparametric estimator. In Theorem
\ref{Theorem 3 Andrews91} this conditions depends on $q$ while in
Theorem \ref{Theorem 3 DK-HAC Nonparametric} it depends on $q$ through
$b_{\theta_{1},T}$ and also on the smoothing over time through $b_{\theta_{2},T}$.
For both HAC and DK-HAC estimators, the condition allows for standard
nonparametric estimators with optimal nonparametric convergence rate. 

\section{\label{Section Monte Carlo}Small-Sample Evaluations}

In this section, we conduct a Monte Carlo analysis to evaluate the
performance of the DK-HAC estimator based on the data-dependent bandwidths
determined via the joint MSE criterion \eqref{Eq. (Joint MSE Criterion)}.
 We consider HAR tests in the linear regression model as well as
HAR tests for  forecast breakdown, i.e., the test of \citet{giacomini/rossi:09}.
The linear regression models have an intercept and a stochastic regressor.
 We focus on the $t$-statistics $t_{r}=\sqrt{T}(\widehat{\beta}^{\left(r\right)}-\beta_{0}^{\left(r\right)})/\sqrt{\widehat{J}_{T}^{\left(r,r\right)}}$
where $\widehat{J}_{T}$ is an estimate of the limit of $\mathrm{Var}(\sqrt{T}(\widehat{\beta}-\beta_{0}))$
and $r=1,\,2$. $t_{1}$ is the $t$-statistic for the parameter associated
to the intercept while $t_{2}$ is associated to the stochastic regressor
$x_{t}$. We omit the discussion of the results concerning to the
$F$-test since they are qualitatively similar. Three basic regression
models are considered. We run a $t$-test on the intercept in model
M1 and a $t$-test on the coefficient of the stochastic regressor
in model M2 and M3.  The models are based on, 
\begin{align}
y_{t} & =\beta_{0}^{\left(1\right)}+\delta+\beta_{0}^{\left(2\right)}x_{t}+e_{t},\qquad\qquad t=1,\ldots,\,T,\label{eq: Model P1}
\end{align}
for the $t$-test on the intercept (i.e., $t_{1}$) and
\begin{align}
y_{t} & =\beta_{0}^{\left(1\right)}+\left(\beta_{0}^{\left(2\right)}+\delta\right)x_{t}+e_{t},\qquad\qquad t=1,\ldots,\,T,\label{eq Model P1 beta2}
\end{align}
for the $t$-test on $\beta_{0}^{\left(2\right)}$ (i.e., $t_{2}$)
where $\delta=0$ under the null. We consider the following models:
\begin{itemize}
\item M1: $e_{t}=0.4e_{t-1}+u_{t},\,u_{t}\sim\mathrm{i.i.d.\,}\mathscr{N}\left(0,\,0.5\right),$
$x_{t}\sim\mathrm{i.i.d.\,}\mathscr{N}\left(1,\,1\right)$, $\beta_{0}^{\left(1\right)}=0$
and $\beta_{0}^{\left(2\right)}=1.$ 
\item M2: $e_{t}=0.4e_{t-1}+u_{t},\,u_{t}\sim\mathrm{i.i.d.\,}\mathscr{N}\left(0,\,1\right),$
$x_{t}\sim\mathrm{i.i.d.\,}\mathscr{N}\left(1,\,1\right)$, and $\beta_{0}^{\left(1\right)}=\beta_{0}^{\left(2\right)}=0.$ 
\item  M3: segmented locally stationary errors $e_{t}=\rho_{t}e_{t-1}+u_{t},\,u_{t}\sim\mathrm{i.i.d.\,}\mathscr{N}\left(0,\,1\right),\,\rho_{t}=\max\{0,\,-1$
$\left(\cos\left(1.5-\cos\left(5t/T\right)\right)\right)\}$\footnote{That is, $\rho_{t}$ varies smoothly between 0 and 0.8071.}
for $t\notin\left(4T/5+1,\,4T/5+h\right)$ and $e_{t}=0.99e_{t-1}+u_{t},\,u_{t}\sim\mathrm{\mathrm{i.i.d.}}$
$\mathscr{N}\left(0,\,1\right)$ for $t\in\left(4T/5+1,\,4T/5+h\right)$
where $h=10$ for $T=200$ and $h=30$ for $T=400$, and $x_{t}=1+0.6x_{t-1}+u_{X,t},\,u_{X,t}\sim\mathrm{i.i.d.\,}\mathscr{N}\left(0,\,1\right)$.
\end{itemize}
Finally, we consider model M4 which we use to investigate the performance
of \citeauthor{giacomini/rossi:09}'s (2009) test for forecast breakdown.
Suppose we want to forecast a variable $y_{t}$ generated by $y_{t}=\beta_{0}^{\left(1\right)}+\beta_{0}^{\left(2\right)}x_{t-1}+e_{t}$
where $x_{t}\sim\mathrm{i.i.d.\,}\mathscr{N}\left(1,\,1.2\right)$
and $e_{t}=0.3e_{t-1}+u_{t}$ with $u_{t}\sim\mathrm{i.i.d.\,}\mathscr{N}\left(0,\,1\right)$.
For a given forecast model and forecasting scheme, the test of \citet{giacomini/rossi:09}
detects a forecast breakdown when the average of the out-of-sample
losses differs significantly from the average of the in-sample losses.
The in-sample is used to obtain estimates of $\beta_{0}^{\left(1\right)}$
and $\beta_{0}^{\left(2\right)}$ which are in turn used to construct
out-of-sample forecasts $\widehat{y}_{t}=\widehat{\beta}_{0}^{\left(1\right)}+\widehat{\beta}_{0}^{\left(2\right)}x_{t-1}$.
We set $\beta_{0}^{\left(1\right)}=\beta_{0}^{\left(2\right)}=1.$
We consider a fixed forecasting scheme. GR's (2009) test statistic
is defined as $t^{\mathrm{GR}}\triangleq\sqrt{T_{n}}\overline{SL}/\sqrt{\widehat{J}_{SL}}$
where $\overline{SL}\triangleq T_{n}^{-1}\sum_{t=T_{m}}^{T-\tau}SL_{t+\tau}$,
$SL_{t+\tau}$ is the surprise loss at time $t+\tau$ (i.e., the difference
between the time $t+\tau$ out-of-sample loss and in-sample loss,
$SL_{t+\tau}=L_{t+\tau}-\overline{L}_{t+\tau}$), $T_{n}$ is the
sample size in the out-of-sample, $T_{m}$ is the sample size in the
in-sample and $\widehat{J}_{SL}$ is a LRV estimator. We restrict
attention to one-step ahead forecasts (i.e., $\tau=1$). Under $H_{1}:\,\mathbb{E}(\overline{SL})\neq0$,
we have $y_{t}=1+x_{t-1}+\delta x_{t-1}\mathbf{1}\left\{ t>T_{1}^{0}\right\} +e_{t}$,
where $T_{1}^{0}=T\lambda_{1}^{0}$ with $\lambda_{1}^{0}=0.7$. Under
this specification there is a break in the coefficient associated
with $x_{t-1}$. Thus, there is a forecast instability or failure
and the test of \citet{giacomini/rossi:09} should reject $H_{0}$.
We set $T_{m}=0.4T$ and $T_{n}=0.6T$.

Throughout our study we consider the following LRV estimators: $\widehat{J}_{T}(\widehat{b}_{1,T},\,\widehat{b}_{2,T})$
with $K_{1}^{\mathrm{opt}},\,K_{2}^{\mathrm{opt}}$ and automatic
bandwidths; the same $\widehat{J}_{T}(\widehat{b}_{1,T},\,\widehat{b}_{2,T})$
with in addition the prewhitening of \citet{casini/perron_PrewhitedHAC};
\citeauthor{casini_hac}'s (2021) DK-HAC with $K_{1}^{\mathrm{opt}},\,K_{2}^{\mathrm{opt}}$
and automatic bandwidths from the sequential method; DK-HAC with
prewhitening; Newey and West's (1987) HAC estimator with the automatic
bandwidth as proposed in \citet{newey/west:94}; the same with the
prewhitening procedure of \citet{andrews/monahan:92}; Newey-West
estimator with the fixed-$b$ method of \citet{Kiefer/vogelsang/bunzel:00};
the Empirical Weighted Cosine (EWC) of \citet{lazarus/lewis/stock:17}.\footnote{We have excluded Andrews's (1991) HAC estimator since its performance
is similar to that of the Newey-West estimator.} \citet{casini/perron_PrewhitedHAC} proposed three methods related
to prewhitening: (1) $\widehat{J}_{T,\mathrm{pw},1}$ uses a stationary
model to whiten the data; (2) $\widehat{J}_{T,\mathrm{pw},\mathrm{SLS}}$
uses a nonstationary model to whiten the data; (3) $\widehat{J}_{T,\mathrm{pw},\mathrm{SLS},\mu}$
is the same as $\widehat{J}_{T,\mathrm{pw},\mathrm{SLS}}$ but it
adds a time-varying intercept in the VAR to whiten the data. 

We set $n_{T}=T^{0.66}$ as explained in \citet{casini_hac} and $n_{2,T}=n_{3,T}=n_{T}.$
Simulation results for models involving ARMA, ARCH and heteroskedastic
errors are not discussed here because the results are qualitatively
equivalent. The significance level is $\alpha=0.05$ throughout.

\subsection{Empirical Sizes of HAR Inference Tests}

Table \ref{Table S1-S2}-\ref{Table S3-S4} report the rejection rates
for model M1-M4. As a general pattern, we confirm previous evidence
that Newey-West's (1987) HAC estimator leads to $t$-tests that are
oversized when the data are stationary and there is substantial dependence
{[}cf. model M1-M2{]}. This is a long-discussed issue in the literature.
Newey-West with prewhitening is often effective in reducing the oversize
problem under stationarity. However, the simulation results below
and in the literature show that the prewhitened Newey-West-based tests
can be oversized when there is high serial dependence. Among the existing
methods, the rejection rates of the Newey-West-based tests with fixed-$b$
are accurate in model M1-M2. Overall, the results in the literature
along with those in \citet{casini_hac} and \citet{casini/perron_PrewhitedHAC}
showed that under stationarity the original fixed-$b$ method of KVB
is the method which is in general the least oversized across different
degrees of dependence among all existing methods. EWC performs similarly
to KVB's fixed-$b$. Among the recently introduced DK-HAC estimators,
Table \ref{Table S1-S2} reports evidence that the non-prewhitened
DK-HAC from \citet{casini_hac} leads to HAR tests that are a bit
oversized whereas the tests based on the new DK-HAC with simultaneous
data-dependent bandwidths, $\widehat{J}_{T}(\widehat{b}_{1,T},\,\widehat{b}_{2,T})$,
are more accurate. The results also show that the tests based on the
prewhitened DK-HAC estimators are competitive with those based on
KVB's fixed-$b$ in controlling the size. In particular, tests based
on $\widehat{J}_{T}(\widehat{b}_{1,T},\,\widehat{b}_{2,T})$ with
prewhitening are more accurate than those using the prewhitened DK-HAC
with sequential data-dependent bandwidths. Since $\widehat{J}_{T,\mathrm{pw},1}$
uses a stationarity VAR model to whiten the data, it works as well
as $\widehat{J}_{T,\mathrm{pw},\mathrm{SLS}}$ and $\widehat{J}_{T,\mathrm{pw},\mathrm{SLS},\mu}$
when stationarity actually holds, as documented in Table \ref{Table S1-S2}. 

Turning to nonstationary data and to the GR test, Table \ref{Table S3-S4}
casts concerns about the finite-sample performance of existing methods
in this context. For both model M3 and M4, existing long-run variance
estimators lead to HAR tests that have either size equal or close
to zero. The methods that use long bandwidths (i.e., many lagged autocovariances)
such as KVB's fixed-$b$ and EWC suffer most from this problem relative
to using the Newey-West estimator. This is demonstrated in \citet{casini/perron_Low_Frequency_Contam_Nonstat:2020}
who showed theoretically that nonstationarity induces positive bias
for each sample autocovariance. That bias is constant across lag orders.
Since existing LRV estimators are weighted sum of sample autocovariances
(or weighted sum of periodogram ordinates), the more lags are included
the larger is the positive bias. Thus, LRV estimators are inflated
and HAR tests have lower rejection rates than the significance level.
As we show below, this mechanism has consequences for power as well.
In model M3-M4, tests based on the non-prewhitened DK-HAC $\widehat{J}_{T}(\widehat{b}_{1,T},\,\widehat{b}_{2,T})$
performs well although tests based on the prewhitened DK-HAC  are
more accurate. $\widehat{J}_{T,\mathrm{pw},1}$ leads to tests that
are slightly less accurate because it uses stationarity and when the
latter is violated its performance is affected. In model M4, KVB's
fixed-$b$ and prewhitened DK-HAC are associated to rejection rates
relatively close to the significance level.

In summary, the prewhitened DK-HAC estimators yield $t$-test in regression
models with rejection rates that are relatively close to the exact
size. The DK-HAC with simultaneous bandwidths developed in Section
\ref{Section Data-Dependent-Bandwidths} performs better (i.e., the
associated null rejection rates are closer to the significance level
and approach it from below) than the corresponding DK-HAC estimators
with sequential bandwidths when the data are stationary. This is in
accordance with our theoretical results.  Also for nonstationary
data the simultaneous bandwidths perform in general better than the
sequential bandwidths, though the margin is smaller. The non-prewhitened
DK-HAC can lead to oversized $t$-test on the intercept if there is
high dependence. Our results confirm the oversize problem induced
by the use of the Newey-West estimators documented in the literature
under stationarity. Fixed-$b$ HAR tests control the size well when
the data are stationary but can be severely undersized under nonstationarity,
a problem that also affects tests based on the Newey-West. Thus, prewhitened
DK-HAC estimators are competitive to fixed-$b$ methods under stationarity
and they perform well also when the data are nonstationary. 

\subsection{Empirical Power of HAR Inference Tests}

For model M1-M4 we report the power results in Table \ref{Table M1 Power}-\ref{Table Power GR}.
The sample size is $T=200$.  For model M1, tests based on the Newey
and West's (1987) HAC and on the non-prewhitened DK-HAC estimators
have the highest power but they were more oversized than the tests
based on other methods. KVB's fixed-$b$ leads to $t$-tests that
sacrifices some power relative to using the prewhitened DK-HAC estimators
while EWC-based tests have lower power locally to $\delta=0$ (i.e.,
$\delta=0.1$ and $0.2$). In model M2, a similar pattern holds. HAR
tests normalized by either classical HAC or DK-HAC estimators have
similarly good power while HAR tests based on KVB's fixed-$b$ have
relatively less power. In model M3, the best power is achieved with
Newey-West's (1987) HAC estimator followed by $\widehat{J}_{T}(\widehat{b}_{1,T},\,\widehat{b}_{2,T})$
and EWC. Using KVB's fixed-$b$ leads to large power losses. In model
M4, it appears that all versions of the classical HAC estimators of
\citet{newey/west:87}, the KVB's fixed-$b$ and EWC lead to $t$-tests
that have, essentially, zero power for all $\delta$. In contrast,
the $t$-test standardized by the DK-HAC estimators have good power.
Among the latter DK-HAC estimators, the ones that use the sequential
bandwidths have slightly higher power but they margin is very small.
This follows from the usual size-power trade-off since the simultaneous
bandwidths led to tests that have more accurate size control.

The severe power problems of tests based on classical HAC estimators,
KVB's fixed-$b$ and EWC can be simply reconciled with the fact that
under the alternative hypotheses the spectrum of $V_{t}$ is not constant.
Existing estimators estimate an average of a time-varying spectrum.
Because of this instability in the spectrum, they overestimate the
dependence in $V_{t}$. \citet{casini/perron_Low_Frequency_Contam_Nonstat:2020}
showed that nonstationarity/misspecification alters the low frequency
components of a time series making the latter appear as more persistent.
Since classical HAC estimators are a weighted sum of an infinite number
of low frequency periodogram ordinates, these estimates tend to be
inflated. Similarly, LRV estimators using long bandwidths are weighted
sum of a large number of sample autocovariances. Each sample autocovariance
is biased upward so that the latter estimates are even more inflated
than the classical HAC estimators. This explains why KVB's fixed-$b$
and EWC HAR tests have large power problems, even though classical
HAC estimators are also affected.

\citet{casini/perron_Low_Frequency_Contam_Nonstat:2020} showed that
the introduction of the smoothing over time in the DK-HAC estimators
avoids such low frequency contamination. This follows because observations
belonging to different regimes do not overlap when computing sample
autocovariances. This guarantees excellent power properties also under
nonstationarity/misspecification or under nonstationary alternative
hypotheses (e.g., GR test discussed above). Simulation evidence suggests
that tests based on the DK-HAC with simultaneous bandwidths are robust
to low frequency contamination and overall performs better than tests
based on the DK-HAC with sequential bandwidths especially with respect
to size control.

\section{\label{Section Conclusions}Conclusions}

We considered the derivation of data-dependent simultaneous bandwidths
for double kernel heteroskedasticity and autocorrelation consistent
(DK-HAC) estimators. We obtained the optimal bandwidths that jointly
minimize the global asymptotic MSE criterion and discussed the trade-off
between bias and variance with respect to smoothing over lagged autocovariances
and over time. We highlighted how the derived MSE bounds are influenced
by nonstationarity unlike the MSE bounds in \citet{andrews:91}. We
compared the DK-HAC estimators with simultaneous bandwidths to the
DK-HAC estimators with bandwidths from the sequential MSE criterion.
The new method leads to HAR tests that performs better in terms of
size control, especially with stationary and close to stationary data.
Finally, we considered long-run variance estimation where the relevant
observations are a function of a nonparametric estimator and established
the validity of the HAC and DK-HAC estimators in this setting.{\footnotesize{}
}\nocite{casini/perron_Oxford_Survey,casini/perron:change-point-spectra,casini/perron_CR_Single_Break,casini/perron_Lap_CR_Single_Inf,casini/perron_Low_Frequency_Contam_Nonstat:2020,casini/perron_PrewhitedHAC,casini/perron_SC_BP_Lap,casini_CR_Test_Inst_Forecast,casini_hac,casini_diss}Hence,
we also extended the consistency results in \citet{andrews:91} and
\citet{newey/west:87} to nonparametric estimation settings.

\newpage{}

\bibliographystyle{elsarticle-harv}
\bibliography{References_JoE}

\begin{thebibliography}{62}
\expandafter\ifx\csname natexlab\endcsname\relax\def\natexlab#1{#1}\fi
\expandafter\ifx\csname url\endcsname\relax
  \def\url#1{\texttt{#1}}\fi
\expandafter\ifx\csname urlprefix\endcsname\relax\def\urlprefix{URL }\fi

\bibitem[{Altissimo and Corradi(2003)}]{altissimo/corradi:2003}
Altissimo, F., Corradi, V., 2003. Strong rules for detecting the number of
  breaks in a time series. Journal of Econometrics 117~(2), 207--244.

\bibitem[{Anderson(1971)}]{anderson:71}
Anderson, {\relax T.W}., 1971. The {S}tastical {A}nalysis of {T}ime {S}eries.
  New {Y}ork: {W}iley.

\bibitem[{Andrews(1991)}]{andrews:91}
Andrews, {\relax D.W.K}., 1991. {H}eteroskedasticity and autocorrelation
  consistent covariance matrix estimation. Econometrica 59~(3), 817--858.

\bibitem[{Andrews and Monahan(1992)}]{andrews/monahan:92}
Andrews, {\relax D.W.K}., Monahan, {\relax J.C}., 1992. An improved
  heteroskedasticity and autocorrelation consistent covariance matrix
  estimator. Econometrica 60~(4), 953--966.

\bibitem[{Bai and Perron(1998)}]{bai/perron:98}
Bai, J., Perron, P., 1998. Estimating and testing linear models with multiple
  structural changes. Econometrica 66~(1), 47--78.

\bibitem[{Belotti et~al.(2021)Belotti, Casini, Catania, Grassi, and
  Perron}]{belotti/casini/catania/grassi/perron_HAC_Sim_Bandws_Supp}
Belotti, F., Casini, A., Catania, L., Grassi, S., Perron, P., 2021. Supplement
  to "{S}imultaneous bandwidths determination for double-kernel {HAC}
  estimators and long-run variance estimation in nonparametric settings".
  Unpublished Manuscript, Department of Economics and Finance, University of
  Rome Tor Vergata.

\bibitem[{Brillinger(1975)}]{brillinger:75}
Brillinger, D., 1975. Time {S}eries {D}ata {A}nalysis and {T}heory. New York:
  Holt, Rinehart and Winston.

\bibitem[{Cai(2007)}]{cai:07}
Cai, Z., 2007. Trending time-varying coefficient time series models with
  serially correlated errors. Journal of Econometrics 136~(1), 163--188.

\bibitem[{Casini(2018)}]{casini_CR_Test_Inst_Forecast}
Casini, A., 2018. Tests for forecast instability and forecast failure under a
  continuous record asymptotic framework. arXiv preprint arXiv:1803.10883.

\bibitem[{Casini(2019)}]{casini_diss}
Casini, A., 2019. Improved methods for statistical inference in the context of
  various types of parameter variation. Ph.D dissertation, Boston University.

\bibitem[{Casini(2021)}]{casini_hac}
Casini, A., 2021. Theory of evolutionary spectra for heteroskedasticity and
  autocorrelation robust inference in possibly misspecified and nonstationary
  models. Unpublished Manuscript, Department of Economics and Finance,
  University of Rome Tor Vergata.

\bibitem[{Casini et~al.(2021)Casini, Deng, and
  Perron}]{casini/perron_Low_Frequency_Contam_Nonstat:2020}
Casini, A., Deng, T., Perron, P., 2021. Theory of low frequency contamination
  from unaccounted nonstationarity: consequences for {HAR} inference.
  Unpublished Manuscript, Department of Economics and Finance, University of
  Rome Tor Vergata.

\bibitem[{Casini and Perron(2019)}]{casini/perron_Oxford_Survey}
Casini, A., Perron, P., 2019. Structural breaks in time series. Oxford Research
  Encyclopedia of Economics and Finance, Oxford University Press.

\bibitem[{Casini and
  Perron(2020{\natexlab{a}})}]{casini/perron_Lap_CR_Single_Inf}
Casini, A., Perron, P., 2020{\natexlab{a}}. Continuous record {L}aplace-based
  inference about the break date in structural change models. Juornal of
  Econometrics forthcoming.

\bibitem[{Casini and Perron(2020{\natexlab{b}})}]{casini/perron_SC_BP_Lap}
Casini, A., Perron, P., 2020{\natexlab{b}}. Generalized {L}aplace inference in
  multiple change-points models. Econometric Theory forthcoming.

\bibitem[{Casini and
  Perron(2021{\natexlab{a}})}]{casini/perron_CR_Single_Break}
Casini, A., Perron, P., 2021{\natexlab{a}}. Continuous record asymptotics for
  change-point models. arXiv preprint arXiv:1803.10881.

\bibitem[{Casini and Perron(2021{\natexlab{b}})}]{casini/perron_PrewhitedHAC}
Casini, A., Perron, P., 2021{\natexlab{b}}. {M}inimax {MSE} bounds and
  nonlinear {VAR} prewhitening for long-run variance estimation under
  nonstattionarity. Unpublished Manuscript, Department of Economics and
  Finance, University of Rome Tor Vergata.

\bibitem[{Chan(2020)}]{chan:2020}
Chan, {\relax K.W}., 2020. Mean-structure and autocorrelation consistent
  covariance matrix estimation. Journal of Business and Economic Statistics,
  forthcoming.

\bibitem[{Chang and Perron(2018)}]{chang/perron:18}
Chang, {\relax S.Y}., Perron, P., 2018. A comparison of alternative methods to
  construct confidence intervals for the estimate of a break date in linear
  regression models. Econometric Reviews 37~(6), 577--601.

\bibitem[{Chen and Hong(2012)}]{chen/hong:12}
Chen, B., Hong, Y., 2012. Testing for smooth structural changes in time series
  models via nonparametric regression. Econometrica 80~(3), 1157--1183.

\bibitem[{Crainiceanu and Vogelsang(2007)}]{crainiceanu/vogelsang:07}
Crainiceanu, {\relax C.M}., Vogelsang, {\relax T.J}., 2007. Nonmonotonic power
  for tests of a mean shift in a time series. Journal of {S}tatistical
  {C}omputation and {S}imulation 77~(6), 457--476.

\bibitem[{Dahlhaus(1997)}]{dahlhaus:96}
Dahlhaus, R., 1997. Fitting time series models to nonstationary processes.
  Annals of Statistics 25~(1), 1--37.

\bibitem[{Dahlhaus(2012)}]{dahlhaus:12}
Dahlhaus, R., 2012. Locally stationary processes. Handbook of Statistics 30,
  351--413.

\bibitem[{Dahlhaus and Giraitis(1998)}]{Dahlhaus/Giraitis:98}
Dahlhaus, R., Giraitis, L., 1998. On the optimal segment length for parameter
  estimates for locally stationary time series. Journal of Time Series Analysis
  19~(6), 629--655.

\bibitem[{de~Jong and Davidson(2000)}]{dejong/davidson:00}
de~Jong, {\relax R.M}., Davidson, J., 2000. Consistency of kernel estimators of
  heteroskedastic and autocorrelated covariance matrices. Econometrica 68~(2),
  407--423.

\bibitem[{Deng and Perron(2006)}]{deng/perron:06}
Deng, A., Perron, P., 2006. A comparison of alternative asymptotic frameworks
  to analyse a structural change in a linear time trend. Econometrics Journal
  9~(3), 423--447.

\bibitem[{Dou(2019)}]{dou:18}
Dou, L., 2019. Optimal {HAR} inference. Unpublished Manuscript, Department of
  Economics, Princeton University.

\bibitem[{Epanechnikov(1969)}]{epanechnikov:69}
Epanechnikov, V., 1969. Non-{P}arametric {E}stimation of a {M}ultivariate
  {P}robability {D}ensity. Theory of Probability and its Applications 14~(1),
  153--158.

\bibitem[{Giacomini and Rossi(2009)}]{giacomini/rossi:09}
Giacomini, R., Rossi, B., 2009. Detecting and predicting forecast breakdowns.
  Review of Economic Studies 76~(2), 669--705.

\bibitem[{Gon{\c c}alves and Vogelsang(2011)}]{Goncalves/vogelsang:11}
Gon{\c c}alves, S., Vogelsang, {\relax T.J}., 2011. Block bootstrap {HAC}
  robust tests: the sophistication of the na\"ive bootstrap. Econometric Theory
  27~(4), 745--791.

\bibitem[{Hamilton(1989)}]{hamilton:89}
Hamilton, {\relax J.D}., 1989. A new approach to the economic analysis of
  nonstationary time series and the business cycle. Econometrica 57~(2),
  357--384.

\bibitem[{Hannan(1970)}]{hannan:70}
Hannan, {\relax E.J}., 1970. Multiple {T}ime {S}eries. New York: Wiley.

\bibitem[{Ibragimov and M{\"u}ller(2010)}]{ibragimov/muller:10}
Ibragimov, R., M{\"u}ller, {\relax U.K}., 2010. t-statistic based correlation
  and heterogeneity robust inference. Journal of Business \& Economic
  Statistics 28~(4), 453--468.

\bibitem[{Jansson(2004)}]{jansson:04}
Jansson, M., 2004. The error in rejection probability of simple autocorrelation
  robust tests. Econometrica 72~(3), 937--946.

\bibitem[{Juhl and Xiao(2009)}]{juhl/xiao:09}
Juhl, T., Xiao, Z., 2009. Testing for changing mean with monotonic power.
  Journal of Econometrics 148~(1), 14--24.

\bibitem[{Kiefer and Vogelsang(2002)}]{Kiefer/vogelsang:02}
Kiefer, N., Vogelsang, {\relax T.J}., 2002. Heteroskedasticity-autocorrelation
  robust standard errors using the {B}artlett kernel without truncation.
  Econometrica 70~(5), 2093--2095.

\bibitem[{Kiefer and Vogelsang(2005)}]{kiefer/vogelsang:05}
Kiefer, N., Vogelsang, {\relax T.J}., 2005. A new asymptotic theory for
  heteroskedasticity-autocorrelation robust tests. Econometric Theory 21~(6),
  1130--1164.

\bibitem[{Kiefer et~al.(2000)Kiefer, Vogelsang, and
  Bunzel}]{Kiefer/vogelsang/bunzel:00}
Kiefer, N., Vogelsang, {\relax T.J}., Bunzel, H., 2000. Simple robust testing
  of regression hypotheses. Econometrica 69~(3), 695--714.

\bibitem[{Kim and Perron(2009)}]{kim/perron:09}
Kim, D., Perron, P., 2009. Assessing the relative power of structural break
  tests using a framework based on the approximate {B}ahadur slope. Journal of
  Econometrics 149~(1), 26--51.

\bibitem[{Lazarus et~al.(2020)Lazarus, Lewis, and
  Stock}]{lazarus/lewis/stock:17}
Lazarus, E., Lewis, {\relax D.J}., Stock, {\relax J.H}., 2020. The size-power
  tradeoff in {HAR} inference. Econometrica, forthcoming.

\bibitem[{Lazarus et~al.(2018)Lazarus, Lewis, Stock, and
  Watson}]{lazarus/lewis/stock/watson:18}
Lazarus, E., Lewis, {\relax D.J}., Stock, {\relax J.H}., Watson, {\relax M.W}.,
  2018. {HAR} inference: recommendations for practice. Journal of Business and
  Economic Statistics 36~(4), 541--559.

\bibitem[{Martins and Perron(2016)}]{martins/perron:16}
Martins, L., Perron, P., 2016. Improved tests for forecast comparisons in the
  presence of instabilities. Journal of Time Series Analysis 37~(5), 650--659.

\bibitem[{M{\"u}ller(2007)}]{muller:07}
M{\"u}ller, {\relax U.K}., 2007. A theory of robust long-run variance
  estimation. Journal of Econometrics 141~(2), 1331--1352.

\bibitem[{M{\"u}ller(2014)}]{mueller:14}
M{\"u}ller, {\relax U.K}., 2014. {HAC} corrections for strongly autocorrelated
  time series. Journal of Business and Economic Statistics 32~(3), 311--322.

\bibitem[{Neumann and {von{ }S}achs(1997)}]{neumann/von_sachs:1997}
Neumann, {\relax M.H}., {von{ }S}achs, R., 1997. Wavelet thresholding in
  anisotropic function classes and application to adaptive estimation of
  evolutionary spectra. Annals of Statistics 25~(1), 38--76.

\bibitem[{Newey and West(1987)}]{newey/west:87}
Newey, {\relax W.K}., West, {\relax K.D}., 1987. A simple positive
  semidefinite, heteroskedastic and autocorrelation consistent covariance
  matrix. Econometrica 55~(3), 703--708.

\bibitem[{Newey and West(1994)}]{newey/west:94}
Newey, {\relax W.K}., West, {\relax K.D}., 1994. Automatic lag selection in
  covariance matrix estimation. Review of Economic Studies 61~(4), 631--653.

\bibitem[{Parzen(1957)}]{parzen:57}
Parzen, E., 1957. On consistent estimates of the spectrum of a stationary time
  series. Annals of Mathematical Statistics 28~(2), 329--348.

\bibitem[{Perron and Yamamoto(2021)}]{perron/yamamoto:18}
Perron, P., Yamamoto, Y., 2021. Testing for changes in forecast performance.
  Journal of Business and Economic Statistics 39~(1), 148--165.

\bibitem[{Phillips(2005)}]{phillips:05}
Phillips, {\relax P.C.B}., 2005. {HAC} estimation by automated regression.
  Econometric Theory 21~(1), 116--142.

\bibitem[{Politis(2011)}]{politis:11}
Politis, {\relax D.M}., 2011. Higher-{O}rder {A}ccurate, {P}ositive
  {S}emidefinite {E}stimation of {L}arge-{S}ample {C}ovariance and {S}pectral
  {D}ensity {M}atrices. Econometric Theory 27~(4), 703--744.

\bibitem[{P\"{o}tscher and Preinerstorfer(2018)}]{potscher/preinerstorfer:18}
P\"{o}tscher, {\relax B.M}., Preinerstorfer, D., 2018. Controlling the size of
  autocorrelation robust tests. Journal of Econometrics 207~(2), 406--431.

\bibitem[{P\"{o}tscher and Preinerstorfer(2019)}]{potscher/preinerstorfer:19}
P\"{o}tscher, {\relax B.M}., Preinerstorfer, D., 2019. Further results on size
  and power of heteroskedasticity and autocorrelation robust tests, with an
  application to trend testing. Electronic Journal of Statistics 13~(2),
  3893--3942.

\bibitem[{Preinerstorfer and P\"{o}tscher(2016)}]{preinerstorfer/potscher:16}
Preinerstorfer, D., P\"{o}tscher, B.~M., 2016. On size and power of
  heteroskedasticity and autocorrelation robust tests. Econometric Theory
  32~(2), 261--358.

\bibitem[{Priestley(1981)}]{priestley:85}
Priestley, {\relax M.B}., 1981. Spectral {A}nalysis and {T}ime {S}eries. Vol. I
  and II. New York: Academic Press.

\bibitem[{Robinson(1998)}]{robinson:98}
Robinson, {\relax P.M}., 1998. Inference-without smoothing in the presence of
  nonparametric autocorrelation. Econometrica 66~(5), 1163--1182.

\bibitem[{Sun(2013)}]{sun:13}
Sun, Y., 2013. Heteroscedasticity and autocorrelation robust {F} test using
  orthonormal series variance estimator. Econometrics Journal 16~(1), 1--26.

\bibitem[{Sun(2014{\natexlab{a}})}]{sun:14a}
Sun, Y., 2014{\natexlab{a}}. Fixed-smoothing asymptotics in a two-step {GMM}
  framework. Econometrica 82~(6), 2327--2370.

\bibitem[{Sun(2014{\natexlab{b}})}]{sun:14}
Sun, Y., 2014{\natexlab{b}}. Let's dix it: fixed-b asymptotics versus small-b
  asymptotics in heteroskedasticity and autocorrelation robust inference.
  Journal of Econometrics 178~(3), 659--677.

\bibitem[{Velasco and Robinson(2001)}]{velasco/robinson:01}
Velasco, C., Robinson, {\relax P.M}., 2001. Edgeworth expansions for spectral
  density estimates and studentized sample mean. Econometric Theory 17~(3),
  497--539.

\bibitem[{Vogelsang(1999)}]{vogeslang:99}
Vogelsang, {\relax T.J}., 1999. Sources of nonmonotonic power when testing for
  a shift in mean of a dynamic time series. Journal of Econometrics 88~(2),
  283--299.

\bibitem[{Zhang and Shao(2013)}]{zhang/shao:13}
Zhang, X., Shao, X., 2013. Fixed-smoothing asymptotics for time series. Annals
  of Statistics 41~(3), 1329--1349.

\end{thebibliography}
\addcontentsline{toc}{section}{References}

\newpage{}

\newpage{}

\clearpage 
\pagenumbering{arabic}
\renewcommand*{\thepage}{A-\arabic{page}}
\appendix

\section{Appendix}

{\footnotesize{}}
\begin{table}[H]
{\footnotesize{}\caption{\label{Table S1-S2}Empirical small-sample size of $t$-tests for
model M1-M2}
}{\footnotesize\par}
\begin{centering}
{\footnotesize{}}%
\begin{tabular}{lcccc}
\hline 
 & \multicolumn{2}{c}{{\footnotesize{}Model M1, $t_{1}$}} & \multicolumn{2}{c}{{\footnotesize{}Model M2, $t_{2}$}}\tabularnewline
{\footnotesize{}$\alpha=0.05$} & {\footnotesize{}$T=200$} & {\footnotesize{}$T=400$} & {\footnotesize{}$T=200$} & {\footnotesize{}$T=400$}\tabularnewline
\hline 
\hline 
{\footnotesize{}$\widehat{J}_{T}(\widehat{b}_{1,T},\,\widehat{b}_{2,T})$} & {\footnotesize{}0.079} & {\footnotesize{}0.059} & {\footnotesize{}0.074} & {\footnotesize{}0.073}\tabularnewline
{\footnotesize{}$\widehat{J}_{T}(\widehat{b}_{1,T},\,\widehat{b}_{2,T})$,
prewhite} & {\footnotesize{}0.042} & {\footnotesize{}0.049} & {\footnotesize{}0.046} & {\footnotesize{}0.063}\tabularnewline
{\footnotesize{}$\widehat{J}_{T}(\widehat{b}_{1,T},\,\widehat{b}_{2,T})$,
prewhite, SLS} & {\footnotesize{}0.055} & {\footnotesize{}0.044} & {\footnotesize{}0.057} & {\footnotesize{}0.055}\tabularnewline
{\footnotesize{}$\widehat{J}_{T}(\widehat{b}_{1,T},\,\widehat{b}_{2,T})$,
prewhite, SLS, $\mu$} & {\footnotesize{}0.057} & {\footnotesize{}0.052} & {\footnotesize{}0.057} & {\footnotesize{}0.056}\tabularnewline
{\footnotesize{}$\widehat{J}_{T}$, Casini (2020)} & {\footnotesize{}0.117} & {\footnotesize{}0.102} & {\footnotesize{}0.083} & {\footnotesize{}0.082}\tabularnewline
{\footnotesize{}$\widehat{J}_{T}$, prewhite, CP} & {\footnotesize{}0.062} & {\footnotesize{}0.055} & {\footnotesize{}0.063} & {\footnotesize{}0.061}\tabularnewline
{\footnotesize{}$\widehat{J}_{T}$, prewhite, SLS, CP} & {\footnotesize{}0.060} & {\footnotesize{}0.059} & {\footnotesize{}0.068} & {\footnotesize{}0.065}\tabularnewline
{\footnotesize{}$\widehat{J}_{T}$, prewhite, SLS, $\mu$, CP } & {\footnotesize{}0.061} & {\footnotesize{}0.060} & {\footnotesize{}0.074} & {\footnotesize{}0.054}\tabularnewline
{\footnotesize{}Newey-West (1987)} & {\footnotesize{}0.101} & {\footnotesize{}0.086} & {\footnotesize{}0.086} & {\footnotesize{}0.085}\tabularnewline
{\footnotesize{}Newey-West (1987), prewhite} & {\footnotesize{}0.076} & {\footnotesize{}0.060} & {\footnotesize{}0.078} & {\footnotesize{}0.080}\tabularnewline
{\footnotesize{}Newey-West (1987), fixed-$b$ (KVB)} & {\footnotesize{}0.058} & {\footnotesize{}0.055} & {\footnotesize{}0.061} & {\footnotesize{}0.059}\tabularnewline
{\footnotesize{}EWC} & {\footnotesize{}0.062} & {\footnotesize{}0.047} & {\footnotesize{}0.059} & {\footnotesize{}0.069}\tabularnewline
\hline 
\end{tabular}{\footnotesize\par}
\par\end{centering}
{\footnotesize{}}~~~~~~~~~~~~~~~~~~~~~~~~~~~~~~%
\noindent\begin{minipage}[t]{1\columnwidth}%
{\scriptsize{}CP  stands for \citet{casini/perron_PrewhitedHAC}.}%
\end{minipage}
\end{table}

{\footnotesize{}}
\begin{table}[H]
{\footnotesize{}\caption{\label{Table S3-S4}Empirical small-sample size of $t_{2}$-tests
for model M3 and of model M4}
}{\footnotesize\par}
\begin{centering}
{\footnotesize{}}%
\begin{tabular}{lcccc}
\hline 
 & \multicolumn{2}{c}{{\footnotesize{}Model M3, $t_{2}$}} & \multicolumn{2}{c}{{\footnotesize{}Model M4, GR test}}\tabularnewline
{\footnotesize{}$\alpha=0.05$} & {\footnotesize{}$T=200$} & {\footnotesize{}$T=400$} & {\footnotesize{}$T=200$} & {\footnotesize{}$T=800$}\tabularnewline
\hline 
\hline 
{\footnotesize{}$\widehat{J}_{T}(\widehat{b}_{1,T},\,\widehat{b}_{2,T})$} & {\footnotesize{}0.064} & {\footnotesize{}0.068} & {\footnotesize{}0.079} & {\footnotesize{}0.061}\tabularnewline
{\footnotesize{}$\widehat{J}_{T}(\widehat{b}_{1,T},\,\widehat{b}_{2,T})$,
prewhite} & {\footnotesize{}0.029} & {\footnotesize{}0.062} & {\footnotesize{}0.054} & {\footnotesize{}0.047}\tabularnewline
{\footnotesize{}$\widehat{J}_{T}(\widehat{b}_{1,T},\,\widehat{b}_{2,T})$,
prewhite, SLS} & {\footnotesize{}0.032} & {\footnotesize{}0.043} & {\footnotesize{}0.062} & {\footnotesize{}0.054}\tabularnewline
{\footnotesize{}$\widehat{J}_{T}(\widehat{b}_{1,T},\,\widehat{b}_{2,T})$,
prewhite, SLS, $\mu$} & {\footnotesize{}0.033} & {\footnotesize{}0.043} & {\footnotesize{}0.073} & {\footnotesize{}0.064}\tabularnewline
{\footnotesize{}$\widehat{J}_{T}$, Casini (2020)} & {\footnotesize{}0.024} & {\footnotesize{}0.027} & {\footnotesize{}0.055} & {\footnotesize{}0.060}\tabularnewline
{\footnotesize{}$\widehat{J}_{T}$, prewhite, CP } & {\footnotesize{}0.015} & {\footnotesize{}0.000} & {\footnotesize{}0.057} & {\footnotesize{}0.042}\tabularnewline
{\footnotesize{}$\widehat{J}_{T}$, prewhite, SLS, CP } & {\footnotesize{}0.063} & {\footnotesize{}0.062} & {\footnotesize{}0.061} & {\footnotesize{}0.055}\tabularnewline
{\footnotesize{}$\widehat{J}_{T}$, prewhite, SLS, $\mu$, CP} & {\footnotesize{}0.073} & {\footnotesize{}0.069} & {\footnotesize{}0.061} & {\footnotesize{}0.056}\tabularnewline
{\footnotesize{}Newey-West (1987)} & {\footnotesize{}0.074} & {\footnotesize{}0.069} & {\footnotesize{}0.000} & {\footnotesize{}0.000}\tabularnewline
{\footnotesize{}Newey-West (1987), prewhite} & {\footnotesize{}0.041} & {\footnotesize{}0.000} & {\footnotesize{}0.000} & {\footnotesize{}0.000}\tabularnewline
{\footnotesize{}Newey-West (1987), fixed-$b$ (KVB)} & {\footnotesize{}0.021} & {\footnotesize{}0.005} & {\footnotesize{}0.061} & {\footnotesize{}0.044}\tabularnewline
{\footnotesize{}EWC} & {\footnotesize{}0.031} & {\footnotesize{}0.010} & {\footnotesize{}0.017} & {\footnotesize{}0.017}\tabularnewline
\hline 
\end{tabular}{\footnotesize\par}
\par\end{centering}
{\footnotesize{}}~~~~~~~~~~~~~~~~~~~~~~~~~~~~~~%
\noindent\begin{minipage}[t]{1\columnwidth}%
{\scriptsize{}CP stands for \citet{casini/perron_PrewhitedHAC}.}%
\end{minipage}
\end{table}
{\footnotesize{} }

{\footnotesize{}}
\begin{table}[H]
{\footnotesize{}\caption{\label{Table M1 Power}Empirical small-sample rejection rates of the
$t_{1}$-test for model M1}
}{\footnotesize\par}
\begin{centering}
{\footnotesize{}}%
\begin{tabular}{lcccc}
\hline 
 & \multicolumn{4}{c}{{\footnotesize{}Model M1, $t_{1}$}}\tabularnewline
{\footnotesize{}$\alpha=0.05$, $T=200$} & {\footnotesize{}$\delta=0.1$} & {\footnotesize{}$\delta=0.2$} & {\footnotesize{}$\delta=0.4$} & {\footnotesize{}$\delta=0.8$}\tabularnewline
\hline 
\hline 
{\footnotesize{}$\widehat{J}_{T}(\widehat{b}_{1,T},\,\widehat{b}_{2,T})$} & {\footnotesize{}0.218} & {\footnotesize{}0.589} & {\footnotesize{}0.980} & {\footnotesize{}1.000}\tabularnewline
{\footnotesize{}$\widehat{J}_{T}(\widehat{b}_{1,T},\,\widehat{b}_{2,T})$,
prewhite} & {\footnotesize{}0.132} & {\footnotesize{}0.465} & {\footnotesize{}0.960} & {\footnotesize{}1.000}\tabularnewline
{\footnotesize{}$\widehat{J}_{T}(\widehat{b}_{1,T},\,\widehat{b}_{2,T})$,
prewhite, SLS} & {\footnotesize{}0.172} & {\footnotesize{}0.553} & {\footnotesize{}0.958} & {\footnotesize{}1.000}\tabularnewline
{\footnotesize{}$\widehat{J}_{T}(\widehat{b}_{1,T},\,\widehat{b}_{2,T})$,
prewhite, SLS, $\mu$} & {\footnotesize{}0.174} & {\footnotesize{}0.544} & {\footnotesize{}0.958} & {\footnotesize{}1.000}\tabularnewline
{\footnotesize{}$\widehat{J}_{T}$, Casini (2020)} & {\footnotesize{}0.291} & {\footnotesize{}0.620} & {\footnotesize{}0.980} & {\footnotesize{}1.000}\tabularnewline
{\footnotesize{}$\widehat{J}_{T}$, prewhite, CP } & {\footnotesize{}0.191} & {\footnotesize{}0.518} & {\footnotesize{}0.949} & {\footnotesize{}1.000}\tabularnewline
{\footnotesize{}$\widehat{J}_{T}$, prewhite, SLS, CP } & {\footnotesize{}0.161} & {\footnotesize{}0.509} & {\footnotesize{}0.969} & {\footnotesize{}1.000}\tabularnewline
{\footnotesize{}$\widehat{J}_{T}$, prewhite, SLS, $\mu$, CP } & {\footnotesize{}0.165} & {\footnotesize{}0.508} & {\footnotesize{}0.970} & {\footnotesize{}1.000}\tabularnewline
{\footnotesize{}Newey-West (1987)} & {\footnotesize{}0.248} & {\footnotesize{}0.629} & {\footnotesize{}0.987} & {\footnotesize{}1.000}\tabularnewline
{\footnotesize{}Newey-West (1987), prewhite} & {\footnotesize{}0.197} & {\footnotesize{}0.576} & {\footnotesize{}0.979} & {\footnotesize{}1.000}\tabularnewline
{\footnotesize{}Newey-West (1987), fixed-$b$ (KVB)} & {\footnotesize{}0.141} & {\footnotesize{}0.373} & {\footnotesize{}0.844} & {\footnotesize{}0.998}\tabularnewline
{\footnotesize{}EWC} & {\footnotesize{}0.150} & {\footnotesize{}0.493} & {\footnotesize{}0.963} & {\footnotesize{}1.000}\tabularnewline
\hline 
\end{tabular}{\footnotesize\par}
\par\end{centering}
{\footnotesize{}}~~~~~~~~~~~~~~~~~~~~~~~~~~~~~~%
\noindent\begin{minipage}[t]{1\columnwidth}%
{\scriptsize{}CP  stands for \citet{casini/perron_PrewhitedHAC}.}%
\end{minipage}
\end{table}

{\footnotesize{}}
\begin{table}[H]
{\footnotesize{}\caption{\label{Table Power M2}Empirical small-sample rejection rates of the
$t_{2}$-tests for model M2}
}{\footnotesize\par}
\begin{centering}
{\footnotesize{}}%
\begin{tabular}{lcccc}
\hline 
 & \multicolumn{4}{c}{}\tabularnewline
{\footnotesize{}$\alpha=0.05$, $T=200$} & {\footnotesize{}$\delta=0.1$} & {\footnotesize{}$\delta=0.2$} & {\footnotesize{}$\delta=0.4$} & {\footnotesize{}$\delta=0.8$}\tabularnewline
\hline 
\hline 
{\footnotesize{}$\widehat{J}_{T}(\widehat{b}_{1,T},\,\widehat{b}_{2,T})$} & {\footnotesize{}0.263} & {\footnotesize{}0.642} & {\footnotesize{}0.988} & {\footnotesize{}1.000}\tabularnewline
{\footnotesize{}$\widehat{J}_{T}(\widehat{b}_{1,T},\,\widehat{b}_{2,T})$,
prewhite} & {\footnotesize{}0.191} & {\footnotesize{}0.532} & {\footnotesize{}0.968} & {\footnotesize{}1.000}\tabularnewline
{\footnotesize{}$\widehat{J}_{T}(\widehat{b}_{1,T},\,\widehat{b}_{2,T})$,
prewhite, SLS} & {\footnotesize{}0.221} & {\footnotesize{}0.592} & {\footnotesize{}0.982} & {\footnotesize{}1.000}\tabularnewline
{\footnotesize{}$\widehat{J}_{T}(\widehat{b}_{1,T},\,\widehat{b}_{2,T})$,
prewhite, SLS, $\mu$} & {\footnotesize{}0.221} & {\footnotesize{}0.597} & {\footnotesize{}0.983} & {\footnotesize{}1.000}\tabularnewline
{\footnotesize{}$\widehat{J}_{T}$, Casini (2020)} & {\footnotesize{}0.276} & {\footnotesize{}0.653} & {\footnotesize{}0.988} & {\footnotesize{}1.000}\tabularnewline
{\footnotesize{}$\widehat{J}_{T}$, prewhite, CP } & {\footnotesize{}0.237} & {\footnotesize{}0.611} & {\footnotesize{}0.986} & {\footnotesize{}1.000}\tabularnewline
{\footnotesize{}$\widehat{J}_{T}$, prewhite, SLS, CP } & {\footnotesize{}0.225} & {\footnotesize{}0.598} & {\footnotesize{}0.982} & {\footnotesize{}1.000}\tabularnewline
{\footnotesize{}$\widehat{J}_{T}$, prewhite, SLS, $\mu$, CP } & {\footnotesize{}0.165} & {\footnotesize{}0.598} & {\footnotesize{}0.988} & {\footnotesize{}1.000}\tabularnewline
{\footnotesize{}Newey-West (1987)} & {\footnotesize{}0.268} & {\footnotesize{}0.332} & {\footnotesize{}0.992} & {\footnotesize{}1.000}\tabularnewline
{\footnotesize{}Newey-West (1987), prewhite} & {\footnotesize{}0.258} & {\footnotesize{}0.374} & {\footnotesize{}0.990} & {\footnotesize{}1.000}\tabularnewline
{\footnotesize{}Newey-West (1987), fixed-$b$ (KVB)} & {\footnotesize{}0.199} & {\footnotesize{}0.463} & {\footnotesize{}0.914} & {\footnotesize{}1.000}\tabularnewline
{\footnotesize{}EWC} & {\footnotesize{}0.193} & {\footnotesize{}0.571} & {\footnotesize{}0.978} & {\footnotesize{}1.000}\tabularnewline
\hline 
\end{tabular}{\footnotesize\par}
\par\end{centering}
{\footnotesize{}}~~~~~~~~~~~~~~~~~~~~~~~~~~~~~~%
\noindent\begin{minipage}[t]{1\columnwidth}%
{\scriptsize{}CP  stands for \citet{casini/perron_PrewhitedHAC}.}%
\end{minipage}
\end{table}

{\footnotesize{}}
\begin{table}[H]
{\footnotesize{}\caption{\label{Table Power M3}Empirical small-sample rejection rates of the
$t_{2}$-test for model M3}
}{\footnotesize\par}
\begin{centering}
{\footnotesize{}}%
\begin{tabular}{lcccc}
\hline 
 & \multicolumn{4}{c}{}\tabularnewline
{\footnotesize{}$\alpha=0.05$, $T=200$} & {\footnotesize{}$\delta=0.1$} & {\footnotesize{}$\delta=0.2$} & {\footnotesize{}$\delta=0.4$} & {\footnotesize{}$\delta=0.8$}\tabularnewline
\hline 
\hline 
{\footnotesize{}$\widehat{J}_{T}(\widehat{b}_{1,T},\,\widehat{b}_{2,T})$} & {\footnotesize{}0.167} & {\footnotesize{}0.429} & {\footnotesize{}0.794} & {\footnotesize{}0.962}\tabularnewline
{\footnotesize{}$\widehat{J}_{T}(\widehat{b}_{1,T},\,\widehat{b}_{2,T})$,
prewhite} & {\footnotesize{}0.112} & {\footnotesize{}0.230} & {\footnotesize{}0.691} & {\footnotesize{}0.921}\tabularnewline
{\footnotesize{}$\widehat{J}_{T}(\widehat{b}_{1,T},\,\widehat{b}_{2,T})$,
prewhite, SLS} & {\footnotesize{}0.104} & {\footnotesize{}0.325} & {\footnotesize{}0.687} & {\footnotesize{}0.912}\tabularnewline
{\footnotesize{}$\widehat{J}_{T}(\widehat{b}_{1,T},\,\widehat{b}_{2,T})$,
prewhite, SLS, $\mu$} & {\footnotesize{}0.104} & {\footnotesize{}0.328} & {\footnotesize{}0.688} & {\footnotesize{}0.912}\tabularnewline
{\footnotesize{}$\widehat{J}_{T}$, Casini (2020)} & {\footnotesize{}0.091} & {\footnotesize{}0.293} & {\footnotesize{}0.687} & {\footnotesize{}0.940}\tabularnewline
{\footnotesize{}$\widehat{J}_{T}$, prewhite, CP } & {\footnotesize{}0.046} & {\footnotesize{}0.228} & {\footnotesize{}0.537} & {\footnotesize{}0.836}\tabularnewline
{\footnotesize{}$\widehat{J}_{T}$, prewhite, SLS, CP } & {\footnotesize{}0.152} & {\footnotesize{}0.381} & {\footnotesize{}0.728} & {\footnotesize{}0.946}\tabularnewline
{\footnotesize{}$\widehat{J}_{T}$, prewhite, SLS, $\mu$, CP } & {\footnotesize{}0.164} & {\footnotesize{}0.395} & {\footnotesize{}0.741} & {\footnotesize{}0.954}\tabularnewline
{\footnotesize{}Newey-West (1987)} & {\footnotesize{}0.212} & {\footnotesize{}0.487} & {\footnotesize{}0.839} & {\footnotesize{}0.973}\tabularnewline
{\footnotesize{}Newey-West (1987), prewhite} & {\footnotesize{}0.154} & {\footnotesize{}0.416} & {\footnotesize{}0.779} & {\footnotesize{}0.950}\tabularnewline
{\footnotesize{}Newey-West (1987), fixed-$b$ (KVB)} & {\footnotesize{}0.101} & {\footnotesize{}0.298} & {\footnotesize{}0.661} & {\footnotesize{}0.906}\tabularnewline
{\footnotesize{}EWC} & {\footnotesize{}0.151} & {\footnotesize{}0.409} & {\footnotesize{}0.793} & {\footnotesize{}0.960}\tabularnewline
\hline 
\end{tabular}{\footnotesize\par}
\par\end{centering}
{\footnotesize{}}~~~~~~~~~~~~~~~~~~~~~~~~~~~~~~%
\noindent\begin{minipage}[t]{1\columnwidth}%
{\scriptsize{}CP  stands for \citet{casini/perron_PrewhitedHAC}.}%
\end{minipage}
\end{table}

{\footnotesize{}}
\begin{table}[H]
{\footnotesize{}\caption{\label{Table Power GR}Empirical small-sample rejection rates of the
GR tests}
}{\footnotesize\par}
\begin{centering}
{\footnotesize{}}%
\begin{tabular}{lcccc}
\hline 
 & \multicolumn{4}{c}{}\tabularnewline
{\footnotesize{}$\alpha=0.05$, $T=800$} & {\footnotesize{}$\delta=0.5$} & {\footnotesize{}$\delta=1$} & {\footnotesize{}$\delta=1.5$} & {\footnotesize{}$\delta=2$}\tabularnewline
\hline 
\hline 
{\footnotesize{}$\widehat{J}_{T}(\widehat{b}_{1,T},\,\widehat{b}_{2,T})$} & {\footnotesize{}0.127} & {\footnotesize{}0.719} & {\footnotesize{}0.982} & {\footnotesize{}1.000}\tabularnewline
{\footnotesize{}$\widehat{J}_{T}(\widehat{b}_{1,T},\,\widehat{b}_{2,T})$,
prewhite} & {\footnotesize{}0.108} & {\footnotesize{}0.681} & {\footnotesize{}0.982} & {\footnotesize{}1.000}\tabularnewline
{\footnotesize{}$\widehat{J}_{T}(\widehat{b}_{1,T},\,\widehat{b}_{2,T})$,
prewhite, SLS} & {\footnotesize{}0.113} & {\footnotesize{}0.730} & {\footnotesize{}0.991} & {\footnotesize{}1.000}\tabularnewline
{\footnotesize{}$\widehat{J}_{T}(\widehat{b}_{1,T},\,\widehat{b}_{2,T})$,
prewhite, SLS, $\mu$} & {\footnotesize{}0.121} & {\footnotesize{}0.719} & {\footnotesize{}0.982} & {\footnotesize{}1.000}\tabularnewline
{\footnotesize{}$\widehat{J}_{T}$, Casini (2020)} & {\footnotesize{}0.139} & {\footnotesize{}0.622} & {\footnotesize{}0.812} & {\footnotesize{}0.915}\tabularnewline
{\footnotesize{}$\widehat{J}_{T}$, prewhite, CP } & {\footnotesize{}0.114} & {\footnotesize{}0.699} & {\footnotesize{}0.965} & {\footnotesize{}0.994}\tabularnewline
{\footnotesize{}$\widehat{J}_{T}$, prewhite, SLS, CP } & {\footnotesize{}0.134} & {\footnotesize{}0.779} & {\footnotesize{}0.989} & {\footnotesize{}0.999}\tabularnewline
{\footnotesize{}$\widehat{J}_{T}$, prewhite, SLS, $\mu$, CP } & {\footnotesize{}0.152} & {\footnotesize{}0.793} & {\footnotesize{}0.989} & {\footnotesize{}0.999}\tabularnewline
{\footnotesize{}Newey-West (1987)} & {\footnotesize{}0.000} & {\footnotesize{}0.000} & {\footnotesize{}0.000} & {\footnotesize{}0.000}\tabularnewline
{\footnotesize{}Newey-West (1987), prewhite} & {\footnotesize{}0.000} & {\footnotesize{}0.000} & {\footnotesize{}0.000} & {\footnotesize{}0.000}\tabularnewline
{\footnotesize{}Newey-West (1987), fixed-$b$ (KVB)} & {\footnotesize{}0.062} & {\footnotesize{}0.042} & {\footnotesize{}0.000} & {\footnotesize{}0.000}\tabularnewline
{\footnotesize{}EWC} & {\footnotesize{}0.050} & {\footnotesize{}0.044} & {\footnotesize{}0.004} & {\footnotesize{}0.000}\tabularnewline
\hline 
\end{tabular}{\footnotesize\par}
\par\end{centering}
{\footnotesize{}}~~~~~~~~~~~~~~~~~~~~~~~~~~~~~~%
\noindent\begin{minipage}[t]{1\columnwidth}%
{\scriptsize{}CP  stands for \citet{casini/perron_PrewhitedHAC}.}%
\end{minipage}
\end{table}
 \newpage{}

\clearpage 
\pagenumbering{arabic}
\renewcommand*{\thepage}{A-\arabic{page}}
\appendix

\clearpage{}

\newpage{}

\pagebreak{}

\section*{}
\addcontentsline{toc}{part}{Supplemental Material}
\begin{center}
\Large{{Supplemental Material} to} 
\end{center}

\begin{center}
\title{\textbf{\Large{Simultaneous Bandwidths Determination for DK-HAC Estimators and Long-Run Variance Estimation in Nonparametric Settings}}} 
\maketitle
\end{center}
\medskip{} 
\medskip{} 
\medskip{} 
\thispagestyle{empty}

\begin{center}
\author{\textsc{\textcolor{MyBlue}{Federico Belotti}}
\quad \quad \quad \quad \quad  
\textsc{\textcolor{MyBlue}{Alessandro Casini}}
\\
\small{University of Rome Tor Vergata} 
\quad \quad 
\small{University of Rome Tor Vergata} 
\\
\normalsize \textsc{\textcolor{MyBlue}{Leopoldo Catania}}
\quad \quad \quad \quad  
\textsc{\textcolor{MyBlue}{Stefano Grassi}}
\quad \quad \quad \quad  
\textsc{\textcolor{MyBlue}{Pierre Perron}}
\\
\quad \small{Aarhus University}
\quad \quad 
\small{University of Rome Tor Vergata}
\quad \quad 
\small{Boston University} 
}

\medskip{}
\medskip{} 
\medskip{} 

\medskip{}
\medskip{} 
\medskip{} 
\medskip{} 
\date{\small{\today}} 
\medskip{} 
\medskip{} 
\medskip{} 
\end{center}
\begin{abstract}
{\footnotesize{}This supplemental material includes the proofs of
the results in the paper. }{\footnotesize\par}
\end{abstract}
\setcounter{page}{0}
\setcounter{section}{0}
\renewcommand*{\theHsection}{\the\value{section}}

\newpage{}

\begin{singlespace} 
\noindent 
\small

\allowdisplaybreaks


\renewcommand{\thepage}{S-\arabic{page}}   
\renewcommand{\thesection}{S.\Alph{section}}   
\renewcommand{\theequation}{S.\arabic{equation}}




\section{\label{Section Mathematical-Appendix}Mathematical Appendix}

In some of the proofs below $\overline{\beta}$ is understood to
be on the line segment joining $\widehat{\beta}$ and $\beta_{0}$.
We discard the degrees of freedom adjustment $T/\left(T-p\right)$
from the derivations since asymptotically it does not play any role.
Similarly, we use $T/n_{T}$ in place of $\left(T-n_{T}\right)/n_{T}$
in the expression for $\widehat{\Gamma}\left(k\right)$. Let $\widetilde{c}_{T}\left(u,\,k\right)$
denote the estimator that uses $\left\{ V_{t,T}\right\} $.

\subsection{Proofs of the Results of Section \ref{Section Simultaneous Bandwiths Determination}}

\subsubsection{Proof of Theorem \ref{Theorem MSE J}}

Part (i) follows from Theorem 3.1 in \citet{casini_hac}. For part
(ii), let $J_{c,T}=\int_{0}^{1}c\left(u,\,0\right)+2\sum_{k=1}^{T-1}\int_{0}^{1}c\left(u,\,k\right)du$
and $\mathcal{T}_{C}\triangleq\left\{ 0,\,n_{T},\ldots,\,T-n_{T},\,T\right\} /\mathcal{T}$.
We begin with the following relationship,
\begin{align*}
\mathbb{E}\left(\widetilde{J}_{T}-J_{T}\right) & =\sum_{k=-T+1}^{T-1}K_{1}\left(b_{1,T}k\right)\mathbb{E}\left(\widetilde{\Gamma}\left(k\right)\right)-J_{c,T}+\left(J_{c,T}-J_{T}\right).
\end{align*}
Using Lemma S.B.4 in \citet{casini_hac}, we have for any $0\leq k\leq T-1$,
\begin{align*}
\mathbb{E} & \left(\frac{n_{T}}{T}\sum_{r=0}^{T/n_{T}}\widetilde{c}_{T}\left(rn_{T}/T,\,k\right)-\int_{0}^{1}c\left(u,\,k\right)du\right)\\
 & =\frac{n_{T}}{T}\sum_{r\in\mathcal{T}_{C}}\left(c\left(rn_{T}/T,\,k\right)+\frac{1}{2}b_{2,T}^{2}\int_{0}^{1}x^{2}K_{2}\left(x\right)dx\frac{\partial^{2}}{\partial^{2}u}c\left(u,\,k\right)+o\left(b_{2,T}^{2}\right)+O\left(\frac{1}{b_{2,T}T}\right)\right)\\
 & \quad+\frac{n_{T}}{T}\sum_{r\notin\mathcal{T}_{C}}\biggl(c\left(rn_{T}/T,\,k\right)+\frac{1}{2}b_{2,T}^{2}\int_{0}^{1}x^{2}K_{2}\left(x\right)dx\\
 & \quad\times\int_{-\pi}^{\pi}\exp\left(i\omega k\right)\left(C_{1}\left(rn_{T}/T,\,\omega\right)+C_{2}\left(rn_{T}/T,\,\omega\right)+C_{3}\left(rn_{T}/T,\,\omega\right)\right)d\omega\\
 & \quad+o\left(b_{2,T}^{2}\right)+O\left(\frac{1}{b_{2,T}T}\right)\biggr)-\int_{0}^{1}c\left(u,\,k\right)du,
\end{align*}
where
\begin{align*}
C_{1}\left(u_{0},\,\omega\right)=2\frac{\partial A_{j}\left(u_{0},\,-\omega\right)}{\partial_{-}u}\frac{\partial A_{j+1}\left(v_{0},\,\omega\right)}{\partial_{+}v} & \qquad C_{2}\left(u_{0},\,\omega\right)=\frac{\partial^{2}A_{j+1}\left(v_{0},\,\omega\right)}{\partial_{+}v^{2}}A_{j}\left(u_{0},\,-\omega\right),\\
C_{3}\left(u_{0},\,\omega\right) & =\frac{\partial^{2}A_{j}\left(u_{0},\,\omega\right)}{\partial_{-}u^{2}}A_{j+1}\left(v_{0},\,\omega\right),
\end{align*}
with $u_{0}=rn_{T}/T$ and $v_{0}=u_{0}-k/2T$. The right-hand side
above is equal to 
\begin{align*}
\frac{n_{T}}{T} & \sum_{r=0}^{T/n_{T}}c\left(rn_{T}/T,\,k\right)-\int_{0}^{1}c\left(u,\,k\right)du\\
 & \quad+\frac{1}{2}b_{2,T}^{2}\int_{0}^{1}x^{2}K_{2}\left(x\right)dx\int_{0}^{1}\frac{\partial^{2}}{\partial^{2}u}c\left(u,\,k\right)du+o\left(b_{2,T}^{2}\right)+O\left(\frac{1}{Tb_{2,T}}\right)\\
 & \quad+\frac{1}{2}b_{2,T}^{2}\int_{0}^{1}x^{2}K_{2}\left(x\right)dx\\
 & \quad\times\int_{0}^{1}\left(\int_{-\pi}^{\pi}\exp\left(i\omega k\right)\left(C_{1}\left(u,\,\omega\right)+C_{2}\left(u,\,\omega\right)+C_{3}\left(u,\,\omega\right)\right)d\omega\mathbf{1}\left\{ Tu\in\mathcal{T}\right\} \right)du\\
 & \quad+o\left(b_{2,T}^{2}\right)+O\left(\frac{1}{Tb_{2,T}}\right)\\
 & =O\left(\frac{n_{T}}{T}\right)+\frac{1}{2}b_{2,T}^{2}\int_{0}^{1}x^{2}K_{2}\left(x\right)dx\int_{0}^{1}\frac{\partial^{2}}{\partial^{2}u}c\left(u,\,k\right)du+o\left(b_{2,T}^{2}\right)+O\left(\frac{1}{Tb_{2,T}}\right),
\end{align*}
where the last equality follows from the convergence of approximations
to Riemann sums and from the fact that $\mathbf{1}\left\{ Tu\in\mathcal{T}\right\} $
has zero Lebesgue measure. This leads to,
\begin{align*}
b_{1,T}^{-q} & \mathbb{E}\left(\widetilde{J}_{T}-J_{c,T}\right)\\
 & =-b_{1,T}^{-q}\sum_{k=-T+1}^{T}\left(1-K_{1}\left(b_{1,T}k\right)\right)\int_{0}^{1}c\left(u,\,k\right)du\\
 & \quad+\frac{1}{2}\frac{b_{2,T}^{2}}{b_{1,T}^{q}}\int_{0}^{1}x^{2}K_{2}\left(x\right)dx\sum_{k=-T+1}^{T}K_{1}\left(b_{1,T}k\right)\int_{0}^{1}\frac{\partial^{2}}{\partial^{2}u}c\left(u,\,k\right)du+O\left(\frac{1}{Tb_{1,T}^{q}b_{2,T}}\right)+O\left(\frac{n_{T}}{Tb_{1,T}^{q}}\right)\\
 & =-b_{1,T}^{-q}\sum_{k=-T+1}^{T}\left(1-K_{1}\left(b_{1,T}k\right)\right)\int_{0}^{1}c\left(u,\,k\right)du\\
 & \quad-\frac{1}{2}b_{2,T}^{2}\int_{0}^{1}x^{2}K_{2}\left(x\right)dxb_{1,T}^{-q}\sum_{k=-T+1}^{T}\left(1-K_{1}\left(b_{1,T}k\right)\right)\int_{0}^{1}\frac{\partial^{2}}{\partial^{2}u}c\left(u,\,k\right)du\\
 & \quad+\frac{1}{2}\frac{b_{2,T}^{2}}{b_{1,T}^{q}}\int_{0}^{1}x^{2}K_{2}\left(x\right)dx\sum_{k=-T+1}^{T}\int_{0}^{1}\frac{\partial^{2}}{\partial^{2}u}c\left(u,\,k\right)du+O\left(\frac{1}{Tb_{1,T}^{q}b_{2,T}}\right)+O\left(\frac{n_{T}}{Tb_{1,T}^{q}}\right)\\
 & =-b_{1,T}^{-q}\sum_{k=-T+1}^{T}\left(1-K_{1}\left(b_{1,T}k\right)\right)\int_{0}^{1}c\left(u,\,k\right)du\\
 & \quad-\frac{1}{2}b_{2,T}^{2}\int_{0}^{1}x^{2}K_{2}\left(x\right)dxO\left(1\right)+\frac{1}{2}\frac{b_{2,T}^{2}}{b_{1,T}^{q}}\int_{0}^{1}x^{2}K_{2}\left(x\right)dx\sum_{k=-T+1}^{T}\int_{0}^{1}\frac{\partial^{2}}{\partial^{2}u}c\left(u,\,k\right)du\\
 & \quad+O\left(\frac{1}{Tb_{1,T}^{q}b_{2,T}}\right)+O\left(\frac{n_{T}}{Tb_{1,T}^{q}}\right),
\end{align*}
 since $\left|\sum_{k=-\infty}^{\infty}\left|k\right|^{q}\int_{0}^{1}\left(\partial^{2}/\partial^{2}u\right)c\left(u,\,k\right)du\right|<\infty$
by Assumption \ref{Assumption A - Dependence}-(i). Since $J_{c,T}-J_{T}=O\left(T^{-1}\right)$
we conclude that, 
\begin{align*}
\lim_{T\rightarrow\infty}b_{1,T}^{-q}\mathbb{E}\left(\widetilde{J}_{T}-J_{T}\right) & =-2\pi K_{1,q}\int_{0}^{1}f^{\left(q\right)}\left(u,\,0\right)du\\
 & \quad+\frac{1}{2}\frac{b_{2,T}^{2}}{b_{1,T}^{q}}\int_{0}^{1}x^{2}K_{2}\left(x\right)dx\sum_{k=-T+1}^{T}\int_{0}^{1}\frac{\partial^{2}}{\partial^{2}u}c\left(u,\,k\right)du,
\end{align*}
using $b_{2,T}^{2}/b_{1,T}^{q}\rightarrow\nu.$ Part (iii) follows
from part (i)-(ii) and the commutation-tensor product formula follows
from the same argument as in Theorem 3.1-(iii) in \citet{casini_hac}.
$\square$

\subsubsection{Proof of Theorem \ref{Theorem 1 -Consistency and Rate}}

The proof follows the same steps as in Theorem 3.2 in \citet{casini_hac}
with references to Theorem 3.1 there replaced by references to Theorem
\ref{Theorem MSE J} here and in addition it uses the fact that $\widetilde{J}_{T}$
is still asymptotically unbiased because
\begin{align*}
\mathbb{E}\left(\widetilde{J}_{T}-J_{T}\right) & =-2\pi K_{1,q}b_{1,T}^{q}\int_{0}^{1}f^{\left(q\right)}\left(u,\,0\right)du\\
 & \quad+\frac{1}{2}b_{2,T}^{2}\int_{0}^{1}x^{2}K_{2}\left(x\right)dx\sum_{k=-T+1}^{T}\int_{0}^{1}\frac{\partial^{2}}{\partial^{2}u}c\left(u,\,k\right)du,\\
 & \rightarrow0.
\end{align*}
 Note that the above result continues to hold even when $q=0$ since
then $K_{1,q}=0.$ $\square$

\subsubsection{Proof of Theorem \ref{Theorem Optimal Kernels}}

Let $\Delta_{2}\triangleq\mathrm{tr}W\left(I_{p^{2}}+C_{pp}\right)I_{p}\otimes I_{p}$.
We focus on the scalar case. The derivations for the multivariate
case are straightforward but tedious and so we omit them. Note that
\begin{align*}
\mathrm{ReMSE} & =\mathbb{E}\left(\left(\frac{\widehat{J}_{T}\left(b_{1,T},\,b_{2,T}\right)}{J}-1\right)^{2}\right)\\
 & =\mathbb{E}\left(\left(\frac{\widehat{J}_{T}\left(b_{1,T},\,b_{2,T}\right)}{J}-\mathbb{E}\left(\frac{\widehat{J}_{T}\left(b_{1,T},\,b_{2,T}\right)}{J}\right)\right)^{2}\right)+\left(\mathbb{E}\left(\frac{\widehat{J}_{T}\left(b_{1,T},\,b_{2,T}\right)}{J}\right)-1\right)^{2}\\
 & =\left(Tb_{2,T}b_{1,T}\right)^{-1}\varpi_{3}+\left(b_{1,T}^{q}\varpi_{1}+b_{2,T}^{2}\varpi_{2}\right)^{2},
\end{align*}
 where $\varpi_{1}=\left(\int_{0}^{1}f\left(u,\,0\right)du\right)^{-1}\Xi_{1,1}\Delta_{1,1,0}$,
$\varpi_{2}=\left(\int_{0}^{1}f\left(u,\,0\right)du\right)^{-1}\Xi_{1,2}\Delta_{1,2}$
and $\varpi_{3}=\Xi_{2}$. Minimizing the right-hand side above with
respect to $b_{1,T}$ and $b_{2,T}$ yields
\begin{align*}
b_{1,T}=T^{-\frac{2}{2+5q}}\left(2q^{-1}\frac{\varpi_{2}}{\varpi_{1}}\right)^{1/q}\left(\frac{\left(\left(2q^{-1}\frac{\varpi_{2}}{\varpi_{1}}\right)^{1/q}\right)^{-1}\varpi_{3}}{\left(2q^{-1}+1\right)4\varpi_{2}^{2}}\right)^{2/\left(2+5q\right)}, & \quad b_{2,T}=\left(\frac{\left(T\left(2q^{-1}\frac{\varpi_{2}}{\varpi_{1}}\right)^{1/q}\right)^{-1}\varpi_{3}}{\left(2q^{-1}+1\right)4\varpi_{2}^{2}}\right)^{q/\left(2+5q\right)}.
\end{align*}
 At this point to solve for $\widetilde{b}_{1,T}^{\mathrm{opt}}$
and $\widetilde{b}_{2,T}^{\mathrm{opt}}$ we guess and verify that
the optimal $K_{1}$ is such that $q=2$. Note that $q=2$ holds for
the Parzen, Tukey-Hanning and the QS kernels. Thus, substituting out
$q=2$, we have 
\begin{align*}
\widetilde{b}_{1,T}^{\mathrm{opt}}=T^{-1/6}\left(\frac{\varpi_{2}}{\varpi_{1}^{5}}\right)^{1/12}\left(\frac{\varpi_{3}}{8}\right)^{1/6}, & \qquad\widetilde{b}_{2,T}^{\mathrm{opt}}=T^{-1/6}\left(\frac{\varpi_{2}^{5}}{\varpi_{1}}\right)^{-1/12}\left(\frac{\varpi_{3}}{8}\right)^{1/6}.
\end{align*}
The optimal relative MSE is then given by 
\begin{align*}
\mathrm{ReMSE}\left(\widetilde{b}_{1,T}^{\mathrm{opt}},\,\widetilde{b}_{2,T}^{\mathrm{opt}}\right) & =\left(Tb_{2,T}b_{1,T}\right)^{-1}\varpi_{3}+\left(b_{1,T}^{2}\varpi_{1}+b_{2,T}^{2}\varpi_{2}\right)^{2}\\
 & =3\left(4\pi\right)^{-1/3}\left(2\pi\right)^{1/3}\left(-K_{1,q}\right)^{1/3}\Delta_{1,1,0}^{1/3}\Delta_{1,2}^{1/3}\Delta_{2}^{2/3}\left(\int_{0}^{1}x^{2}K_{2}\left(x\right)dx\right)^{1/3}\\
 & \quad\times\left(\int K_{1}^{2}\left(y\right)dy\int_{0}^{1}K_{2}^{2}\left(x\right)dx\right)^{2/3}.
\end{align*}
Recall that $K_{1,2}=-\left(d^{2}K_{1}\left(x\right)/dx^{2}\right)|_{x=0}/2$.
 The spectral window generator of $K_{1}$ is defined as $K_{1,\mathrm{SWG}}\left(\omega\right)=\left(2\pi\right)^{-1}\int_{-\infty}^{\infty}K_{1}\left(x\right)e^{-ix\omega}dx.$
It follows from \citeauthor{priestley:85} (1981, Ch. 6) that for
$K_{1}^{\mathrm{QS}}$ we have 
\begin{align*}
K_{1,\mathrm{SWG}}^{\mathrm{QS}}\left(\omega\right) & =\begin{cases}
\frac{3}{4\pi}\left(1-\left(\omega/\pi\right)\right), & \left|\omega\right|\leq\pi\\
0 & \left|\omega\right|>\pi
\end{cases}.
\end{align*}
We also have the following properties: $K_{1,2}=\int_{-\infty}^{\infty}\omega^{2}K_{1,\mathrm{SWG}}\left(\omega\right)d\omega$,
$K_{1}\left(0\right)=\int_{-\infty}^{\infty}K_{1,\mathrm{SWG}}\left(\omega\right)d\omega$,
and $\int_{-\infty}^{\infty}K_{1}^{2}\left(x\right)dx=\int_{-\infty}^{\infty}K_{1,\mathrm{SWG}}^{2}\left(\omega\right)d\omega$.
We now minimize $\mathrm{ReMSE}\left(\widetilde{b}_{1,T}^{\mathrm{opt}},\,\widetilde{b}_{2,T}^{\mathrm{opt}}\right)$
with respect to $K_{1}$ and $K_{2}$ under the restrictions that
$\int_{0}^{1}K_{2}\left(x\right)dx=1$ and that (a) $\int_{-\infty}^{\infty}K_{1,\mathrm{SWG}}\left(\omega\right)d\omega=1$,
(b) $K_{1,\mathrm{SWG}}\left(\omega\right)\geq0,\,\forall\,\omega\in\mathbb{R}$,
and (c) $K_{1,\mathrm{SWG}}\left(\omega\right)=K_{1,\mathrm{SWG}}\left(-\omega\right),\,\forall\,\omega\in\mathbb{R}$.
This is equivalent to minimizing (1) $(\int_{0}^{1}K_{2}^{2}\left(x\right)dx)^{2}$
$(\int_{0}^{1}x^{2}K_{2}\left(x\right)dx)$ subject to $\int_{0}^{1}K_{2}\left(x\right)dx=1$,
and (2) $\int_{-\infty}^{\infty}\omega^{2}\widetilde{K}_{1}\left(\omega\right)d\omega\left(\int_{-\infty}^{\infty}\widetilde{K}_{1}^{2}\left(\omega\right)d\omega\right)^{2}$
subject to (a)-(c). Using a calculus of variations, \citeauthor{priestley:85}
(1981, Ch. 7) showed that $K_{2}^{\mathrm{opt}}=6x\left(1-x\right)$
for $x\in\left[0,\,1\right]$ solves (1) and that $K_{1}^{\mathrm{QS}}$
solves (2). Since the equivalence between the optimization problem
(1)-(2) and our problem is independent of $q$, this verifies that
our guess was correct because $q=2$ for $K_{1}^{\mathrm{QS}}$. Therefore,
\begin{align*}
\widetilde{b}_{1,T}^{\mathrm{opt}} & =0.46T^{-1/6}\left(\Delta_{1,1,0}/\int_{0}^{1}f\left(u,\,0\right)du\right)^{-5/12}\left(\Delta_{1,2}/\int_{0}^{1}f\left(u,\,0\right)du\right)^{1/12},\qquad\mathrm{and}\\
\widetilde{b}_{2,T}^{\mathrm{opt}} & =3.56T^{-1/6}\left(\Delta_{1,1,0}/\int_{0}^{1}f\left(u,\,0\right)du\right)^{1/12}\left(\Delta_{1,2}/\int_{0}^{1}f\left(u,\,0\right)du\right)^{-5/12}.
\end{align*}
 The result for $\widetilde{b}_{1,T}^{\mathrm{opt}}$ and $\widetilde{b}_{2,T}^{\mathrm{opt}}$
for the multivariate case follows from the matrix-form of the above
expressions. $\square$ 

\subsection{Proofs of the Results of Section \ref{Section Data-Dependent-Bandwidths}}

\subsubsection{Proof of Theorem \ref{Theorem 3 Andrews 91}}

Without loss of generality, we assume that $V_{t}$ is a scalar. The
constant $C<\infty$ may vary from line to line. We begin with the
proof of part (ii) because it becomes then simpler to prove part (i).
By Theorem \ref{Theorem 1 -Consistency and Rate}-(ii), $\sqrt{Tb_{\theta_{1},T}b_{\theta_{2},T}}(\widehat{J}_{T}(b_{\theta_{1},T},\,b_{\theta_{2},T})-J_{T})=O_{\mathbb{P}}\left(1\right)$.
It remains to establish the second result of Theorem \ref{Theorem 3 Andrews 91}-(ii).
Let  $S_{T}=\left\lfloor b_{\theta_{1},T}^{-r}\right\rfloor $ where
\begin{align*}
r\in\left(\max\left\{ 1,\,\left(\overline{b}-1/2\right)/\left(\overline{b}-1\right),\,2/\left(l-1\right),\,\left(b-2\right)/\left(b-1\right)\right\} ,\,5/4\right) & ,
\end{align*}
  with $b>1+1/q$ and $\overline{b}>3$. We will use the following
decomposition 
\begin{align}
\widehat{J}_{T}(\widehat{b}_{1,T},\,\widehat{b}_{2,T})-\widehat{J}_{T}(b_{\theta_{1},T},\,b_{\theta_{2},T}) & =(\widehat{J}_{T}(\widehat{b}_{1,T},\,\widehat{b}_{2,T})-\widehat{J}_{T}(b_{\theta_{1},T},\,\widehat{b}_{2,T}))\label{eq (Decomposition J_T proof of Theorem 3 Andrews91)}\\
 & \quad+(\widehat{J}_{T}(b_{\theta_{1},T},\,\widehat{b}_{2,T})-\widehat{J}_{T}(b_{\theta_{1},T},\,b_{\theta_{2},T})).\nonumber 
\end{align}
Let 
\begin{align*}
N_{1} & \triangleq\left\{ -S_{T},\,-S_{T}+1,\ldots,\,-1,\,1,\ldots,\,S_{T}-1,\,S_{T}\right\} ,\\
N_{2} & \triangleq\left\{ -T+1,\ldots,\,-S_{T}-1,\,S_{T}+1,\ldots,\,T-1\right\} .
\end{align*}
Let us consider the first term above, 
\begin{align}
T^{1/3} & (\widehat{J}_{T}(\widehat{b}_{1,T},\,\widehat{b}_{2,T})-\widehat{J}_{T}(b_{\theta_{1},T},\,\widehat{b}_{2,T}))\label{Eq. 23}\\
 & =T^{1/3}\sum_{k\in N_{1}}(K_{1}(\widehat{b}_{1,T}k)-K_{1}(b_{\theta_{1},T}k))\widehat{\Gamma}\left(k\right)+T^{1/3}\sum_{k\in N_{2}}K_{1}(\widehat{b}_{1,T}k)\widehat{\Gamma}\left(k\right)\nonumber \\
 & \quad-T^{1/3}\sum_{k\in N_{2}}K_{1}(b_{\theta_{1},T}k)\widehat{\Gamma}\left(k\right)\nonumber \\
 & \triangleq A_{1,T}+A_{2,T}-A_{3,T}.\nonumber 
\end{align}
 We first show that $A_{1,T}\overset{\mathbb{P}}{\rightarrow}0$.
Let $A_{1,1,T}$ denote $A_{1,T}$ with the summation restricted over
positive integers $k$. Let $\widetilde{n}_{T}=\inf\{T/n_{3,T},\,\sqrt{n_{2,T}}\}$.
We can use the Lipschitz condition on $K_{1}\left(\cdot\right)\in\boldsymbol{K}_{3}$
to yield, 
\begin{align}
\left|A_{1,1,T}\right| & \leq T^{1/3}\sum_{k=1}^{S_{T}}C_{2}\left|\widehat{b}_{1,T}-b_{\theta_{1},T}\right|k\left|\widehat{\Gamma}\left(k\right)\right|\label{Eq. 24}\\
 & \leq C\widetilde{n}_{T}\left|\widehat{\phi}_{1}^{1/24}-\phi_{1,\theta^{*}}^{1/24}\right|\left(\widehat{\phi}_{1}\phi_{1,\theta^{*}}\right)^{-1/24}T^{1/3-1/6}\widetilde{n}_{T}^{-1}\sum_{k=1}^{S_{T}}k\left|\widehat{\Gamma}\left(k\right)\right|,\nonumber 
\end{align}
for some $C<\infty$. By Assumption \ref{Assumption E-F-G}-(ii)
($\widetilde{n}_{T}|\widehat{\phi}_{1}-\phi_{1,\theta^{*}}|=O_{\mathbb{P}}\left(1\right)$)
and using the delta method it suffices to show that $B_{1,T}+B_{2,T}+B_{3,T}\overset{\mathbb{P}}{\rightarrow}0$
where 
\begin{align}
B_{1,T} & =T^{1/6}\widetilde{n}_{T}^{-1}\sum_{k=1}^{S_{T}}k\left|\widehat{\Gamma}\left(k\right)-\widetilde{\Gamma}\left(k\right)\right|,\label{Eq. A.25 Andrews 91}\\
B_{2,T} & =T^{1/6}\widetilde{n}_{T}^{-1}\sum_{k=1}^{S_{T}}k\left|\widetilde{\Gamma}\left(k\right)-\Gamma_{T}\left(k\right)\right|,\qquad\mathrm{and}\nonumber \\
B_{3,T} & =T^{1/6}\widetilde{n}_{T}^{-1}\sum_{k=1}^{S_{T}}k\left|\Gamma_{T}\left(k\right)\right|,\nonumber 
\end{align}
with $\Gamma_{T}\left(k\right)\triangleq\left(n_{T}/T\right)\sum_{r=0}^{\left\lfloor T/n_{T}\right\rfloor }c\left(rn_{T}/T,\,k\right).$
By a mean-value expansion, we have 
\begin{align}
B_{1,T} & \leq T^{1/6}\widetilde{n}_{T}^{-1}T^{-1/2}\sum_{k=1}^{S_{T}}k\left|\left(\frac{\partial}{\partial\beta'}\widehat{\Gamma}\left(k\right)|_{\beta=\overline{\beta}}\right)\sqrt{T}\left(\widehat{\beta}-\beta_{0}\right)\right|\label{Eq. A.26}\\
 & \leq CT^{1/6-1/2}T^{2r/6}\widetilde{n}_{T}^{-1}\sup_{k\geq1}\left\Vert \frac{\partial}{\partial\beta'}\widehat{\Gamma}\left(k\right)|_{\beta=\overline{\beta}}\right\Vert \sqrt{T}\left\Vert \widehat{\beta}-\beta_{0}\right\Vert \nonumber \\
 & \leq CT^{1/6-1/2+r/3}\widetilde{n}_{T}^{-1}\sup_{k\geq1}\left\Vert \frac{\partial}{\partial\beta'}\widehat{\Gamma}\left(k\right)|_{\beta=\overline{\beta}}\right\Vert \sqrt{T}\left\Vert \widehat{\beta}-\beta_{0}\right\Vert \overset{\mathbb{P}}{\rightarrow}0,\nonumber 
\end{align}
since $\widetilde{n}_{T}/T^{1/3}\rightarrow\infty$, $r<2$, $\sqrt{T}||\widehat{\beta}-\beta_{0}||=O_{\mathbb{P}}\left(1\right)$,
and $\sup_{k\geq1}||\left(\partial/\partial\beta'\right)\widehat{\Gamma}\left(k\right)|_{\beta=\overline{\beta}}||=O_{\mathbb{P}}\left(1\right)$
using (S.28) in \citet{casini_hac} and Assumption \ref{Assumption B}-(ii,iii).
In addition, 
\begin{align}
\mathbb{E}\left(B_{2,T}^{2}\right) & \leq\mathbb{E}\left(T^{2/3}\widetilde{n}_{T}^{-2}\sum_{k=1}^{S_{T}}\sum_{j=1}^{S_{T}}kj\left|\widetilde{\Gamma}\left(k\right)-\Gamma_{T}\left(k\right)\right|\left|\widetilde{\Gamma}\left(j\right)-\Gamma_{T}\left(j\right)\right|\right)\label{Eq. A.27}\\
 & \leq b_{\theta_{2},T}^{-1}T^{2/3-1-2/3}S_{T}^{4}\sup_{k\geq1}Tb_{\theta_{2},T}\mathrm{Var}\left(\widetilde{\Gamma}\left(k\right)\right)\nonumber \\
 & \leq b_{\theta_{2},T}^{-1}T^{-1}T^{4r/6}\sup_{k\geq1}Tb_{\theta_{2},T}\mathrm{Var}\left(\widetilde{\Gamma}\left(k\right)\right)\rightarrow0,\nonumber 
\end{align}
 given that $\sup_{k\geq1}Tb_{\theta_{2},T}\mathrm{Var}(\widetilde{\Gamma}(k))=O\left(1\right)$
using Lemma S.B.5 in \citet{casini_hac} and $r<5/4$. Assumption
\ref{Assumption E-F-G}-(iii) and $\sum_{k=1}^{\infty}k^{1-l}<\infty$
for $l>2$ yield, 
\begin{align}
B_{3,T} & \leq T^{1/6}\widetilde{n}_{T}^{-1}C_{3}\sum_{k=1}^{\infty}k^{1-l}\rightarrow0,\label{Eq. A.28}
\end{align}
 where we have used the fact that $\widetilde{n}_{T}/T^{3/10}\rightarrow\infty$.
 Combining \eqref{Eq. 24}-\eqref{Eq. A.28} we deduce that $A_{1,1,T}\overset{\mathbb{P}}{\rightarrow}0$.
The same argument applied to $A_{1,T}$ where the summation now extends
over negative integers $k$ gives $A_{1,T}\overset{\mathbb{P}}{\rightarrow}0$.
Next, we show that $A_{2,T}\overset{\mathbb{P}}{\rightarrow}0$. Again,
we use the notation $A_{2,1,T}$ (resp., $A_{2,2,T}$) to denote $A_{2,T}$
with the summation over positive (resp., negative) integers. Let $A_{2,1,T}=L_{1,T}+L_{2,T}+L_{3,T}$,
where 
\begin{align}
L_{1,T}=L_{1,T}^{A}+L_{1,T}^{B} & =T^{1/3}\left(\sum_{k=S_{T}+1}^{\left\lfloor D_{T}T^{1/3}\right\rfloor }+\sum_{k=\left\lfloor D_{T}T^{1/3}\right\rfloor +1}^{T-1}\right)K_{1}\left(\widehat{b}_{1,T}k\right)\left(\widehat{\Gamma}\left(k\right)-\widetilde{\Gamma}\left(k\right)\right),\label{Eq. A.29}\\
L_{2,T}=L_{2,T}^{A}+L_{2,T}^{B} & =T^{1/3}\left(\sum_{k=S_{T}+1}^{\left\lfloor D_{T}T^{1/3}\right\rfloor }+\sum_{k=\left\lfloor D_{T}T^{1/3}\right\rfloor +1}^{T-1}\right)K_{1}\left(\widehat{b}_{1,T}k\right)\left(\widetilde{\Gamma}\left(k\right)-\Gamma_{T}\left(k\right)\right),\nonumber \\
\mathrm{and}\qquad L_{3,T} & =T^{1/3}\sum_{k=S_{T}+1}^{T-1}K_{1}\left(\widehat{b}_{1,T}k\right)\Gamma_{T}\left(k\right).\nonumber 
\end{align}
We apply a mean-value expansion and use $\sqrt{T}(\widehat{\beta}-\beta_{0})=O_{\mathbb{P}}\left(1\right)$
to obtain
\begin{align}
\left|L_{1,T}^{A}\right| & =T^{1/3-1/2}\sum_{k=S_{T}+1}^{\left\lfloor D_{T}T^{1/3}\right\rfloor }C_{1}\left(\widehat{b}_{1,T}k\right)^{-b}\left|\left(\frac{\partial}{\partial\beta'}\widehat{\Gamma}\left(k\right)\right)|_{\beta=\overline{\beta}}\sqrt{T}\left(\widehat{\beta}-\beta_{0}\right)\right|\label{Eq. (30)}\\
 & =T^{1/3-1/2+b/6}\sum_{k=S_{T}+1}^{\left\lfloor D_{T}T^{1/3}\right\rfloor }C_{1}k^{-b}\left|\left(\frac{\partial}{\partial\beta'}\widehat{\Gamma}\left(k\right)\right)|_{\beta=\overline{\beta}}\sqrt{T}\left(\widehat{\beta}-\beta_{0}\right)\right|\nonumber \\
 & =T^{1/3-1/2+b/6+r\left(1-b\right)/6}\left|\left(\frac{\partial}{\partial\beta'}\widehat{\Gamma}\left(k\right)\right)|_{\beta=\overline{\beta}}\sqrt{T}\left(\widehat{\beta}-\beta_{0}\right)\right|\nonumber \\
 & =T^{1/3-1/2+b/6+r\left(1-b\right)/6}O_{\mathbb{P}}\left(1\right)O_{\mathbb{P}}\left(1\right),\nonumber 
\end{align}
 which goes to zero since $r>1$ and 
\begin{align*}
\left|L_{1,T}^{B}\right| & =T^{1/3-1/2}\sum_{k=\left\lfloor D_{T}T^{1/3}\right\rfloor +1}^{T-1}C_{1}\left(\widehat{b}_{1,T}k\right)^{-b}\left|\left(\frac{\partial}{\partial\beta'}\widehat{\Gamma}\left(k\right)\right)|_{\beta=\overline{\beta}}\sqrt{T}\left(\widehat{\beta}-\beta_{0}\right)\right|\\
 & =CT^{1/3-1/2+b/6}\sum_{k=\left\lfloor D_{T}T^{1/3}\right\rfloor +1}^{T-1}C_{1}k^{-b}\left|\left(\frac{\partial}{\partial\beta'}\widehat{\Gamma}\left(k\right)\right)|_{\beta=\overline{\beta}}\sqrt{T}\left(\widehat{\beta}-\beta_{0}\right)\right|\\
 & =CD_{T}^{1-b}T^{1/3-1/2+b/6+\left(1-b\right)/3}\left|\left(\frac{\partial}{\partial\beta'}\widehat{\Gamma}\left(k\right)\right)|_{\beta=\overline{\beta}}\sqrt{T}\left(\widehat{\beta}-\beta_{0}\right)\right|\overset{\mathbb{P}}{\rightarrow}0,
\end{align*}
 given that $1-b<0.$ Let us now consider $L_{2,T}$. We have 
\begin{align}
\left|L_{2,T}^{A}\right| & =T^{1/3}\sum_{k=S_{T}+1}^{\left\lfloor D_{T}T^{1/3}\right\rfloor }C_{1}\left(\widehat{b}_{1,T}k\right)^{-b}\left|\widetilde{\Gamma}\left(k\right)-\Gamma_{T}\left(k\right)\right|\label{Eq. 31}\\
 & =C_{1}\left(\widehat{\phi}_{1}\right)^{-b/24}T^{1/3+b/6-1/2}b_{\theta_{2},T}^{-1/2}\left(\sum_{k=S_{T}+1}^{\left\lfloor D_{T}T^{1/3}\right\rfloor }k^{-b}\right)\nonumber \\
 & \quad\times\sqrt{Tb_{\theta_{2},T}}\left|\widetilde{\Gamma}\left(k\right)-\Gamma_{T}\left(k\right)\right|.\nonumber 
\end{align}
Note that 
\begin{align}
\mathbb{E} & \left(T^{1/3+b/6-1/2}b_{\theta_{2},T}^{-1/2}\sum_{k=S_{T}+1}^{\left\lfloor D_{T}T^{1/3}\right\rfloor }k^{-b}\sqrt{Tb_{\theta_{2},T}}\left|\widetilde{\Gamma}\left(k\right)-\Gamma_{T}\left(k\right)\right|\right)^{2}\label{Eq. 32 Andrews 91}\\
 & \leq CT^{2/3+b/3-1}b_{\theta_{2},T}^{-1}\left(\sum_{k=S_{T}+1}^{\left\lfloor D_{T}T^{1/3}\right\rfloor }k^{-b}\sqrt{Tb_{\theta_{2},T}}\left(\mathrm{Var}\left(\widetilde{\Gamma}\left(k\right)\right)\right)^{1/2}\right)^{2}\nonumber \\
 & =T^{2/3+b/3-1}b_{\theta_{2},T}^{-1}\left(\sum_{k=S_{T}+1}^{\left\lfloor D_{T}T^{1/3}\right\rfloor }k^{-b}\right)^{2}O\left(1\right)\nonumber \\
 & =T^{2/3+b/3-1}b_{\theta_{2},T}^{-1}S_{T}^{2\left(1-b\right)}O\left(1\right)\rightarrow0,\nonumber 
\end{align}
    since $r>\left(b-1/2\right)/\left(b-1\right)$, $b>3$, and
$Tb_{\theta_{2},T}\mathrm{Var}(\widetilde{\Gamma}\left(k\right))=O\left(1\right)$
as above. Next, 
\begin{align*}
\left|L_{2,T}^{B}\right| & =T^{1/3}\sum_{k=\left\lfloor D_{T}T^{1/3}\right\rfloor +1}^{T-1}C_{1}\left(\widehat{b}_{1,T}k\right)^{-b}\left|\widetilde{\Gamma}\left(k\right)-\Gamma_{T}\left(k\right)\right|\\
 & =C_{1}\left(\widehat{\phi}_{1}\right)^{-b/24}T^{1/3+b/6-1/2}b_{\theta_{2},T}^{-1/2}\left(\sum_{k=S_{T}+1}^{T-1}k^{-b}\right)\\
 & \quad\times\sqrt{Tb_{\theta_{2},T}}\left|\widetilde{\Gamma}\left(k\right)-\Gamma_{T}\left(k\right)\right|,
\end{align*}
 and 
\begin{align}
\mathbb{E} & \left(T^{1/3+b/6-1/2}b_{\theta_{2},T}^{-1/2}\sum_{k=\left\lfloor D_{T}T^{1/3}\right\rfloor +1}^{T-1}k^{-b}\sqrt{Tb_{\theta_{2},T}}\left|\widetilde{\Gamma}\left(k\right)-\Gamma_{T}\left(k\right)\right|\right)^{2}\label{Eq. 32 Andrews 91-2}\\
 & =T^{2/3+b/3-1}b_{\theta_{2},T}^{-1}\left(\sum_{k=\left\lfloor D_{T}T^{1/3}\right\rfloor +1}^{T-1}k^{-b}\right)^{2}O\left(1\right)\nonumber \\
 & =T^{2/3+b/3-1}b_{\theta_{2},T}^{-1}T^{2\left(1-b\right)/3}D_{T}^{2}O\left(1\right)\rightarrow0,\nonumber 
\end{align}
since $2b>3$ given $b>1+1/q$ and $q\leq2$. The cross-product term
involving 
\begin{align*}
\left(\sum_{k=S_{T}+1}^{\left\lfloor D_{T}T^{1/3}\right\rfloor }\sum_{j=\left\lfloor D_{T}T^{1/3}\right\rfloor +1}^{T-1}\right)K_{1}\left(\widehat{b}_{1,T}k\right)K_{1}\left(\widehat{b}_{1,T}j\right) & \left(\widetilde{\Gamma}\left(j\right)-\Gamma_{T}\left(j\right)\right)\left(\widetilde{\Gamma}\left(j\right)-\Gamma_{T}\left(j\right)\right),
\end{align*}
can be treated in a similar fashion. Combining \eqref{Eq. 31}-\eqref{Eq. 32 Andrews 91-2}
yields $L_{2,T}\overset{\mathbb{P}}{\rightarrow}0$. Let us turn
to $L_{3,T}$. By Assumption \ref{Assumption E-F-G}-(iii) and $\left|K_{1}\left(\cdot\right)\right|\leq1$,
we have, 
\begin{align}
\left|L_{3,T}\right| & \leq T^{1/3}\sum_{k=S_{T}}^{T-1}C_{3}k^{-l}\leq T^{1/3}C_{3}S_{T}^{1-l}\leq C_{3}T^{1/3}T^{-r\left(l-1\right)/6}\rightarrow0,\label{Eq. (33)}
\end{align}
 since $r>2/\left(l-1\right)$.  In view of \eqref{Eq. A.29}-\eqref{Eq. (33)}
we deduce that $A_{2,1,T}\overset{\mathbb{P}}{\rightarrow}0$. Applying
the same argument to $A_{2,2,T}$, we have $A_{2,T}\overset{\mathbb{P}}{\rightarrow}0$.
Using similar arguments, one has $A_{3,T}\overset{\mathbb{P}}{\rightarrow}0$.
It remains to show that $T^{1/3}(\widehat{J}_{T}(b_{\theta_{1},T},\,\widehat{b}_{2,T})-\widehat{J}_{T}(b_{\theta_{1},T},\,b_{\theta_{2},T}))\overset{\mathbb{P}}{\rightarrow}0$.
Let $\widehat{c}_{\theta_{2},T}\left(rn_{T}/T,\,k\right)$ denote
the estimator that uses $b_{\theta_{2},T}$ in place of $\widehat{b}_{2,T}.$
We have for $k\geq0,$ 
\begin{align}
\widehat{c}_{T} & \left(rn_{T}/T,\,k\right)-\widehat{c}_{\theta_{2},T}\left(rn_{T}/T,\,k\right)\nonumber \\
 & =\left(Tb_{\theta_{2},T}\right)^{-1}\sum_{s=k+1}^{T}\left(K_{2}\left(\frac{\left(\left(r+1\right)n_{T}-\left(s-k/2\right)\right)/T}{\widehat{b}_{2,T}}\right)-K_{2}\left(\frac{\left(\left(r+1\right)n_{T}-\left(s-k/2\right)\right)/T}{b_{\theta_{2},T}}\right)\right)\widehat{V}_{s}\widehat{V}{}_{s-k}\nonumber \\
 & \quad+O_{\mathbb{P}}\left(1/Tb_{2,b_{\theta_{2},T}}\right).\label{eq (c_hat - c_theta2)}
\end{align}
Given Assumption \ref{Assumption E-F-G}-(ii,v) and using the delta
method, we have for $s\in\left\{ Tu-\left\lfloor Tb_{\theta_{2},T}\right\rfloor ,\ldots,\,Tu+\left\lfloor Tb_{\theta_{2},T}\right\rfloor \right\} $
\begin{align}
K_{2} & \left(\frac{\left(Tu-\left(s-k/2\right)\right)/T}{\widehat{b}_{2,T}}\right)-K_{2}\left(\frac{\left(Tu-\left(s-k/2\right)\right)/T}{b_{\theta_{2},T}}\right)\label{Eq. K2-K2 for part (ii)}\\
 & \leq C_{4}\left|\frac{Tu-\left(s-k/2\right)}{T\widehat{b}_{2,T}}-\frac{Tu-\left(s-k/2\right)}{Tb_{\theta_{2},T}}\right|\nonumber \\
 & \leq C_{4}T^{-5/6}\widetilde{n}_{T}^{-1}\widetilde{n}_{T}\left|\widehat{\phi}_{2}^{-1/24}-\phi_{2,\theta^{*}}^{-1/24}\right|\left|Tu-\left(s-k/2\right)\right|\nonumber \\
 & \leq CT^{-5/6}\widetilde{n}_{T}^{-1}O_{\mathbb{P}}\left(1\right)\left|Tu-\left(s-k/2\right)\right|.\nonumber 
\end{align}
 Therefore, 
\begin{align*}
T^{1/3} & \left(\widehat{J}_{T}\left(b_{\theta_{1},T},\,\widehat{b}_{2,T}\right)-\widehat{J}_{T}\left(b_{\theta_{1},T},\,b_{\theta_{2},T}\right)\right)\\
 & =T^{1/3}\sum_{k=-T+1}^{T-1}K_{1}\left(b_{\theta_{1},T}k\right)\frac{n_{T}}{T}\sum_{r=0}^{\left\lfloor T/n_{T}\right\rfloor }\left(\widehat{c}\left(rn_{T}/T,\,k\right)-\widehat{c}_{\theta_{2},T}\left(rn_{T}/T,\,k\right)\right)\\
 & \leq T^{1/3}C\sum_{k=-T+1}^{T-1}\bigl|K_{1}\left(b_{\theta_{1},T}k\right)\bigr|\frac{n_{T}}{T}\sum_{r=0}^{\left\lfloor T/n_{T}\right\rfloor }\frac{1}{Tb_{\theta_{2},T}}\\
 & \quad\times\sum_{s=k+1}^{T}\left|K_{2}\left(\frac{\left(\left(r+1\right)n_{T}-\left(s-k/2\right)\right)/T}{\widehat{b}_{2,T}}\right)-K_{2}\left(\frac{\left(\left(r+1\right)n_{T}-\left(s-k/2\right)\right)/T}{b_{\theta_{2},T}}\right)\right|\\
 & \quad\times\left|\left(\widehat{V}_{s}\widehat{V}_{s-k}-V_{s}V_{s-k}\right)+\left(V_{s}V_{s-k}-\mathbb{E}\left(V_{s}V_{s-k}\right)\right)+\mathbb{E}\left(V_{s}V_{s-k}\right)\right|\\
 & \triangleq H_{1,T}+H_{2,T}+H_{3,T}.
\end{align*}
We have to show that $H_{1,T}+H_{2,T}+H_{3,T}\overset{\mathbb{P}}{\rightarrow}0$.
Let $H_{1,1,T}$ (resp. $H_{1,2,T}$) be defined as $H_{1,T}$ but
with the sum over $k$ be restricted to $k=1,\ldots,\,S_{T}$ (resp.
$k=S_{T}+1,\ldots,\,T$). By a mean-value expansion,  using \eqref{Eq. K2-K2 for part (ii)},
\begin{align*}
\left|H_{1,1,T}\right| & \leq CT^{1/3-1/2}\sum_{k=1}^{S_{T}}\left|K_{1}\left(b_{\theta_{1},T}k\right)\right|\frac{n_{T}}{T}\sum_{r=0}^{\left\lfloor T/n_{T}\right\rfloor }\frac{1}{Tb_{\theta_{2},T}}\\
 & \quad\times\sum_{s=k+1}^{T}\left|K_{2}\left(\frac{\left(\left(r+1\right)n_{T}-\left(s-k/2\right)\right)/T}{\widehat{b}_{2,T}}\right)-K_{2}\left(\frac{\left(\left(r+1\right)n_{T}-\left(s-k/2\right)\right)/T}{b_{\theta_{2},T}}\right)\right|\\
 & \quad\times\left\Vert V_{s}\left(\overline{\beta}\right)\frac{\partial}{\partial\beta}V_{s-k}\left(\overline{\beta}\right)+V_{s-k}\left(\overline{\beta}\right)\frac{\partial}{\partial\beta}V_{s}\left(\overline{\beta}\right)\right\Vert \sqrt{T}\left\Vert \widehat{\beta}-\beta_{0}\right\Vert \\
 & \leq Cb_{\theta_{2},T}^{-1}T^{1/3-1/2-1/3}S_{T}\frac{n_{T}}{T}\sum_{r=0}^{\left\lfloor T/n_{T}\right\rfloor }\left(CO_{\mathbb{P}}\left(1\right)\right)\\
 & \quad\times\left(\left(T^{-1}\sum_{s=1}^{T}\sup_{\beta\in\Theta}V_{s}^{2}\left(\beta\right)\right)^{2}\left(T^{-1}\sum_{s=1}^{T}\sup_{\beta\in\Theta}\left\Vert \frac{\partial}{\partial\beta}V_{s}\left(\beta\right)\right\Vert ^{2}\right)^{1/2}\right)\sqrt{T}\left\Vert \widehat{\beta}-\beta_{0}\right\Vert ,
\end{align*}
 where we have used the fact that $\widetilde{n}_{T}/T^{1/3}\rightarrow\infty.$
Using Assumption \ref{Assumption B} the right-hand side above is
\begin{align*}
C & T^{-1/2}b_{\theta_{2},T}^{-1}S_{T}\frac{n_{T}}{T}\sum_{r=0}^{\left\lfloor T/n_{T}\right\rfloor }O_{\mathbb{P}}\left(1\right)\overset{\mathbb{P}}{\rightarrow}0,
\end{align*}
since $r<2$. Next, 
\begin{align*}
\left|H_{1,2,T}\right| & \leq CT^{1/3-1/2}\sum_{k=S_{T}+1}^{T-1}\left(b_{\theta_{1},T}k\right)^{-b}\frac{n_{T}}{T}\sum_{r=0}^{\left\lfloor T/n_{T}\right\rfloor }\frac{1}{Tb_{\theta_{2},T}}\\
 & \quad\times\sum_{s=k+1}^{T}\left|K_{2}\left(\frac{\left(\left(r+1\right)n_{T}-\left(s-k/2\right)\right)/T}{\widehat{b}_{2,T}}\right)-K_{2}\left(\frac{\left(\left(r+1\right)n_{T}-\left(s-k/2\right)\right)/T}{b_{\theta_{2},T}}\right)\right|\\
 & \quad\times\left\Vert V_{s}\left(\overline{\beta}\right)\frac{\partial}{\partial\beta}V_{s-k}\left(\overline{\beta}\right)+V_{s-k}\left(\overline{\beta}\right)\frac{\partial}{\partial\beta}V_{s}\left(\overline{\beta}\right)\right\Vert \sqrt{T}\left\Vert \widehat{\beta}-\beta_{0}\right\Vert \\
 & \leq Cb_{\theta_{2},T}^{-1}T^{1/3-1/2-1/3}b_{\theta_{1},T}^{-b}\sum_{k=S_{T}+1}^{T-1}k^{-b}\frac{n_{T}}{T}\sum_{r=0}^{\left\lfloor T/n_{T}\right\rfloor }O_{\mathbb{P}}\left(1\right)\\
 & \quad\times\left(\left(T^{-1}\sum_{s=1}^{T}\sup_{\beta\in\Theta}V_{s}^{2}\left(\beta\right)\right)^{2}\left(T^{-1}\sum_{s=1}^{T}\sup_{\beta\in\Theta}\left\Vert \frac{\partial}{\partial\beta}V_{s}\left(\beta\right)\right\Vert ^{2}\right)^{1/2}\right)\sqrt{T}\left\Vert \widehat{\beta}-\beta_{0}\right\Vert \\
 & \leq Cb_{\theta_{2},T}^{-1}T^{-1/2}b_{\theta_{1},T}^{-b}\sum_{k=S_{T}+1}^{T-1}k^{-b}O_{\mathbb{P}}\left(1\right)\\
 & \leq Cb_{\theta_{2},T}^{-1}T^{-1/2}b_{\theta_{1},T}^{-b}S_{T}^{1-b}O_{\mathbb{P}}\left(1\right)\\
 & \leq Cb_{\theta_{2},T}^{-1}T^{-1/2}b_{\theta_{1},T}^{-b}b_{\theta_{1},T}^{-r\left(1-b\right)}O_{\mathbb{P}}\left(1\right)\\
 & \leq Cb_{\theta_{2},T}^{-1}T^{-1/2}b_{\theta_{1},T}^{-b}T^{r\left(1-b\right)/6}O_{\mathbb{P}}\left(1\right)\rightarrow0,
\end{align*}
 since $r>\left(b-2\right)/\left(b-1\right).$ This shows $H_{1,T}\overset{\mathbb{P}}{\rightarrow}0$.
Let $H_{2,1,T}$ (resp. $H_{2,2,T}$) be defined as $H_{2,T}$ but
with the sum over $k$ be restricted to $k=1,\ldots,\,S_{T}$ (resp.
$k=S_{T}+1,\ldots,\,T$). Using $\left|K_{1}\left(\cdot\right)\right|\leq1$
we have,
\begin{align}
\mathbb{E}\left(H_{2,1,T}^{2}\right) & \leq CT^{2/3}\sum_{k=1}^{S_{T}}\sum_{j=1}^{S_{T}}K_{1}\left(b_{\theta_{1},T}k\right)K_{1}\left(b_{\theta_{1},T}j\right)\left(\frac{n_{T}}{T}\right)^{2}\sum_{r_{1}=0}^{\left\lfloor T/n_{T}\right\rfloor }\sum_{r_{2}=0}^{\left\lfloor T/n_{T}\right\rfloor }\frac{1}{\left(Tb_{\theta_{2},T}\right)^{2}}\label{Eq. (H21) =00003D 0 part (ii)}\\
 & \quad\times\sum_{s=k+1}^{T}\sum_{t=j+1}^{T}\left|K_{2}\left(\frac{\left(\left(r_{1}+1\right)n_{T}-\left(s-k/2\right)\right)/T}{\widehat{b}_{2,T}}\right)-K_{2}\left(\frac{\left(\left(r_{1}+1\right)n_{T}-\left(s-k/2\right)\right)/T}{b_{\theta_{2},T}}\right)\right|\nonumber \\
 & \quad\times\left|K_{2}\left(\frac{\left(\left(r_{2}+1\right)n_{T}-\left(t-j/2\right)\right)/T}{\widehat{b}_{2,T}}\right)-K_{2}\left(\frac{\left(\left(r_{2}+1\right)n_{T}-\left(t-j/2\right)\right)/T}{b_{\theta_{2},T}}\right)\right|\nonumber \\
 & \quad\times\left|\mathbb{E}\left(V_{s}V_{s-k}-\mathbb{E}\left(V_{s}V_{s-k}\right)\right)\left(V_{t}V_{t-k}-\mathbb{E}\left(V_{t}V_{t-k}\right)\right)\right|\nonumber \\
 & \leq CT^{2/3}S_{T}^{2}\widetilde{n}_{T}^{-2}\left(Tb_{\theta_{2},T}\right)^{-1}\sup_{k\geq1}Tb_{\theta_{2},T}\mathrm{Var}\left(\widetilde{\Gamma}\left(k\right)\right)O_{\mathbb{P}}\left(1\right)\nonumber \\
 & \leq CT^{2/3-2/3-1+2r/6}O_{\mathbb{P}}\left(b_{\theta_{2},T}^{-1}\right)\rightarrow0,\nonumber 
\end{align}
where we have used Lemma S.B.5 in \citet{casini_hac} and $r<3$.
Turning to $H_{2,2,T},$ 
\begin{align}
\mathbb{E}\left(H_{2,2,T}^{2}\right) & \leq CT^{2/3-2/3}\left(Tb_{\theta_{2},T}\right)^{-1}b_{\theta_{1},T}^{-2b}\left(\sum_{k=S_{T}+1}^{T-1}k^{-b}\sqrt{Tb_{\theta_{2},T}}\left(\mathrm{Var}\left(\widetilde{\Gamma}\left(k\right)\right)\right)^{1/2}O\left(1\right)\right)^{2}\label{Eq. (H22) =00003D0 part (ii)}\\
 & \leq CT^{-1}b_{\theta_{2},T}^{-1}b_{\theta_{1},T}^{-2b}\left(\sum_{k=S_{T}+1}^{T-1}k^{-b}\sqrt{Tb_{\theta_{2},T}}\left(\mathrm{Var}\left(\widetilde{\Gamma}\left(k\right)\right)\right)^{1/2}\right)^{2}\nonumber \\
 & \leq CT^{-1}b_{\theta_{2},T}^{-1}b_{\theta_{1},T}^{-2b}\left(\sum_{k=S_{T}+1}^{T-1}k^{-b}O\left(1\right)\right)^{2}\nonumber \\
 & \leq CT^{-1}b_{\theta_{2},T}^{-1}b_{\theta_{1},T}^{-2b}S_{T}^{2\left(1-b\right)}\rightarrow0,\nonumber 
\end{align}
since $r>(b-5/2)/\left(b-1\right).$ Eq. \eqref{Eq. (H21) =00003D 0 part (ii)}
and \eqref{Eq. (H22) =00003D0 part (ii)} yield $H_{2,T}\overset{\mathbb{P}}{\rightarrow}0.$
 Let $H_{3,1,T}$ (resp. $H_{3,2,T}$) be defined as $H_{3,T}$ but
with the sum over $k$ be restricted to $k=1,\ldots,\,S_{T}$ (resp.
$k=S_{T}+1,\ldots,\,T$). Given $\left|K_{1}\left(\cdot\right)\right|\leq1$
and \eqref{Eq. K2-K2 for part (ii)}, we have  
\begin{align*}
\left|H_{3,1,T}\right| & \leq CT^{1/3}\widetilde{n}_{T}^{-1}\sum_{k=1}^{S_{T}}\left|\Gamma_{T}\left(k\right)\right|\leq CT^{1/3}\widetilde{n}_{T}^{-1}\sum_{k=1}^{\infty}k^{-l}\rightarrow0,
\end{align*}
since $\sum_{k=1}^{\infty}k^{-l}<\infty$ for $l>1$ and $\widetilde{n}_{T}/T^{1/3}\rightarrow\infty$.
Finally, 
\begin{align*}
\left|H_{3,2,T}\right| & \leq CT^{1/3}\widetilde{n}_{T}^{-1}\sum_{k=S_{T}+1}^{T-1}\left|\Gamma_{T}\left(k\right)\right|\leq CT^{1/3}\widetilde{n}_{T}^{-1}\sum_{k=S_{T}+1}^{T-1}k^{-l}\\
 & \leq CT^{1/3}\widetilde{n}_{T}^{-1}S_{T}^{1-l}\leq CT^{1/3}\widetilde{n}_{T}^{-1}T^{r\left(1-l\right)/6}\rightarrow0,
\end{align*}
since $l>1$ and $\widetilde{n}_{T}/T^{1/3}\rightarrow\infty.$ This
completes the proof of part (ii).

We now move to part (i). For some $\phi_{1,\theta^{*}},\,\phi_{2,\theta^{*}}\in\left(0,\,\infty\right)$,
$\widehat{J}_{T}\left(b_{\theta_{1},T},\,b_{\theta_{2},T}\right)-J_{T}=o_{\mathbb{P}}\left(1\right)$
by Theorem \ref{Theorem 1 -Consistency and Rate}-(i) since $O\left(b_{\theta_{1},T}\right)=O\left(b_{\theta_{2},T}\right)$
and $\sqrt{T}b_{1,T}\rightarrow\infty$ hold.  Hence, it remains
to show $\widehat{J}_{T}(b_{\theta_{1},T},\,b_{\theta_{2},T})-\widehat{J}_{T}(\widehat{b}_{1,T},\,\widehat{b}_{2,T})=o_{\mathbb{P}}\left(1\right)$.
Note that this result differs from the result of part (ii) only because
the scale factor $T^{1/3}$ does not appear, Assumption \ref{Assumption E-F-G}-(ii)
is replaced by part (i) of the same assumption and Assumption \ref{Assumption E-F-G}-(iii)
is not imposed. Let
\begin{align*}
r\in\left(\max\left\{ \left(2b-5\right)/2\left(b-1\right)\right\} ,\,1\right) & ,
\end{align*}
with $b>1+1/q$ and let $S_{T}$ be defined as in part (ii). We will
use the decomposition in \eqref{eq (Decomposition J_T proof of Theorem 3 Andrews91)},
and $N_{1}$ and $N_{2}$ as defined after \eqref{eq (Decomposition J_T proof of Theorem 3 Andrews91)}.
Let $A_{1,T},\,A_{2,T}$ and $A_{3,T}$ be as in \eqref{Eq. 23}
without the scale factor $T^{1/3}$. Proceeding as in \eqref{Eq. 24},
\begin{align}
\left|A_{1,1,T}\right| & \leq\sum_{k=1}^{S_{T}}C_{2}\left|\widehat{b}_{1,T}-b_{\theta_{1},T}\right|k\left|\widehat{\Gamma}\left(k\right)\right|\label{Eq. 24-1}\\
 & \leq C\left|\widehat{\phi}_{1}{}^{1/24}-\phi_{1,\theta^{*}}^{1/24}\right|\left(\widehat{\phi}_{1}\phi_{1,\theta^{*}}\right)^{-1/24}T^{-1/6}\sum_{k=1}^{S_{T}}k\left|\widehat{\Gamma}\left(k\right)\right|,\nonumber 
\end{align}
for some $C<\infty$. By Assumption \ref{Assumption E-F-G}-(i),
\begin{align*}
\left|\widehat{\phi}_{1}{}^{1/24}-\phi_{1,\theta^{*}}^{1/24}\right|\left(\widehat{\phi}_{1}\phi_{1,\theta^{*}}\right)^{-1/24} & =O_{\mathbb{P}}\left(1\right).
\end{align*}
Then, it suffices to show that $B_{1,T}+B_{2,T}+B_{3,T}\overset{\mathbb{P}}{\rightarrow}0$,
where 
\begin{align}
B_{1,T} & =T^{-1/6}\sum_{k=1}^{S_{T}}k\left|\widehat{\Gamma}\left(k\right)-\widetilde{\Gamma}\left(k\right)\right|\label{Eq. A.25 Andrews 91-1}\\
B_{2,T} & =T^{-1/6}\sum_{k=1}^{S_{T}}k\left|\widetilde{\Gamma}\left(k\right)-\Gamma_{T}\left(k\right)\right|\nonumber \\
B_{3,T} & =T^{-1/6}\sum_{k=1}^{S_{T}}k\left|\Gamma_{T}\left(k\right)\right|.\nonumber 
\end{align}
By a mean-value expansion, we have 
\begin{align}
B_{1,T} & \leq T^{-1/6}T^{-1/2}\sum_{k=1}^{S_{T}}k\left|\left(\frac{\partial}{\partial\beta'}\widehat{\Gamma}\left(k\right)|_{\beta=\overline{\beta}}\right)\sqrt{T}\left(\widehat{\beta}-\beta_{0}\right)\right|\label{Eq. A.26-1}\\
 & \leq CT^{-1/6}T^{2r/6}T^{-1/2}\sup_{k\geq1}\left\Vert \frac{\partial}{\partial\beta}\widehat{\Gamma}\left(k\right)|_{\beta=\overline{\beta}}\right\Vert \sqrt{T}\left\Vert \widehat{\beta}-\beta_{0}\right\Vert ,\nonumber 
\end{align}
 since $r<2$, and $\sup_{k\geq1}||\left(\partial/\partial\beta\right)\widehat{\Gamma}\left(k\right)|_{\beta=\overline{\beta}}||=O_{\mathbb{P}}\left(1\right)$
using (S.28) in \citet{casini_hac} and Assumption \ref{Assumption B}-(ii,iii).
In addition, 
\begin{align}
\mathbb{E}\left(B_{2,T}^{2}\right) & \leq\mathbb{E}\left(T^{-1/3}\sum_{k=1}^{S_{T}}\sum_{j=1}^{S_{T}}kj\left|\widetilde{\Gamma}\left(k\right)-\Gamma_{T}\left(k\right)\right|\left|\widetilde{\Gamma}\left(j\right)-\Gamma_{T}\left(j\right)\right|\right)\label{Eq. A.27-1}\\
 & \leq\mathbb{E}\left(T^{-1/3}\sum_{k=1}^{S_{T}}\sum_{j=1}^{S_{T}}kj\left|\widetilde{\Gamma}\left(k\right)-\Gamma_{T}\left(k\right)\right|\left|\widetilde{\Gamma}\left(j\right)-\Gamma_{T}\left(j\right)\right|\right)\nonumber \\
 & \leq T^{-1/3-5/6}S_{T}^{4}\sup_{k\geq1}Tb_{\theta_{2},T}\mathrm{Var}\left(\widetilde{\Gamma}\left(k\right)\right)\nonumber \\
 & \leq T^{-1/3-5/6}T^{4r/6}\sup_{k\geq1}Tb_{\theta_{2},T}\mathrm{Var}\left(\widetilde{\Gamma}\left(k\right)\right)\rightarrow0,\nonumber 
\end{align}
 given that $\sup_{k\geq1}Tb_{\theta_{2},T}\mathrm{Var}(\widetilde{\Gamma}\left(k\right))=O\left(1\right)$
by Lemma S.B.5 in \citet{casini_hac} and $r<7/4$. The bound in equation
\eqref{Eq. A.28} is replaced by,
\begin{align}
B_{3,T} & \leq T^{-1/6}S_{T}\sum_{k=1}^{\infty}\left|\Gamma_{T}\left(k\right)\right|\leq T^{\left(r-1\right)/6}O_{\mathbb{P}}\left(1\right)\rightarrow0,\label{Eq. A.28-1}
\end{align}
using Assumption \ref{Assumption A - Dependence}-(i) since $r<1$.
This gives $A_{1,T}\overset{\mathbb{P}}{\rightarrow}0$. Next, we
show that $A_{2,T}\overset{\mathbb{P}}{\rightarrow}0$. As above,
let $A_{2,1,T}=L_{1,T}+L_{2,T}+L_{3,T}$ where each summand is defined
as in \eqref{Eq. A.29} without the factor $T^{1/3}$. We have
\begin{align}
\left|L_{1,T}\right| & =T^{-1/2}\sum_{k=S_{T}+1}^{T-1}C_{1}\left(\widehat{b}_{1,T}k\right)^{-b}\left|\left(\frac{\partial}{\partial\beta'}\widehat{\Gamma}\left(k\right)\right)|_{\beta=\overline{\beta}}\sqrt{T}\left(\widehat{\beta}-\beta_{0}\right)\right|\label{Eq. (30)-2}\\
 & =T^{-1/2+b/6}\sum_{k=S_{T}+1}^{T-1}C_{1}k^{-b}\left|\left(\frac{\partial}{\partial\beta'}\widehat{\Gamma}\left(k\right)\right)|_{\beta=\overline{\beta}}\sqrt{T}\left(\widehat{\beta}-\beta_{0}\right)\right|\nonumber \\
 & =T^{-1/2+b/6+r\left(1-b\right)/6}\left|\left(\frac{\partial}{\partial\beta'}\widehat{\Gamma}\left(k\right)\right)|_{\beta=\overline{\beta}}\sqrt{T}\left(\widehat{\beta}-\beta_{0}\right)\right|\nonumber \\
 & =T^{-1/2+b/6+r\left(1-b\right)/6}O\left(1\right)O_{\mathbb{P}}\left(1\right),\nonumber 
\end{align}
 which converges to zero since $r>\left(b-3\right)/\left(b-1\right)$.
The bound for $L_{2,T}$ is given by 
\begin{align}
\left|L_{2,T}\right| & =\sum_{k=S_{T}+1}^{T-1}C_{1}\left(\widehat{b}_{1,T}k\right)^{-b}\left|\widetilde{\Gamma}\left(k\right)-\Gamma_{T}\left(k\right)\right|\label{Eq. 31-1}\\
 & =C_{1}\widehat{\phi}_{1}^{-b/24}T^{b/6-1/2}b_{\theta_{2},T}^{-1/2}\left(\sum_{k=S_{T}+1}^{T-1}k^{-b}\right)\sqrt{Tb_{\theta_{2},T}}\left|\widetilde{\Gamma}\left(k\right)-\Gamma_{T}\left(k\right)\right|,\nonumber 
\end{align}
and the bound in \eqref{Eq. 32 Andrews 91} is replaced by, 
\begin{align}
\mathbb{E} & \left(T^{b/6-1/2}b_{\theta_{2},T}^{-1/2}\sum_{k=S_{T}}^{T-1}k^{-b}\sqrt{Tb_{\theta_{2},T}}\left|\widetilde{\Gamma}\left(k\right)-\Gamma_{T}\left(k\right)\right|\right)^{2}\label{Eq. 32 Andrews 91-1}\\
 & \leq T^{b/3-1}b_{\theta_{2},T}^{-1}\left(\sum_{k=S_{T}}^{T-1}k^{-b}\sqrt{Tb_{\theta_{2},T}}\left(\mathrm{Var}\left(\widetilde{\Gamma}\left(k\right)\right)\right)^{1/2}\right)^{2}\nonumber \\
 & =T^{b/3-1}b_{\theta_{2},T}^{-1}\left(\sum_{k=S_{T}}^{T-1}k^{-b}\right)^{2}O\left(1\right)\nonumber \\
 & =T^{b/3-1+1/6}S_{T}^{2\left(1-b\right)}O\left(1\right)\rightarrow0,\nonumber 
\end{align}
 since $r>\left(2b-5\right)/2\left(b-1\right)$ and $Tb_{2,T}\mathrm{Var}(\widetilde{\Gamma}\left(k\right))=O\left(1\right)$
as above. Equations \eqref{Eq. 31-1}-\eqref{Eq. 32 Andrews 91-1}
combine to yield $L_{2,T}\overset{\mathbb{P}}{\rightarrow}0$ since
$\widehat{\phi}_{1}=O_{\mathbb{P}}\left(1\right)$. The bound for
$L_{3,T}$ is given by
\begin{align}
\left|\sum_{k=S_{T}+1}^{T-1}K_{1}\left(\widehat{b}_{1,T}k\right)\Gamma_{T}\left(k\right)\right| & \leq\sum_{k=S_{T}+1}^{T-1}\frac{n_{T}}{T}\sum_{r=0}^{\left\lfloor T/n_{T}\right\rfloor }\left|c\left(rn_{T}/T,\,k\right)\right|\label{Eq. (33)-1}\\
 & \leq\sum_{k=S_{T}+1}^{T-1}\sup_{u\in\left[0,\,1\right]}\left|c\left(u,\,k\right)\right|\rightarrow0.\nonumber 
\end{align}
Equations \eqref{Eq. (30)-2}-\eqref{Eq. (33)-1} imply $A_{2,1,T}\overset{\mathbb{P}}{\rightarrow}0$.
Thus, as in the proof of part (ii), we have $A_{2,T}\overset{\mathbb{P}}{\rightarrow}0$
and $A_{3,T}\overset{\mathbb{P}}{\rightarrow}0$. It remains to show
that $(\widehat{J}_{T}(b_{\theta_{1},T},\,\widehat{b}_{2,T})-\widehat{J}_{T}(b_{\theta_{1},T},\,b_{\theta_{2},T}))\overset{\mathbb{P}}{\rightarrow}0$.
Let $\widehat{c}_{\theta_{2},T}\left(rn_{T}/T,\,k\right)$ be defined
as in part (ii). We have \eqref{eq (c_hat - c_theta2)} and \eqref{Eq. K2-K2 for part (ii)}
is replaced by 
\begin{align}
K_{2} & \left(\frac{\left(\left(r+1\right)n_{T}-\left(s-k/2\right)\right)/T}{\widehat{b}_{2,T}}\right)-K_{2}\left(\frac{\left(\left(r+1\right)n_{T}-\left(s-k/2\right)\right)/T}{b_{\theta_{2},T}}\right)\label{Eq. K2-K2 for part (i)}\\
 & \leq C_{4}\left|\frac{Tu-\left(s-k/2\right)}{T\widehat{b}_{2,T}}-\frac{Tu-\left(s-k/2\right)}{Tb_{\theta_{2},T}}\right|\nonumber \\
 & \leq C_{4}T^{-1}\left|\frac{Tu-\left(s-k/2\right)\left(\widehat{b}_{2,T}-b_{\theta_{2},T}\right)}{\widehat{b}_{2,T}b_{\theta_{2},T}}\right|\nonumber \\
 & \leq C_{4}T^{-5/6}\left(\frac{1}{\widehat{\phi}_{2}\phi_{2,\theta^{*}}}\right)^{1/24}\left(\widehat{\phi}_{2}^{1/24}-\phi_{2,\theta^{*}}^{1/24}\right)\left(Tu-\left(s-k/2\right)\right),\nonumber 
\end{align}
for $s\in\left\{ Tu-\left\lfloor Tb_{\theta_{2},T}\right\rfloor ,\ldots,\,Tu+\left\lfloor Tb_{\theta_{2},T}\right\rfloor \right\} $.
Therefore, 
\begin{align}
\widehat{J}_{T} & \left(b_{\theta_{1},T},\,\widehat{b}_{2,T}\right)-\widehat{J}_{T}\left(b_{\theta_{1},T},\,b_{\theta_{2},T}\right)\label{Eq. (H1+H2+H3)}\\
 & =\sum_{k=-T+1}^{T-1}K_{1}\left(b_{\theta_{1},T}k\right)\frac{n_{T}}{T}\sum_{r=0}^{\left\lfloor T/n_{T}\right\rfloor }\left(\widehat{c}\left(rn_{T}/T,\,k\right)-\widehat{c}_{\theta_{2},T}\left(rn_{T}/T,\,k\right)\right)\nonumber \\
 & \leq C\sum_{k=-T+1}^{T-1}K_{1}\left(b_{\theta_{1},T}k\right)\nonumber \\
 & \quad\times\frac{n_{T}}{T}\sum_{r=0}^{\left\lfloor T/n_{T}\right\rfloor }\frac{1}{Tb_{\theta_{2},T}}\sum_{s=k+1}^{T}\nonumber \\
 & \quad\left|K_{2}\left(\frac{\left(\left(r+1\right)n_{T}-\left(s-k/2\right)\right)/T}{\widehat{b}_{2,T}}\right)-K_{2}\left(\frac{\left(\left(r+1\right)n_{T}-\left(s-k/2\right)\right)/T}{b_{\theta_{2},T}}\right)\right|\nonumber \\
 & \quad\times\left(\left|\widehat{V}_{s}\widehat{V}_{s-k}-V_{s}V_{s-k}\right|+\left|V_{s}V_{s-k}-\mathbb{E}\left(V_{s}V_{s-k}\right)\right|+\left|\mathbb{E}\left(V_{s}V_{s-k}\right)\right|\right)\nonumber \\
 & \triangleq H_{1,T}+H_{2,T}+H_{3,T}.\nonumber 
\end{align}
We have to show that $H_{1,T}+H_{2,T}+H_{3,T}\overset{\mathbb{P}}{\rightarrow}0$.
By a mean-value expansion,  using \eqref{Eq. K2-K2 for part (i)},
\begin{align*}
\left|H_{1,T}\right| & \leq CT^{-1/2}\sum_{k=-T+1}^{T-1}\left|K_{1}\left(b_{\theta_{1},T}k\right)\right|\\
 & \quad\times\frac{n_{T}}{T}\sum_{r=0}^{\left\lfloor T/n_{T}\right\rfloor }\frac{1}{Tb_{\theta_{2},T}}\sum_{s=k+1}^{T}\\
 & \quad\left|K_{2}\left(\frac{\left(\left(r+1\right)n_{T}-\left(s-k/2\right)\right)/T}{\widehat{b}_{2,T}}\right)-K_{2}\left(\frac{\left(\left(r+1\right)n_{T}-\left(s-k/2\right)\right)/T}{b_{\theta_{2},T}}\right)\right|\\
 & \quad\times\left\Vert V_{s}\left(\overline{\beta}\right)\frac{\partial}{\partial\beta}V_{s-k}\left(\overline{\beta}\right)+V_{s-k}\left(\overline{\beta}\right)\frac{\partial}{\partial\beta}V_{s}\left(\overline{\beta}\right)\right\Vert \sqrt{T}\left\Vert \widehat{\beta}-\beta_{0}\right\Vert \\
 & \leq Cb_{\theta_{2},T}^{-1}T^{-1/2}\sum_{k=-T+1}^{T-1}\left|K_{1}\left(b_{\theta_{1},T}k\right)\right|\\
 & \quad\times\frac{n_{T}}{T}\sum_{r=0}^{\left\lfloor T/n_{T}\right\rfloor }CO_{\mathbb{P}}\left(1\right)\\
 & \quad\times\left(\left(T^{-1}\sum_{s=1}^{T}\sup_{\beta\in\Theta}V_{s}^{2}\left(\beta\right)\right)^{2}\left(T^{-1}\sum_{s=1}^{T}\sup_{\beta\in\Theta}\left\Vert \frac{\partial}{\partial\beta}V_{s}\left(\beta\right)\right\Vert ^{2}\right)^{1/2}\right)\sqrt{T}\left\Vert \widehat{\beta}-\beta_{0}\right\Vert .
\end{align*}
Using Assumption \ref{Assumption B} the right-hand side above is
\begin{align*}
C & T^{-1/2}b_{\theta_{2},T}^{-1}b_{\theta_{1},T}^{-1}b_{\theta_{1},T}\sum_{k=-T+1}^{T-1}\left|K_{1}\left(b_{\theta_{1},T}k\right)\right|\frac{n_{T}}{T}\sum_{r=0}^{\left\lfloor T/n_{T}\right\rfloor }O_{\mathbb{P}}\left(1\right)\overset{\mathbb{P}}{\rightarrow}0,
\end{align*}
 since $T^{-1/2}b_{\theta_{1},T}^{-1}b_{\theta_{2},T}^{-1}\rightarrow0$.
This shows $H_{1,T}\overset{\mathbb{P}}{\rightarrow}0$. Let $H_{2,1,T}$
(resp. $H_{2,2,T}$) be defined as $H_{2,T}$ but with the sum over
$k$ be restricted to $k=1,\ldots,\,S_{T}$ (resp. $k=S_{T}+1,\ldots,\,T$).
We have 
\begin{align}
\mathbb{E}\left(H_{2,1,T}^{2}\right) & \leq\sum_{k=1}^{S_{T}}\sum_{j=1}^{S_{T}}K_{1}\left(b_{\theta_{1},T}k\right)K_{1}\left(b_{\theta_{1},T}j\right)\label{Eq. (H21) =00003D 0}\\
 & \quad\times\left(\frac{n_{T}}{T}\right)^{2}\sum_{r_{1}=0}^{\left\lfloor T/n_{T}\right\rfloor }\sum_{r_{2}=0}^{\left\lfloor T/n_{T}\right\rfloor }\frac{1}{\left(Tb_{\theta_{2},T}\right)^{2}}\sum_{s=k+1}^{T}\sum_{t=j+1}^{T}\nonumber \\
 & \quad\times\left|K_{2}\left(\frac{\left(\left(r_{1}+1\right)n_{T}-\left(s-k/2\right)\right)/T}{\widehat{b}_{2,T}}\right)-K_{2}\left(\frac{\left(\left(r_{1}+1\right)n_{T}-\left(s-k/2\right)\right)/T}{b_{\theta_{2},T}}\right)\right|\nonumber \\
 & \quad\times\left|K_{2}\left(\frac{\left(\left(r_{2}+1\right)n_{T}-\left(t-j/2\right)\right)/T}{\widehat{b}_{2,T}}\right)-K_{2}\left(\frac{\left(\left(r_{2}+1\right)n_{T}-\left(t-j/2\right)\right)/T}{b_{\theta_{2},T}}\right)\right|\nonumber \\
 & \quad\times\left|\left(V_{s}V_{s-k}-\mathbb{E}\left(V_{s}V_{s-k}\right)\right)\left(V_{t}V_{t-k}-\mathbb{E}\left(V_{t}V_{t-k}\right)\right)\right|\nonumber \\
 & \leq CS_{T}^{2}\left(Tb_{\theta_{2},T}\right)^{-1}\sup_{k\geq1}Tb_{\theta_{2},T}\mathrm{Var}\left(\widetilde{\Gamma}\left(k\right)\right)O_{\mathbb{P}}\left(1\right)\nonumber \\
 & \leq CT^{r/3}O_{\mathbb{P}}\left(T^{-1}b_{\theta_{2},T}^{-1}\right)\rightarrow0,\nonumber 
\end{align}
where we have used Lemma S.B.5 in \citet{casini_hac}, \eqref{Eq. K2-K2 for part (i)}
and $r<5/2$. Turning to $H_{2,2,T},$ 
\begin{align}
\mathbb{E}\left(H_{2,2,T}^{2}\right) & \leq\left(Tb_{\theta_{2},T}\right)^{-1}b_{\theta_{1},T}^{-2b}\left(\sum_{k=S_{T}+1}^{T-1}k^{-b}\sqrt{Tb_{\theta_{2},T}}\left(\mathrm{Var}\left(\widetilde{\Gamma}\left(k\right)\right)\right)^{1/2}O\left(1\right)\right)^{2}\label{Eq. (H22) =00003D0}\\
 & \leq T^{-1}b_{\theta_{2},T}^{-1}b_{\theta_{1},T}^{-2b}\left(\sum_{k=S_{T}+1}^{T-1}k^{-b}\sqrt{Tb_{\theta_{2},T}}\left(\mathrm{Var}\left(\widetilde{\Gamma}\left(k\right)\right)\right)^{1/2}\right)^{2}\nonumber \\
 & \leq T^{-1}b_{\theta_{2},T}^{-1}b_{\theta_{1},T}^{-2b}\left(\sum_{k=S_{T}+1}^{T-1}k^{-b}O\left(1\right)\right)^{2}\nonumber \\
 & \leq T^{-1}b_{\theta_{2},T}^{-1}b_{\theta_{1},T}^{-2b}S_{T}^{2\left(1-b\right)}\rightarrow0,\nonumber 
\end{align}
since $r>\left(2b-5\right)/2\left(b-1\right).$ Eq. \eqref{Eq. (H21) =00003D 0}
and \eqref{Eq. (H22) =00003D0} yield $H_{2,T}\overset{\mathbb{P}}{\rightarrow}0.$
 Given $\left|K_{1}\left(\cdot\right)\right|\leq1$ and \eqref{Eq. K2-K2 for part (i)},
we have 
\begin{align*}
\left|H_{3,T}\right| & \leq C\sum_{k=-\infty}^{\infty}\left|\Gamma_{T}\left(k\right)\right|o_{\mathbb{P}}\left(1\right)\rightarrow0.
\end{align*}
This concludes the proof of part (i).

The result of part (iii) follows from the same argument as in Theorem
\ref{Theorem 1 -Consistency and Rate}-(iii) with references to Theorem
\ref{Theorem 1 -Consistency and Rate}-(i,ii) changed to Theorem\textit{
}\ref{Theorem 3 Andrews 91}-(i,ii). $\square$

\subsection{Proof of the Results in Section \ref{Section LRV with nonparametric rate of convergence}}

\subsubsection{Proof of Theorem \ref{Theorem 3 Andrews91}}

We begin with the following lemma which extends Theorem 1 in \citet{andrews:91}
to the present setting. Let $\widetilde{J}_{\mathrm{Cla},T}$ denote
the estimator that uses $\{V_{t}(\beta_{0})\}$ in place of $\{\widehat{V}_{t}\}$
and let $\widetilde{J}_{\mathrm{Cla},T}\left(\beta\right)$ denote
the estimator calculated using $\{V_{t}(\beta)\}$. Let $b_{\mathrm{Cla},\theta_{1},T}=(qK_{1,q}^{2}\alpha_{\theta}T/\int K_{1}^{2}\left(y\right)dy)^{-1/\left(2q+1\right)}$
where $\alpha_{\theta}\in\left(0,\,\infty\right)$. 
\begin{lem}
\label{Lemma: Theorem 1 in Andrews91}Suppose $K_{1}\left(\cdot\right)\in\boldsymbol{K}_{1}$
and $b_{1,T}\rightarrow0$. 

(i) If Assumption \ref{Assumption A - Dependence Andrews91}-\ref{Assumption B Andrews91}
hold and $T^{\vartheta}b_{1,T}\rightarrow\infty$, then $\widehat{J}_{\mathrm{Cla},T}-J_{T}\overset{\mathbb{P}}{\rightarrow}0$
and $\widehat{J}_{\mathrm{Cla},T}-\widetilde{J}_{\mathrm{Cla},T}\overset{\mathbb{P}}{\rightarrow}0$. 

(ii) If Assumption \ref{Assumption B Andrews91}-\ref{Assumption C Andrews 91 - Andrews91}
hold, $T^{1/2-2\vartheta}b_{1,T}^{-1/2}\rightarrow0$, $T^{-2\vartheta}b_{1,T}^{-1}\rightarrow0$
and $Tb_{1,T}^{2q+1}\rightarrow\gamma\in\left(0,\,\infty\right)$
for some $q\in\left(0,\,\infty\right)$ for which $K_{1,q},\,||f^{\left(q\right)}||<\infty$,
then $\sqrt{Tb_{1,T}}(\widehat{J}_{\mathrm{Cla,}T}-J_{T})=O_{\mathbb{P}}\left(1\right)$
and $\sqrt{Tb_{1,T}}(\widehat{J}_{\mathrm{Cla},T}-\widetilde{J}_{\mathrm{Cla},T})\overset{\mathbb{P}}{\rightarrow}0$. 

(iii) Under the conditions of part (b),
\begin{align*}
\lim_{T\rightarrow\infty}\mathrm{MSE}\left(Tb_{1,T},\,\widehat{J}_{\mathrm{Cla},T},\,W\right) & =\lim_{T\rightarrow\infty}\mathrm{MSE}\left(Tb_{1,T},\,\widetilde{J}_{\mathrm{Cla},T},\,W\right)\\
 & =4\pi^{2}\left(\gamma K_{1,q}^{2}\left(\mathrm{vec}f^{\left(q\right)}\right)'W\mathrm{vec}f^{\left(q\right)}+\int K_{1}^{2}\left(y\right)dy\,\mathrm{tr}W\left(I+C_{pp}\right)f\otimes f\right).
\end{align*}
\end{lem}
\noindent\textit{Proof of Lemma \ref{Lemma: Theorem 1 in Andrews91}}.
Using the same arguments as in Theorem 1 of \citet{andrews:91}, we
have $\widetilde{J}_{\mathrm{Cla},T}-J_{T}=o_{\mathbb{P}}\left(1\right)$
and $\sqrt{Tb_{1,T}}(\widetilde{J}_{\mathrm{Cla},T}-J_{T})=O_{\mathbb{P}}\left(1\right)$.
Thus, Lemma \ref{Lemma: Theorem 1 in Andrews91}-(i,ii) hold if the
second result stated in each of these parts holds. The latter hold
if and only if they hold with $\widehat{J}_{\mathrm{Cla},T}-\widetilde{J}_{\mathrm{Cla},T}$
replaced by $a'\widehat{J}_{\mathrm{Cla},T}a-a'\widetilde{J}_{\mathrm{Cla},T}a$
for arbitrary $a\in\mathbb{R}^{p}$. Thus, it is sufficient to consider
the scalar case. We now show that $T^{\vartheta}b_{1,T}(\widehat{J}_{\mathrm{Cla},T}-\widetilde{J}_{\mathrm{Cla},T})=O_{\mathbb{P}}\left(1\right)$
provided $b_{1,T}\rightarrow0$ and Assumption \ref{Assumption B Andrews91}
holds. This yields the second result of Lemma \ref{Lemma: Theorem 1 in Andrews91}-(i).
A mean-value expansion of $\widetilde{J}_{\mathrm{Cla},T}(\widehat{\beta}_{\mathrm{np}})(=\widehat{J}_{\mathrm{Cla},T})$
about $\beta_{0}$ yields 
\begin{align}
T^{\vartheta}b_{1,T}\left(\widehat{J}_{\mathrm{Cla},T}-\widetilde{J}_{\mathrm{Cla},T}\right) & =b_{1,T}\frac{\partial}{\partial\beta'}\widetilde{J}_{\mathrm{Cla},T}\left(\bar{\beta}\right)T^{\vartheta}\left(\widehat{\beta}_{\mathrm{np}}-\beta_{0}\right)\label{Eq. (A.9) in Andrews91}\\
 & =b_{1,T}\sum_{k=-T+1}^{T-1}K_{1}\left(b_{1,T}k\right)\frac{\partial}{\partial\beta'}\widehat{\Gamma}_{\mathrm{Cla}}\left(k\right)|_{\beta=\bar{\beta}}T^{\vartheta}\left(\widehat{\beta}_{\mathrm{np}}-\beta_{0}\right).\nonumber 
\end{align}
 \citet{andrews:91} showed that
\begin{align}
\sup_{k\geq1}\left\Vert \frac{\partial}{\partial\beta'}\widehat{\Gamma}_{\mathrm{Cla}}\left(k\right)\right\Vert |_{\beta=\bar{\beta}} & =O_{\mathbb{P}}\left(1\right).\label{Eq. (A.10) Andrews91}
\end{align}
 This result, Assumption \ref{Assumption B Andrews91}-(i), and the
fact that $b_{1,T}\sum_{k=-T+1}^{T-1}\left|K_{1}\left(b_{1,T}k\right)\right|\rightarrow\int_{-\infty}^{\infty}\left|K_{1}\left(y\right)\right|dy<\infty$
imply that the right-hand side of \eqref{Eq. (A.9) in Andrews91}
is $O_{\mathbb{P}}\left(1\right)$ and Lemma \ref{Lemma: Theorem 1 in Andrews91}-(i)
follows because $T^{\vartheta}b_{1,T}\rightarrow\infty.$

Next we show that $\sqrt{Tb_{1,T}}(\widehat{J}_{\mathrm{Cla},T}-\widetilde{J}_{\mathrm{Cla},T})=o_{\mathbb{P}}\left(1\right)$
under the assumptions of Lemma \ref{Lemma: Theorem 1 in Andrews91}-(ii).
A second-order Taylor expansion gives 
\begin{align*}
\sqrt{Tb_{1,T}} & \left(\widehat{J}_{\mathrm{Cla},T}-\widetilde{J}_{\mathrm{Cla},T}\right)\\
 & =\left[\sqrt{b_{1,T}}\frac{\partial}{\partial\beta'}\widetilde{J}_{\mathrm{Cla},T}\left(\beta_{0}\right)\right]T^{1/2-\vartheta}T^{\vartheta}\left(\widehat{\beta}_{\mathrm{np}}-\beta_{0}\right)\\
 & \quad+\frac{1}{2}\sqrt{T}\left(\widehat{\beta}_{\mathrm{np}}-\beta_{0}\right)'\left[\sqrt{b_{1,T}}\frac{\partial^{2}}{\partial\theta\partial\theta'}\widetilde{J}_{T}\left(\overline{\beta}\right)/\sqrt{T}\right]\sqrt{T}\left(\widehat{\beta}_{\mathrm{np}}-\beta_{0}\right)\\
 & =G_{T}T^{1/2-\vartheta}T^{\vartheta}\left(\widehat{\beta}_{\mathrm{np}}-\beta_{0}\right)+\frac{1}{2}T^{1/2-\vartheta}T^{\vartheta}\left(\widehat{\beta}_{\mathrm{np}}-\beta_{0}\right)'H{}_{T}T^{1/2-\vartheta}T^{\vartheta}\left(\widehat{\beta}_{\mathrm{np}}-\beta_{0}\right),
\end{align*}
where $G_{T}$ $\left(\in\mathbb{R}^{p}\right)$ and $H_{T}$ $\left(\in\mathbb{R}^{p}\right)$
are defined implicitly. Assumption \ref{Assumption B Andrews91}-(ii,iii),
\ref{Assumption C Andrews 91 - Andrews91}-(ii) and simple manipulations
yield
\begin{align}
T^{1-2\vartheta}\left\Vert H_{T}\right\Vert  & =T^{\left(1-2\vartheta\right)}\left(\frac{b_{1,T}}{T}\right)^{1/2}\sum_{k=-T+1}^{T-1}\left|K_{1}\left(b_{1,T}k\right)\right|\frac{1}{T}\sum_{t=\left|k\right|+1}^{T}\sup_{\beta\in\Theta}\left\Vert \frac{\partial^{2}}{\partial\beta\partial\beta'}V_{t}\left(\beta\right)V_{t-\left|k\right|}\left(\beta\right)\right\Vert \label{Eq. (12) in Andrews91}\\
 & =T^{\left(1-2\vartheta\right)}\left(\frac{1}{Tb_{1,T}}\right)^{1/2}\left(b_{1,T}\sum_{k=-T+1}^{T-1}\left|K_{1}\left(b_{1,T}k\right)\right|\right)O_{\mathbb{P}}\left(1\right)=o_{\mathbb{P}}\left(1\right),\nonumber 
\end{align}
 since $T^{1/2-2\vartheta}b_{1,T}^{-1/2}\rightarrow0.$  \citet{andrews:91}
showed that $G_{T}=o_{\mathbb{P}}\left(1\right)$. Thus, we have to
show that $T^{1/2-\vartheta}G_{T}=o_{\mathbb{P}}\left(1\right).$
Let 
\begin{align*}
D_{T} & =\sqrt{b_{1,T}}\sum_{k=-T+1}^{T-1}K_{1}\left(b_{1,T}k\right)\frac{1}{T}\sum_{t=\left|k\right|+1}^{T}\left(V_{t}+V_{t-\left|k\right|}\right).
\end{align*}
Using eq. (A.13) in \citet{andrews:91} we have 
\begin{align*}
T^{1-2\vartheta}\mathbb{E}\left(D_{T}^{2}\right) & \leq T^{1-2\vartheta}b_{1,T}\sum_{k=-T+1}^{T-1}\sum_{j=-T+1}^{T-1}\left|K_{1}\left(b_{1,T}k\right)K_{1}\left(b_{1,T}j\right)\right|\frac{4}{T^{2}}\sum_{s=1}^{T}\sum_{t=1}^{T}\left|\mathbb{E}\left(V_{s}V_{t}\right)\right|\\
 & \leq T^{1-2\vartheta}\frac{1}{Tb_{1,T}}\left(b_{1,T}\sum_{k=-T+1}^{T-1}\left|K_{1}\left(b_{1,T}k\right)\right|\right)^{2}\sum_{u=-T+1}^{T-1}\left|\Gamma\left(u\right)\right|\\
 & =T^{1-2\vartheta}\frac{1}{Tb_{1,T}}O_{\mathbb{P}}\left(1\right),
\end{align*}
since $T^{-2\vartheta}b_{1,T}^{-1}\rightarrow0.$ This concludes
the proof of part (ii). The proof of part (iii) of the lemma follows
the same argument as in the corresponding proof in \citet{andrews:91}.
$\square$

\bigskip{}

\noindent\textit{Proof of Theorem \ref{Theorem 3 Andrews91}}. By
Lemma \ref{Lemma: Theorem 1 in Andrews91}-(i) $\widehat{J}_{\mathrm{Cla},T}(b_{\mathrm{Cla},\theta_{1},T})-J_{T}=o_{\mathbb{P}}\left(1\right)$,
since $q>\left(1/\vartheta-1\right)/2$ implies $T^{\vartheta}b_{1,T}\rightarrow\infty$.
Hence, it suffices to show that $\widehat{J}_{\mathrm{Cla},T}(\widehat{b}_{\mathrm{Cla},1,T})-\widehat{J}_{\mathrm{Cla},T}(b_{\mathrm{Cla},\theta_{1},T})=o_{\mathbb{P}}\left(1\right)$.
Let $S_{T}=\bigl\lfloor(b_{\mathrm{Cla},\theta_{1},T})^{-r}\bigr\rfloor$
with 
\begin{align*}
r & \in\left(\max\left(\left(b-q-1/2\right)/\left(b-1\right),\,1-\left(2q-1\right)/\left(2b-2\right)\right),\,\min\left(1,\,3/4+q/2\right)\right).
\end{align*}
 We have
\begin{align}
\widehat{J}_{\mathrm{Cla},T}(\widehat{b}_{\mathrm{Cla},1,T})-\widehat{J}_{\mathrm{Cla},T}(b_{\mathrm{Cla},\theta_{1},T}) & =2\sum_{k=1}^{S_{T}}\left(K_{1}\left(\widehat{b}_{\mathrm{Cla},1,T}k\right)-K_{1}\left(b_{\mathrm{Cla},\theta_{1},T}k\right)\right)\widehat{\Gamma}_{\mathrm{Cla}}\left(k\right)\label{Eq. (A.23) Andrews91}\\
 & \quad+2\sum_{k=S_{T}+1}^{T-1}K_{1}\left(\widehat{b}_{\mathrm{Cla},1,T}k\right)\widehat{\Gamma}_{\mathrm{Cla}}\left(k\right)\nonumber \\
 & \quad-2\sum_{k=S_{T}+1}^{T-1}K_{1}\left(b_{\mathrm{Cla},\theta_{1},T}k\right)\widehat{\Gamma}_{\mathrm{Cla}}\left(k\right)\nonumber \\
 & =2A_{1,T}+2A_{2,T}-2A_{3,T}.\nonumber 
\end{align}
 We show $A_{1,T}\overset{\mathbb{P}}{\rightarrow}0$ as follows.
Using the Lipschitz condition on $K_{1}\left(\cdot\right)$, 
\begin{align}
\left|A_{1,T}\right| & \leq\sum_{k=1}^{S_{T}}C_{2}\left|\widehat{b}_{\mathrm{Cla},1,T}-b_{\mathrm{Cla},\theta_{1},T}\right|k\left|\widehat{\Gamma}_{\mathrm{Cla}}\left(k\right)\right|\label{Eq. (A.24) Andrews91}\\
 & \leq C\left|\widehat{\alpha}\left(q\right)^{1/\left(2q+1\right)}-\alpha_{\theta}^{1/\left(2q+1\right)}\right|\left(\widehat{\alpha}\left(q\right)\alpha_{\theta}\right)^{-1/\left(2q+1\right)}T^{-1/\left(2q+1\right)}\sum_{k=1}^{S_{T}}k\left|\widehat{\Gamma}_{\mathrm{Cla}}\left(k\right)\right|,\nonumber 
\end{align}
 for some constant $C<\infty$. By Assumption \ref{Assumption E Andrews91},
\begin{align*}
\left|\widehat{\alpha}\left(q\right)^{1/\left(2q+1\right)}-\alpha_{\theta}^{1/\left(2q+1\right)}\right|\left(\widehat{\alpha}\left(q\right)\alpha_{\theta}\right)^{-1/\left(2q+1\right)} & =O_{\mathbb{P}}\left(1\right),
\end{align*}
and so it suffices to show that $B_{1,T}+B_{2,T}+B_{3,T}\overset{\mathbb{P}}{\rightarrow}0$,
where
\begin{align}
B_{1,T} & =T^{-1/\left(2q+1\right)}\sum_{k=1}^{S_{T}}k\left|\widehat{\Gamma}_{\mathrm{Cla}}\left(k\right)-\widetilde{\Gamma}_{\mathrm{Cla}}\left(k\right)\right|\label{Eq. (A.25) Andrews91}\\
B_{2,T} & =T^{-1/\left(2q+1\right)}\sum_{k=1}^{S_{T}}k\left|\widetilde{\Gamma}_{\mathrm{Cla}}\left(k\right)-\Gamma_{T}\left(k\right)\right|\nonumber \\
B_{3,T} & =T^{-1/\left(2q+1\right)}\sum_{k=1}^{S_{T}}k\left|\Gamma_{T}\left(k\right)\right|.\nonumber 
\end{align}
 By a mean-value expansion, we have 
\begin{align}
B_{1,T} & \leq T^{-1/\left(2q+1\right)-\vartheta}S_{T}\sum_{k=1}^{S_{T}}\left|\left(\frac{\partial}{\partial\beta'}\widehat{\Gamma}_{\mathrm{Cla}}\left(k\right)|_{\beta=\overline{\beta}}\right)T^{\vartheta}\left(\widehat{\beta}_{\mathrm{np}}-\beta_{0}\right)\right|\label{Eq. (A.26) Andrews91}\\
 & \leq CT^{-1/\left(2q+1\right)-\vartheta+2r/\left(2q+1\right)}\sup_{k\geq1}\left\Vert \frac{\partial}{\partial\beta}\widehat{\Gamma}_{\mathrm{Cla}}\left(k\right)|_{\beta=\overline{\beta}}\right\Vert T^{\vartheta}\left\Vert \widehat{\beta}_{\mathrm{np}}-\beta_{0}\right\Vert \overset{\mathbb{P}}{\rightarrow}0,\nonumber 
\end{align}
since $r<3/4+q/2$, $T^{\vartheta}||\widehat{\beta}_{\mathrm{np}}-\beta_{0}||=O_{\mathbb{P}}\left(1\right)$
by Assumption \ref{Assumption B Andrews91}, and $\sup_{k\geq1}||\left(\partial/\partial\beta\right)\widehat{\Gamma}_{\mathrm{Cla}}\left(k\right)|_{\beta=\overline{\beta}}||=O_{\mathbb{P}}\left(1\right)$
by \eqref{Eq. (A.10) Andrews91} and Assumption \ref{Assumption B Andrews91}-(ii,iii).
\citet{andrews:91} showed that $\mathbb{E}\left(B_{2,T}^{2}\right)\rightarrow0$
if $r<3/4+q/2$ and $B_{3,T}\rightarrow0$ if $r<1$. Altogether,
this yields $A_{1,T}\overset{\mathbb{P}}{\rightarrow}0$. Let $A_{2,T}=L_{1,T}+L_{2,T}+L_{3,T}$,
where 
\begin{align*}
L_{1,T} & =\sum_{k=S_{T}+1}^{T-1}K_{1}\left(\widehat{b}_{\mathrm{Cla},1,T}k\right)\left(\widehat{\Gamma}_{\mathrm{Cla}}\left(k\right)-\widetilde{\Gamma}_{\mathrm{Cla}}\left(k\right)\right)\\
L_{2,T} & =\sum_{k=S_{T}+1}^{T-1}K_{1}\left(\widehat{b}_{\mathrm{Cla},1,T}k\right)\left(\widetilde{\Gamma}_{\mathrm{Cla}}\left(k\right)-\Gamma_{T}\left(k\right)\right)\\
L_{3,T} & =\sum_{k=S_{T}+1}^{T-1}K_{1}\left(\widehat{b}_{\mathrm{Cla},1,T}k\right)\Gamma_{T}\left(k\right).
\end{align*}
We now show $A_{2,T}\overset{\mathbb{P}}{\rightarrow}0$. By a mean-value
expansion and the definition of $\boldsymbol{K}_{3,\mathrm{Cla}}$,
\begin{align}
\left|L_{1,T}\right| & =T^{-\vartheta}\sum_{k=S_{T}+1}^{T-1}C_{1}\left(\widehat{b}_{\mathrm{Cla},1,T}k\right)^{-b}\left|\left(\frac{\partial}{\partial\beta'}\widehat{\Gamma}_{\mathrm{Cla}}\left(k\right)\right)|_{\beta=\overline{\beta}}T^{\vartheta}\left(\widehat{\beta}_{\mathrm{np}}-\beta_{0}\right)\right|\label{Eq. (A.30) Andrews91}\\
 & =T^{-\vartheta+b/\left(2q+1\right)}\left(\sum_{k=S_{T}+1}^{\infty}k^{-b}\right)O_{\mathbb{P}}\left(1\right)\nonumber \\
 & =T^{-\vartheta+b/\left(2q+1\right)-\left(b-1\right)r/\left(2q+1\right)}O_{\mathbb{P}}\left(1\right)\rightarrow0,\nonumber 
\end{align}
 where the second equality uses \eqref{Eq. (A.10) Andrews91} and
Assumption \ref{Assumption B Andrews91}, and the convergence to zero
follows from $r>\left(b-q-1/2\right)/\left(b-1\right)$. Further,
\citet{andrews:91} showed that $L_{2,T}=o_{\mathbb{P}}\left(1\right)$
if $r>1-\left(2q-1\right)/\left(2b-2\right)$. Using $\left|K_{1}\left(\cdot\right)\right|\leq1$
we have $L_{3,T}\leq\sum_{k=S_{T}+1}^{T-1}\left|\Gamma\left(k\right)\right|\rightarrow0$.
Thus, $A_{2,T}\overset{\mathbb{P}}{\rightarrow}0$. and an analogous
argument yields $A_{3,T}\overset{\mathbb{P}}{\rightarrow}0$. Combined
with $A_{1,T}\overset{\mathbb{P}}{\rightarrow}0$, the proof of Theorem
\ref{Theorem 3 Andrews91} is completed. $\square$

\subsubsection{Proof of Theorem \ref{Theorem 3 DK-HAC Nonparametric}}

We begin with the following lemma which extends Theorem \ref{Theorem 1 -Consistency and Rate}
to the present setting.

\begin{lem}
\label{Lemma 1 -Consistency and Rate- Nonparametric}Suppose $K_{1}\left(\cdot\right)\in\boldsymbol{K}_{1}$,
$K_{2}\left(\cdot\right)\in\boldsymbol{K}_{2}$, $b_{1,T},\,b_{2,T}\rightarrow0$,\textbf{
}$n_{T}\rightarrow\infty,\,n_{T}/Tb_{1,T}\rightarrow0,$ and $1/Tb_{1,T}b_{2,T}\rightarrow0$.
We have:

(i) If Assumption \ref{Assumption Smothness of A (for HAC)}-\ref{Assumption A - Dependence}
and \ref{Assumption B' Nonparametric} hold, $T^{\vartheta}b_{1,T}\rightarrow\infty$,
$b_{2,T}/b_{1,T}\rightarrow\nu\in[0,\,\infty)$ then $\widehat{J}_{T}-J_{T}\overset{\mathbb{P}}{\rightarrow}0$
and $\widehat{J}_{T}-\widetilde{J}_{T}\overset{\mathbb{P}}{\rightarrow}0$. 

(ii) If Assumption \ref{Assumption Smothness of A (for HAC)}, \ref{Assumption B}-\ref{Assumption C Andrews 91}
hold, $T^{1/2-2\vartheta}b_{1,T}^{-1/2}\rightarrow0$, $T^{-2\vartheta}\left(b_{1,T}b_{2,T}\right)^{-1}\rightarrow0,$
$n_{T}/Tb_{1,T}^{q}\rightarrow0$, $1/Tb_{1,T}^{q}b_{2,T}\rightarrow0$,
$b_{2,T}^{2}/b_{1,T}^{q}\rightarrow\nu\in[0,\,\infty)$ and $Tb_{1,T}^{2q+1}b_{2,T}\rightarrow\gamma\in\left(0,\,\infty\right)$
for some $q\in[0,\,\infty)$ for which $K_{1,q},\,||\int_{0}^{1}f^{\left(q\right)}\left(u,\,0\right)du||\in[0,\,\infty)$,
then $\sqrt{Tb_{1,T}b_{2,T}}(\widehat{J}_{T}-J_{T})=O_{\mathbb{P}}\left(1\right)$
and $\sqrt{Tb_{1,T}}(\widehat{J}_{T}-\widetilde{J}_{T})=o_{\mathbb{P}}\left(1\right).$ 

(iii) Under the conditions of part (ii) with $\nu\in\left(0,\,\infty\right)$,
\begin{align*}
\lim_{T\rightarrow\infty}\mathrm{MSE}\left(Tb_{1,T}b_{2,T},\,\widehat{J}_{T},\,W_{T}\right)=\lim_{T\rightarrow\infty}\mathrm{MSE}\left(Tb_{1,T}b_{2,T},\,\widetilde{J}_{T},\,W\right) & .
\end{align*}
\end{lem}
\noindent\textit{Proof of Lemma \ref{Lemma 1 -Consistency and Rate- Nonparametric}.
}As in Theorem \ref{Theorem 1 -Consistency and Rate} $\widetilde{J}_{T}-J_{T}=o_{\mathbb{P}}\left(1\right)$.
Proceeding as in Theorem \ref{Theorem 1 -Consistency and Rate}-(ii),
we first show that $T^{\vartheta}b_{1,T}(\widehat{J}_{T}-\widetilde{J}_{T})=O_{\mathbb{P}}\left(1\right)$
under Assumption \ref{Assumption B' Nonparametric}. A mean-value
expansion of $\widehat{J}_{T}$ about $\beta_{0}$ yields 
\begin{align}
T^{\vartheta}b_{1,T}(\widehat{J}_{T}-\widetilde{J}_{T}) & =b_{1,T}\frac{\partial}{\partial\beta'}\widetilde{J}_{T}(\bar{\beta})T^{\vartheta}(\widehat{\beta}_{\mathrm{np}}-\beta_{0})\nonumber \\
 & =b_{1,T}\sum_{k=-T+1}^{T-1}K_{1}\left(b_{1,T}k\right)\frac{\partial}{\partial\beta'}\widehat{\Gamma}\left(k\right)|_{\beta=\bar{\beta}}T^{\vartheta}(\widehat{\beta}_{\mathrm{np}}-\beta_{0}).\label{eq (A.9) Andrews}
\end{align}
 Using (S.28) in \citet{casini_hac} we have
\begin{align*}
b_{1,T} & \sum_{k=T+1}^{T-1}K_{1}\left(b_{1,T}k\right)\frac{\partial}{\partial\beta'}\widehat{\Gamma}\left(k\right)|_{\beta=\bar{\beta}}T^{\vartheta}\left(\widehat{\beta}_{\mathrm{np}}-\beta_{0}\right)\\
 & \leq b_{1,T}\sum_{k=-T+1}^{T-1}\left|K_{1}\left(b_{1,T}k\right)\right|O_{\mathbb{P}}\left(1\right)\\
 & =O_{\mathbb{P}}\left(1\right),
\end{align*}
where the last equality uses $b_{1,T}\sum_{k=-T+1}^{T-1}\left|K_{1}\left(b_{1,T}k\right)\right|\rightarrow\int\left|K_{1}\left(y\right)\right|dy<\infty.$
This concludes the proof of part (i) of Lemma \ref{Lemma 1 -Consistency and Rate- Nonparametric}
because $T^{\vartheta}b_{1,T}\rightarrow\infty$. The next step is
to show that $\sqrt{Tb_{1,T}}(\widehat{J}_{T}-\widetilde{J}_{T})=o_{\mathbb{P}}\left(1\right)$
under the assumptions of Lemma \ref{Lemma 1 -Consistency and Rate- Nonparametric}-(ii).
A second-order Taylor expansion gives
\begin{align*}
\sqrt{Tb_{1,T}}\left(\widehat{J}_{T}-\widetilde{J}_{T}\right) & =\left[\sqrt{b_{1,T}}\frac{\partial}{\partial\beta'}\widetilde{J}_{T}\left(\beta_{0}\right)\right]\sqrt{T}\left(\widehat{\beta}_{\mathrm{np}}-\beta_{0}\right)\\
 & \quad+\frac{1}{2}T^{1/2-\vartheta}T^{\vartheta}\left(\widehat{\beta}_{\mathrm{np}}-\beta_{0}\right)'\left[\sqrt{b_{1,T}}\frac{\partial^{2}}{\partial\beta\partial\beta'}\widetilde{J}_{T}\left(\overline{\beta}\right)/\sqrt{T}\right]T^{1/2-\vartheta}T^{\vartheta}\left(\widehat{\beta}_{\mathrm{np}}-\beta_{0}\right)\\
 & \triangleq G_{T}'T^{1/2-\vartheta}T^{\vartheta}\left(\widehat{\beta}_{\mathrm{np}}-\beta_{0}\right)+\frac{1}{2}T^{1/2-\vartheta}T^{\vartheta}\left(\widehat{\beta}_{\mathrm{np}}-\beta_{0}\right)'H_{T}T^{1/2-\vartheta}T^{\vartheta}\left(\widehat{\beta}_{\mathrm{np}}-\beta_{0}\right).
\end{align*}
Using Assumption \ref{Assumption C Andrews 91}-(ii), \citet{casini_hac}
showed that 
\begin{align*}
\biggl\Vert & \frac{\partial^{2}}{\partial\beta\partial\beta'}\widehat{c}\left(rn_{T}/T,\,k\right)\biggr\Vert\biggl|_{\beta=\bar{\beta}}=O_{\mathbb{P}}\left(1\right),
\end{align*}
 and thus, 
\begin{align*}
T^{1-2\vartheta}\left\Vert H_{T}\right\Vert  & \leq T^{1-2\vartheta}\left(\frac{b_{1,T}}{T}\right)^{1/2}\sum_{k=-T+1}^{T-1}\left|K_{1}\left(b_{1,T}k\right)\right|\sup_{\beta\in\Theta}\left\Vert \frac{\partial^{2}}{\partial\beta\partial\beta'}\widehat{\Gamma}\left(k\right)\right\Vert \\
 & \leq T^{1-2\vartheta}\left(\frac{b_{1,T}}{T}\right)^{1/2}\sum_{k=-T+1}^{T-1}\left|K_{1}\left(b_{1,T}k\right)\right|O_{\mathbb{P}}\left(1\right)\\
 & \leq T^{1-2\vartheta}\left(\frac{1}{Tb_{1,T}}\right)^{1/2}b_{1,T}\sum_{k=-T+1}^{T-1}\left|K_{1}\left(b_{1,T}k\right)\right|O_{\mathbb{P}}\left(1\right)=o_{\mathbb{P}}\left(1\right),
\end{align*}
 since $T^{1/2-2\vartheta}b_{1,T}^{-1/2}\rightarrow0$. Next, we
want to show that $T^{1/2-\vartheta}G_{T}=o_{\mathbb{P}}\left(1\right)$.
Following \citet{casini_hac}, it is sufficient to prove $\mathbb{E}\left(A_{3}^{2}\right)\rightarrow0$
where
\begin{align*}
A_{3} & =T^{1/2-\vartheta}\sqrt{b_{1,T}}\sum_{k=-T+1}^{T-1}\left|K_{1}\left(b_{1,T}k\right)\right|\frac{n_{T}}{T}\sum_{r=0}^{T/n_{T}}\frac{1}{Tb_{2,T}}\\
 & \quad\times\sum_{s=k+1}^{T}\left|K_{2}^{*}\left(\frac{\left(\left(r+1\right)n_{T}-\left(s-k/2\right)\right)/T}{b_{2,T}}\right)\right|\left|\left(V_{s}+V_{s-k}\right)\right|.
\end{align*}
Using the same steps as in \citet{casini_hac},
\begin{align*}
\mathbb{E}\left(A_{3}^{2}\right) & \leq T^{1-2\vartheta}\frac{1}{Tb_{1,T}b_{2,T}}\left(b_{1,T}\sum_{k=-T+1}^{T-1}\left|K_{1}\left(b_{1,T}k\right)\right|\right)^{2}\int_{0}^{1}K_{2}^{2}\left(x\right)dx\int_{0}^{1}\sum_{h=-\infty}^{\infty}\left|c\left(u,\,h\right)\right|du=o\left(1\right),
\end{align*}
since $T^{-2\vartheta}\left(b_{1,T}b_{2,T}\right)^{-1}\rightarrow0$.
 This implies $G_{T}=o_{\mathbb{P}}\left(1\right)$. It follows that
$\sqrt{Tb_{1,T}}(\widehat{J}_{T}-\widetilde{J}_{T})=o_{\mathbb{P}}\left(1\right)$
which concludes the proof of part (ii) because $\sqrt{Tb_{1,T}b_{2,T}}(\widetilde{J}_{T}-J_{T})=O_{\mathbb{P}}\left(1\right)$
by Theorem \ref{Theorem MSE J}-(iii). Part (iii) follows from the
same argument used in the  proof of Theorem 3.2-(iii) in \citet{casini_hac}.
$\square$

\bigskip{}

\noindent\textit{Proof of Theorem \ref{Theorem 3 DK-HAC Nonparametric}}.
Without loss of generality, we assume that $V_{t}$ is a scalar. The
constant $C<\infty$ may vary from line to line. By Lemma \ref{Lemma 1 -Consistency and Rate- Nonparametric}-(i),
$\widehat{J}_{T}(b_{\theta_{1},T},\,b_{\theta_{2},T})-J_{T}=o_{\mathbb{P}}\left(1\right)$.
It remains to establish $\widehat{J}_{T}(\widehat{b}_{1,T},\,\widehat{b}_{2,T})-\widehat{J}_{T}(b_{\theta_{1},T},\,b_{\theta_{2},T})=o_{\mathbb{P}}\left(1\right).$
Let
\begin{align*}
r\in\left(\max\left\{ \left(2b-5\right)/2\left(b-1\right),\,\left(b-6\vartheta\right)/\left(b-1\right)\right\} ,\,\min\left(\left(3+6\vartheta\right)/2,\,7/4\right)\right) & ,
\end{align*}
  and $S_{T}=\bigl\lfloor b_{\theta_{1},T}^{-r}\bigr\rfloor$. We
will use the decomposition \eqref{eq (Decomposition J_T proof of Theorem 3 Andrews91)}
and $N_{1}$ and $N_{2}$ as defined after \eqref{eq (Decomposition J_T proof of Theorem 3 Andrews91)}.
Let us consider the first term on the right-hand side of \eqref{eq (Decomposition J_T proof of Theorem 3 Andrews91)},
\begin{align}
\widehat{J}_{T} & (\widehat{b}_{1,T},\,\widehat{b}_{2,T})-\widehat{J}_{T}(b_{\theta_{1},T},\,\widehat{b}_{2,T})\label{Eq. 23-2}\\
 & =\sum_{k\in N_{1}}(K_{1}(\widehat{b}_{1,T}k)-K_{1}(b_{\theta_{1},T}k))\widehat{\Gamma}\left(k\right)+\sum_{k\in N_{2}}K_{1}(\widehat{b}_{1,T}k)\widehat{\Gamma}\left(k\right)\nonumber \\
 & \quad-\sum_{k\in N_{2}}K_{1}(b_{\theta_{1},T}k)\widehat{\Gamma}\left(k\right)\nonumber \\
 & \triangleq A_{1,T}+A_{2,T}-A_{3,T}.\nonumber 
\end{align}
 We first show that $A_{1,T}\overset{\mathbb{P}}{\rightarrow}0$.
Let $A_{1,1,T}$ denote $A_{1,T}$ with the summation restricted over
positive integers $k$. Let $\widetilde{n}_{T}=\inf\{T/n_{3,T},\,\sqrt{n_{2,T}}\}$.
We can use the Lipschitz condition on $K_{1}\left(\cdot\right)\in\boldsymbol{K}_{3}$
to yield, 
\begin{align}
\left|A_{1,1,T}\right| & \leq\sum_{k=1}^{S_{T}}C_{1}\left|\widehat{b}_{1,T}-b_{\theta_{1},T}\right|k\left|\widehat{\Gamma}\left(k\right)\right|\label{Eq. 24-2}\\
 & \leq C\left|\widehat{\phi}_{1}^{1/24}-\phi_{1,\theta^{*}}^{1/24}\right|\left(\widehat{\phi}_{1}\phi_{1,\theta^{*}}\right)^{-1/24}T^{-1/6}\sum_{k=1}^{S_{T}}k\left|\widehat{\Gamma}\left(k\right)\right|,\nonumber 
\end{align}
for some $C<\infty$. By Assumption \ref{Assumption E-F-G}-(i)
($|\widehat{\phi}_{1}^{1/24}-\phi_{1,\theta^{*}}^{1/24}|\left(\widehat{\phi}_{1}\phi_{1,\theta^{*}}\right)^{-1/24}=O_{\mathbb{P}}\left(1\right)$)
and so it suffices to show that $B_{1,T}+B_{2,T}+B_{3,T}\overset{\mathbb{P}}{\rightarrow}0$
where 
\begin{align}
B_{1,T} & =T^{-1/6}\sum_{k=1}^{S_{T}}k\left|\widehat{\Gamma}\left(k\right)-\widetilde{\Gamma}\left(k\right)\right|,\label{Eq. A.25 Andrews 91-2}\\
B_{2,T} & =T^{-1/6}\sum_{k=1}^{S_{T}}k\left|\widetilde{\Gamma}\left(k\right)-\Gamma_{T}\left(k\right)\right|,\qquad\mathrm{and}\nonumber \\
B_{3,T} & =T^{-1/6}\sum_{k=1}^{S_{T}}k\left|\Gamma_{T}\left(k\right)\right|.\nonumber 
\end{align}
 By a mean-value expansion, we have 
\begin{align}
B_{1,T} & \leq T^{-\vartheta-1/6}\sum_{k=1}^{S_{T}}k\left|\left(\frac{\partial}{\partial\beta'}\widehat{\Gamma}\left(k\right)|_{\beta=\overline{\beta}}\right)T^{\vartheta}\left(\widehat{\beta}_{\mathrm{np}}-\beta_{0}\right)\right|\label{Eq. A.26-2}\\
 & \leq CT^{-\vartheta-1/6}T^{2r/6}\sup_{k\geq1}\left\Vert \frac{\partial}{\partial\beta'}\widehat{\Gamma}\left(k\right)|_{\beta=\overline{\beta}}\right\Vert T^{\vartheta}\left\Vert \widehat{\beta}_{\mathrm{np}}-\beta_{0}\right\Vert \nonumber \\
 & \leq CT^{-\vartheta-1/6+r/3}\sup_{k\geq1}\left\Vert \frac{\partial}{\partial\beta'}\widehat{\Gamma}\left(k\right)|_{\beta=\overline{\beta}}\right\Vert O_{\mathbb{P}}\left(1\right)\overset{\mathbb{P}}{\rightarrow}0,\nonumber 
\end{align}
since $r<\left(3+6\vartheta\right)/2$, $\sqrt{T}||\widehat{\beta}_{\mathrm{np}}-\beta_{0}||=O_{\mathbb{P}}\left(1\right)$,
and $\sup_{k\geq1}||\left(\partial/\partial\beta'\right)\widehat{\Gamma}\left(k\right)|_{\beta=\overline{\beta}}||=O_{\mathbb{P}}\left(1\right)$
using (S.28) in \citet{casini_hac} and Assumption \ref{Assumption B' Nonparametric}-(i-iii).
In addition, 
\begin{align}
\mathbb{E}\left(B_{2,T}^{2}\right) & \leq\mathbb{E}\left(T^{-1/3}\sum_{k=1}^{S_{T}}\sum_{j=1}^{S_{T}}kj\left|\widetilde{\Gamma}\left(k\right)-\Gamma_{T}\left(k\right)\right|\left|\widetilde{\Gamma}\left(j\right)-\Gamma_{T}\left(j\right)\right|\right)\label{Eq. A.27-2}\\
 & \leq b_{\theta_{2},T}^{-1}T^{-1/3-1}S_{T}^{4}\sup_{k\geq1}Tb_{\theta_{2},T}\mathrm{Var}\left(\widetilde{\Gamma}\left(k\right)\right)\nonumber \\
 & \leq b_{\theta_{2},T}^{-1}T^{-4/3}T^{4r/6}\sup_{k\geq1}Tb_{\theta_{2},T}\mathrm{Var}\left(\widetilde{\Gamma}\left(k\right)\right)\rightarrow0,\nonumber 
\end{align}
 given that $\sup_{k\geq1}Tb_{\theta_{2},T}\mathrm{Var}(\widetilde{\Gamma}(k))=O\left(1\right)$
using Lemma S.B.5 in \citet{casini_hac} and $r<7/4$. Assumption
\ref{Assumption E-F-G}-(iii) and $\sum_{k=1}^{\infty}k^{1-l}<\infty$
for $l>2$ yield, 
\begin{align}
B_{3,T} & \leq T^{-1/6}C_{3}\sum_{k=1}^{\infty}k^{1-l}\rightarrow0.\label{Eq. A.28-2}
\end{align}
  Combining \eqref{Eq. 24-2}-\eqref{Eq. A.28-2} we deduce that
$A_{1,1,T}\overset{\mathbb{P}}{\rightarrow}0$. The same argument
applied to $A_{1,T}$ where the summation now extends over negative
integers $k$ gives $A_{1,T}\overset{\mathbb{P}}{\rightarrow}0$.
Next, we show that $A_{2,T}\overset{\mathbb{P}}{\rightarrow}0$. Again,
we use the notation $A_{2,1,T}$ (resp., $A_{2,2,T}$) to denote $A_{2,T}$
with the summation over positive (resp., negative) integers. Let $A_{2,1,T}=L_{1,T}+L_{2,T}+L_{3,T}$,
where 

\begin{align}
L_{1,T} & =\sum_{k=S_{T}+1}^{T-1}K_{1}\left(\widehat{b}_{1,T}k\right)\left(\widehat{\Gamma}\left(k\right)-\widetilde{\Gamma}\left(k\right)\right),\label{Eq. A.29-2}\\
L_{2,T} & =\sum_{k=S_{T}+1}^{T-1}K_{1}\left(\widehat{b}_{1,T}k\right)\left(\widetilde{\Gamma}\left(k\right)-\Gamma_{T}\left(k\right)\right),\quad\mathrm{and}\nonumber \\
L_{3,T} & =\sum_{k=S_{T}+1}^{T-1}K_{1}\left(\widehat{b}_{1,T}k\right)\Gamma_{T}\left(k\right).\nonumber 
\end{align}
We have
\begin{align}
\left|L_{1,T}\right| & =T^{-\vartheta}\sum_{k=S_{T}+1}^{T-1}C_{1}\left(\widehat{b}_{1,T}k\right)^{-b}\left|\left(\frac{\partial}{\partial\beta'}\widehat{\Gamma}\left(k\right)\right)|_{\beta=\overline{\beta}}T^{\vartheta}\left(\widehat{\beta}_{\mathrm{np}}-\beta_{0}\right)\right|\label{Eq. (30)-2-1}\\
 & =T^{-\vartheta+b/6}\sum_{k=S_{T}+1}^{T-1}C_{1}k^{-b}\left|\left(\frac{\partial}{\partial\beta'}\widehat{\Gamma}\left(k\right)\right)|_{\beta=\overline{\beta}}T^{\vartheta}\left(\widehat{\beta}_{\mathrm{np}}-\beta_{0}\right)\right|\nonumber \\
 & =T^{-\vartheta+b/6+r\left(1-b\right)/6}\left|\left(\frac{\partial}{\partial\beta'}\widehat{\Gamma}\left(k\right)\right)|_{\beta=\overline{\beta}}T^{\vartheta}\left(\widehat{\beta}_{\mathrm{np}}-\beta_{0}\right)\right|\nonumber \\
 & =T^{-\vartheta+b/6+r\left(1-b\right)/6}O\left(1\right)O_{\mathbb{P}}\left(1\right),\nonumber 
\end{align}
 which converges to zero since $r>\left(b-6\vartheta\right)/\left(b-1\right)$.
The bound for $L_{2,T}$ remains the same as in the proof of Theorem
\ref{Theorem 3 Andrews 91}-(i) since $r>\left(2b-5\right)/2\left(b-1\right)$.
Similarly, $L_{3,T}\rightarrow0$ as in the proof of the aforementioned
theorem.  Altogether, we have $A_{2,T}\overset{\mathbb{P}}{\rightarrow}0$
and using the same steps $A_{3,T}\overset{\mathbb{P}}{\rightarrow}0$.
It remains to show that $(\widehat{J}_{T}(b_{\theta_{1},T},\,\widehat{b}_{2,T})-\widehat{J}_{T}(b_{\theta_{1},T},\,b_{\theta_{2},T}))\overset{\mathbb{P}}{\rightarrow}0$.
The proof is different from the proof of the same result in Theorem
\eqref{Theorem 3 Andrews 91}-(i) because Assumption \ref{Assumption B' Nonparametric}
replaces Assumption \ref{Assumption B}. We have \eqref{Eq. K2-K2 for part (i)}
and we have to show  $H_{1,T}+H_{2,T}+H_{3,T}\overset{\mathbb{P}}{\rightarrow}0$,
where $H_{i,T}$ ($i=1,\,2,\,3$) is defined in \eqref{Eq. (H1+H2+H3)}.
We have 
\begin{align*}
\left|H_{1,T}\right| & \leq CT^{-\vartheta}\sum_{k=-T+1}^{T-1}\left|K_{1}\left(b_{\theta_{1},T}k\right)\right|\\
 & \quad\times\frac{n_{T}}{T}\sum_{r=0}^{\left\lfloor T/n_{T}\right\rfloor }\frac{1}{Tb_{\theta_{2},T}}\sum_{s=k+1}^{T}\left|K_{2}\left(\frac{\left(\left(r+1\right)n_{T}-\left(s-k/2\right)\right)/T}{\widehat{b}_{2,T}}\right)-K_{2}\left(\frac{\left(\left(r+1\right)n_{T}-\left(s-k/2\right)\right)/T}{b_{\theta_{2},T}}\right)\right|\\
 & \quad\times\left\Vert V_{s}\left(\overline{\beta}\right)\frac{\partial}{\partial\beta}V_{s-k}\left(\overline{\beta}\right)+V_{s-k}\left(\overline{\beta}\right)\frac{\partial}{\partial\beta}V_{s}\left(\overline{\beta}\right)\right\Vert T^{\vartheta}\left\Vert \widehat{\beta}_{\mathrm{np}}-\beta_{0}\right\Vert \\
 & \leq Cb_{\theta_{2},T}^{-1}T^{-\vartheta}\sum_{k=-T+1}^{T-1}\left|K_{1}\left(b_{\theta_{1},T}k\right)\right|\\
 & \quad\times\frac{n_{T}}{T}\sum_{r=0}^{\left\lfloor T/n_{T}\right\rfloor }\left(CO_{\mathbb{P}}\left(1\right)\right)\left(\left(T^{-1}\sum_{s=1}^{T}\sup_{\beta\in\Theta}V_{s}^{2}\left(\beta\right)\right)^{2}\left(T^{-1}\sum_{s=1}^{T}\sup_{\beta\in\Theta}\left\Vert \frac{\partial}{\partial\beta}V_{s}\left(\beta\right)\right\Vert ^{2}\right)^{1/2}\right)T^{\vartheta}\left\Vert \widehat{\beta}_{\mathrm{np}}-\beta_{0}\right\Vert .
\end{align*}
Using Assumption \ref{Assumption B} the right-hand side above is
\begin{align*}
C & T^{-\vartheta}b_{\theta_{2},T}^{-1}b_{\theta_{1},T}^{-1}b_{\theta_{1},T}\sum_{k=-T+1}^{T-1}\left|K_{1}\left(b_{\theta_{1},T}k\right)\right|\frac{n_{T}}{T}\sum_{r=0}^{\left\lfloor T/n_{T}\right\rfloor }O_{\mathbb{P}}\left(1\right)\overset{\mathbb{P}}{\rightarrow}0,
\end{align*}
 since $T^{-\vartheta}b_{\theta_{1},T}^{-1}b_{\theta_{2},T}^{-1}\rightarrow0$.
This shows $H_{1,T}\overset{\mathbb{P}}{\rightarrow}0$. The proof
of $H_{2,T}+H_{3,T}\overset{\mathbb{P}}{\rightarrow}0$ remains the
same as that of Theorem \ref{Theorem 3 Andrews 91}-(i) because it
does not depend on $\widehat{\beta}_{\mathrm{np}}.$ $\square$

\clearpage{}

\end{singlespace}
\end{document}